\documentclass[12pt,a4paper]{article}
\usepackage{cite}
\def\dfrac#1#2{{\displaystyle{#1\over#2}}}
\def\vec#1#2{ \left( \begin{array}{c} {#1} \\ #2 \end{array} \right)}

\def\Rup#1{\mbox{$R\!\!\!\!\raise4pt\hbox{$^{{^{#1}}}$} $}}
\def\Rupp#1#2{\mbox{$\hbox{$R^{#1}$}\!\!\!\!\!\!
\!\raise4pt\hbox{$^{{^{#2}}}$} $}}
\def\pd#1#2{\displaystyle{\frac{\partial\,{\displaystyle{#1}}}{\partial
{\displaystyle{#2}}}}}

\begin{document}
\title{\bf Group analysis of hydrodynamic-type systems}
\author{\bf M.B. Sheftel$^*$}
\date{}
\maketitle
Subj-class: Mathematical Physics
\begin{center}
$^*$ Feza G\"{u}rsey Institute, PO Box 6, Cengelkoy, 81220
Istanbul, Turkey and Department of higher mathematics,
North-Western State Technical University, Millionnaya Str. 5,
191186, St. Petersburg, Russia. \hspace{3mm}
{\bf E-mail:} sheftel@gursey.gov.tr
\begin{abstract}
We study point and higher symmetries for the hydrodynamic-type systems with
two independent variables $t$ and $x$ with and without explicit dependence of
the equations on $t,x$. We consider those systems which possess an
infinite group of the hydrodynamic symmetries, establish existence conditions
for this property and, using it, derive linearizing transformations for these
systems. The recursion operators for symmetries are obtained and used for
constructing infinite series of exact solutions of the studied equations.
Higher symmetries, {\it i.e.} the Lie-B\"{a}cklund transformation groups,
are also studied and the interrelation between the existence conditions for
higher symmetries and recursion operators is established. More complete
results are obtained for two-component systems, though $n$-component systems
are also studied. In particular, we consider Hamiltonian and semi-Hamiltonian
systems.
\end{abstract}
\end{center}

\section{Introduction}

   In this paper we review our results of application of the group
analysis of differential equations to systems of the hydrodynamic type. This
concept includes a systematic study of point and higher symmetries,
recursion operators, Hamiltonian structures and the usage of
these results in order to obtain exact analytic solutions of these
equations.  Originally B.A.  Dubrovin and S.P. Novikov meant by
hydrodynamic-type systems quasi-linear systems of first order partial
differential equations which possessed Hamiltonian structure
\cite{dubnov83,dubnov89}. Here we consider a more general class of
hydrodynamic-type equations which specifically includes semi-Hamiltonian
equations of S.P. Tsarev \cite{tsar85,tsar91} and explicitly dependent on $t$
(or $x$) equations which are rich in symmetries. As a consequence of the last
property they admit linearization and as soon as existence conditions for
recursion operators are satisfied infinite series of their exact solutions
can be obtained.  Thus, they have as good integrability properties as the
Hamiltonian equations.

  Hydrodynamic-type systems describe various physical phenomena: gas
dynamics and hydrodynamics, magnetic hydrodynamics \cite{rozhd},
nonlinear-elasticity and phase-transition models \cite{olvnut},
chromatography and electrophoresis equations from physical chemistry and
biology \cite{pavl,fertsar}.

  Applications of another kind are obtained by a representation of
physically interesting higher order equations as integrability
conditions of hydrodynamic-type systems: the Euler and Poisson
equations of nonlinear acoustics \cite{gumnut},
the Born-Infeld equation of nonlinear electrodynamics \cite{anov}, systems of
relativistic-string equations \cite{men'}.  Modern
applications of hydrodynamic-type systems arise in the theory of averaging
nonlinear soliton equations \cite{dubnov83,dubnov89}.

  An attractive feature of hydrodynamic-type systems is the
possibility of a geometric formulation of their properties by close
analogy with the Hamilton-Jacobi  and  eikonal equations of mechanics
and optics.

Symmetry group analysis of these systems leads naturally and algorithmically
to associated differential-geometric structures: metric, connection,
curvature, curvilinear orthogonal coordinate systems and their
transformations. If the Hamiltonian structure exists, then it turns out to
be merely an aspect of this geometrical theory \cite{dubnov83,tsar85,tsar91}.
Thus, here we can see clearly a relation of
transformation groups to geometry which S. Lie and F. Klein meant by their
Erlangen program.

A purely differential-geometric theory does not solve the problem of
integrating \linebreak
hydrodynamic-type equations. We present here another approach to
this problem based on a systematic study of higher symmetries and
recursion operators \cite{sh93a,sh93b,sh94a,sh94b}. The essence of this
approach is the following. We pick out a class of those hydrodynamic-type
systems which possess an infinite set of hydrodynamic symmetries depending on
arbitrary solutions of a linear system of partial differential equations.
Formulae for the corresponding invariant solutions determine a linearizing
transformation, which reduces the problem of obtaining solutions of a
nonlinear hydrodynamic-type system to a problem of solving a linear system.
Then we may construct recursion operators which map symmetries again into
symmetries if we assume in addition their existence conditions.
They give rise to
recursions of hydrodynamic symmetries and, as a
consequence, lead to recursions for solutions of the corresponding
linear system. Thus, we obtain the recursion formulae which allow us to
multiply solutions of the original nonlinear hydrodynamic-type system. The
action of recursion operators on the hydrodynamic symmetries explicitly
dependent on $x,t$ gives rise to infinite series of higher symmetries.
It turns  out to be possible to find the corresponding invariant solutions
explicitly. Hence we obtain infinite series of exact solutions of the
hydrodynamic-type system. They are analogous to the similarity solutions
which are well-known in gas and fluid mechanics and therefore we may expect
that they correspond to a physically interesting behavior.

Symmetries, recursions, Hamiltonian structures and
exact solutions of two-component spatially one-dimensional
hydrody\-namic-type systems were studied by the author mainly
before S.P. Tsarev's publications \cite{sh82,sh83,sh86,sh93b}. Symmetries,
linearizing transformations and geometric theory of the multi-component
hydrodynamic-type systems explicitly dependent on x, t have been constructed
by the author in 1989 \cite{sh89a,sh89b}. Higher symmetries,
higher conservation laws, their interrelations and the theory of recursion
operators for multi-component hydrodynamic-type systems have been constructed
by the author in 1993, 1994 \cite{sh93a,sh94a,sh94b,sh95}.
Finally in the recent article \cite{grshwi} we further developed the theory
of integrability of the diagonal hydrodynamic-type systems with explicit
$t$ or $x$ dependence and presented a non-trivial example of such integrable
system. There we also clarified and formulated precisely the concept of
the hydrodynamic symmetries which are first order symmetries though being
neither point, nor contact ones.

 In section 2 we study two-component diagonal hydrodynamic-type systems with
explicit dependence on $t$ or $x$. We find an infinite group of their
hydrodynamic symmetries and establish their existence conditions. By
considering the corresponding invariant solutions we derive linearizing
transformations for these systems. Second order symmetries and first and
second order recursion operators are obtained together with their existence
conditions which gives rise to infinite series of exact solutions of the
studied systems.

 In section 3 the generalized gas-dynamics equations are studied which
include in particular the equations of the one-dimensional isoentropic gas
dynamics. We obtain all their hydrodynamic and higher symmetries up to the
third order inclusively. We construct first order recursion operators which
lead naturally to the Lax representation of these equations. The latter is
used for obtaining the explicit formulae for infinite series of their
invariant solutions.

 In section 4 a particular class of the two-component systems which may
depend explicitly on $t$ and possess
the Hamiltonian structure, the so-called separable Hamiltonian systems, is
studied. We obtain their hydrodynamic and higher symmetries, linearizing
transformation, second order recursion operator, the Lax representation and
explicit formulas for the infinite series of exact (invariant) solutions.

  In section 5 we consider the $n$-component diagonal hydrodynamic-type
systems with no explicit dependence on $t$ and $x$ which possess an
infinite-dimensional group of the hydrodynamic symmetries, the so-called
semi-Hamiltonian systems. We obtain their hydrodynamic symmetries together
with their existence conditions and the associated differential-geometric
structure. A linearizing transformation, first and second order recursion
operators, higher symmetries and infinite series of exact solutions are
constructed.

  Finally, in section 6 the $n$-component diagonal hydrodynamic systems with
explicit dependence on $t$ or $x$ are studied. The hydrodynamic symmetries
and linearizing transformations are obtained.

\section{Symmetries, recursions and invariant solutions \protect\newline for
two-component hydrody\-na\-mic-type systems}

  Here we study a two-component diagonal hydrodynamic-type system
which may depend explicitly on time or space variable t or x. It
is linear homogeneous in derivatives of the unknowns $s(x, t ),
r(x,t)$
\begin{equation} s_t=\phi(s,r,t) s_x, \qquad
r_t=\psi(s,r,t)r_x
\label{2.1}
\end{equation}
where the subscripts denote partial derivatives with respect to $t$ and
$x.$ Functions $\phi$ and $\psi$ are real-valued and smooth and satisfy
the following nondegeneracy conditions
\begin{equation}
\phi \neq
\psi, \qquad \phi_r(s,r,t)\psi_s(s,r,t) \neq 0
\label{2.2}
\end{equation}

Symmetries of system (\ref{2.1}) are generated by the Lie equations for
one-parameter Lie-B\"{a}cklund groups in a canonical representation
\cite{andibr,ibreng}
\begin{equation}
\begin{array}{c}
s_{\tau}=f(x,t,s,r,s_x,r_x,\dots,s_x^{(N)},r_x^{(N)}), \\
r_{\tau}=g(x,t,s,r,s_x,r_x,\dots,s_x^{(N)},r_x^{(N)}),
\end{array}
\label{2.3}
\end{equation}
\begin{equation}
x_{\tau}=t_{\tau}=0,
\label{2.4}
\end{equation}
which are compatible with system (\ref{2.1}). Here
$ s_x^{(N)}=\partial^N s/\partial x^N, \quad $ $r_x^{(N)}=\partial^N
r/\partial x^N $
and $\tau$ is a parameter of a symmetry group: $ s=s(x,t,\tau),
r=r(x,t,\tau).$

Compatibility conditions for systems (\ref{2.1}) and (\ref{2.3})
$ s_{\tau t}= s_{t \tau }, r_{\tau t}= r_{t \tau }$ take the form of
the determining equations for characteristics $(f,g)$ of symmetries
\begin{equation}
\begin{array}{c}
D_t[f] |_{(\ref{2.1})} -\phi D_x[f] -s_x ( \phi_s f + \phi_r g) =0, \\
D_t[g] |_{(\ref{2.1})} -\psi D_x[g] -r_x ( \psi_s f + \psi_r g) =0,
\end{array}
\label{2.5}
\end{equation}
where $D_t$ and $D_x$ are operators of the total derivatives with respect to
$t$ and $x$
\begin{equation}
\begin{array}{rl}
D_x=& \pd{ }{x} + s_x \pd{ }{s} + r_x \pd{ }{r} +
\sum\limits_{k=1}^{\infty}
\left(  s_x^{(k+1)} \pd{ }{ s_x^{(k)} } + r_x^{(k+1)} \pd{ }{ r_x^{(k)}}
\right), \\
D_t|_{(\ref{2.1})}=& \pd{ }{t} + \phi s_x \pd{ }{s} + \psi r_x \pd{ }{r}+
\\      & +\sum\limits_{k=1}^{\infty} \left(D_x^k[\phi s_x]
\pd{ }{ s_x^{(k)} } +
D_x^k[\psi r_x] \pd{ }{ r_x^{(k)} } \right)
\end{array}
\label{2.6}
\end{equation}
and $D_t$ is calculated with the use of system (\ref{2.1}).

\subsection{Hydrodynamic symmetries of diagonal systems with \protect\\
explicit temporal dependence}

For hydrodynamic symmetries put $N=1$ in the Lie equations (\ref{2.3})
\begin{equation}
s_{\tau}=f(x,t,s,r,s_x,r_x), \qquad
r_{\tau}=g(x,t,s,r,s_x,r_x),
\label{2.7}
\end{equation}
and in determining equations (\ref{2.5}). Analysis of equations (\ref{2.5})
produces the following results.
Define the functions
\begin{equation}
\Phi_r(s,r,t)   = \phi_r(s,r,t) / (\phi -\psi), \quad
\Theta_s(s,r,t)= \psi_r(s,r,t) / (\psi-\phi),
\label{2.8}
\end{equation}

\begin{equation}
\begin{array}{c}
\hat{\Phi}(s,r,t)   = b(s)\Phi_s(s,r,t)  + d(r) \Phi_r(s,r,t)
+\Phi_0(s),   \\
\hat{\Theta}(s,r,t)= b(s)\Theta_s(s,r,t) + d(r) \Theta_r(s,r,t)
+\Theta_0(r),
\end{array}
\label{2.9}
\end{equation}

\begin{equation}
\begin{array}{c}
\hat{\phi}(s,r,t)   = b(s)\phi_s(s,r,t)  + d(r) \phi_r(s,r,t),   \\
\hat{\psi}(s,r,t)= b(s)\psi_s(s,r,t) + d(r) \psi_r(s,r,t).
\end{array}
\label{2.10}
\end{equation}

\underline{\bf Definition 2.1.}  We call system (\ref{2.1}) a generic system
(with respect to hydrodynamic symmetries) if its coefficients $\phi,\psi$ do
not satisfy the constraints
\begin{equation}
\phi_t=\beta(t)\phi^2+\varepsilon(t) \phi + \lambda(t), \quad
\psi_t=\beta(t)\psi ^2+\varepsilon(t) \psi + \lambda(t)
\label{2.11}
\end{equation}
with arbitrary smooth functions $\beta(t),\varepsilon(t)
,\lambda(t).$

Results of the group analysis are formulated as the following basic
theorem \cite{sh93b}.

{\bf Theorem 2.1.}  A diagonal two-component generic hydrodynamic-type
system (\ref{2.1}) which may explicitly depend on $t$ possesses an
infinite set of hydrodynamic symmetries with a functional arbitrariness
iff the following two conditions are satisfied:

1. Coefficients $\phi,\psi$ of system (\ref{2.1}) satisfy the equalities
\begin{equation}
\Phi_{rt}=\beta\phi_r, \quad
\Theta_{st}=\beta\psi_s,
\label{2.12}
\end{equation}
where $\beta$ is arbitrary (real) constant, the functions
$\Phi(s,r,t),\Theta(s,r,t)$ are defined by equations (\ref{2.8}) and partial
derivatives with respect to $t$ are taken  at constant values of $s$ and
$r$;

2. There exist four functions of one variable
$b(s),d(r),\Phi_0(s),\Theta_0(r)$ which satisfy the equalities
\begin{equation}
\hat{\Phi}_r=\Phi_r(\hat{\Phi}-\hat{\Theta}), \quad
\hat{\Theta}_s=\Theta_s(\hat{\Theta} -\hat{\Phi})
\label{2.13}
\end{equation}
where the functions $\hat{\Phi}(s,r,t),\hat{\Theta}(s,r,t)$ are
defined by the formulae (\ref{2.9}).

These symmetries are generated by the Lie equations
\begin{equation}
\begin{array}{c}
s_{\tau}=f=\tilde\phi(x,t,s,r) s_x +b(s), \\
r_{\tau}=g=\tilde\psi(x,t,s,r) r_x +d(r),
\end{array}
\label{2.14}
\end{equation}
where functions $\tilde\phi, \tilde\psi$ are defined by the following
formulae

if $\beta \neq 0$
\begin{equation}
\begin{array}{c}
\tilde\phi(x,t,s,r)=a(s,r) \exp\left\{ \beta\left[
x+\int\limits_0^t\phi(s,r,t)dt\right]    \right\}+
\dfrac{1}{\beta}
\hat{\Phi}(s,r,t),  \\
\tilde\psi(x,t,s,r)= c(s,r) \exp\left\{ \beta\left[
x+\int\limits_0^t\psi(s,r,t)dt\right]    \right\}+
\dfrac{1}{\beta}
\hat{\Theta}(s,r,t),
\end{array}
\label{2.15}
\end{equation}

and if $\beta = 0$
\begin{equation}
\begin{array}{rl}
\tilde\phi(x,t,s,r)= & a(s,r) +\int\limits_0^t\hat{\phi}(s,r,t)dt-\\
    & -\hat{\Phi}(s,r) \left[ x +
\int\limits_0^t\phi(s,r,t)dt\right], \\
\tilde\psi(x,t,s,r)= & c(s,r) +\int\limits_0^t\hat{\psi}(s,r,t)dt-\\
    & -\hat{\Theta}(s,r) \left[ x +
\int\limits_0^t\psi(s,r,t)dt\right].
\end{array}
\label{2.16}
\end{equation}
Here the integrals with respect to $t$ are taken at constant values of
$s$
and $r$ and the functions $\hat{\phi},\hat{\psi}$ are defined by the
formulae (\ref{2.10}). The functions $a(s,r),c(s,r)$ constitute arbitrary
smooth solution of the linear system of equations
\begin{equation}
\begin{array}{c}
a_r(s,r)=\Phi_r(s,r,0)(a-c), \\
c_s(s,r)=\Theta_s(s,r,0)(c-a).
\end{array}
\label{2.17}
\end{equation}

{\it Remark 2.1.} We can use the freedom in the definition (\ref{2.8}) of the
functions $\Phi,\Theta$ to transform equations (\ref{2.12}) in the condition
1 of Theorem 2.1 to a more simple form
\begin{equation}
\Phi_t(s,r,t)=\beta\phi(s,r,t), \Theta_t(s,r,t)=\beta\psi(s,r,t).
\label{2.18}
\end{equation}

{\it Remark 2.2.} A solution manifold of linear system (\ref{2.17}) is
locally parametrized by two arbitrary functions $c_1(s),c_2(r)$ of one
variable. They determine a functional arbitrariness in the definitions
(\ref{2.15}), (\ref{2.16}) of hydrodynamic symmetries (\ref{2.14}) for system
(\ref{2.1}).

System (\ref{2.17}) always has a trivial solution $a(s,r)=c(s,r)=c_0=const.$

The condition 2 of Theorem 2.1 always has a trivial solution
\begin{equation}
\begin{array}{c}
 b(s)=d(r)=0, \qquad \Phi_0(s)=\Theta_0(r)=c_0=const,\\
\hat{\Phi}=\hat{\Theta}=c_0, \qquad \hat{\phi}=\hat{\psi}=0.
\end{array}
\label{2.19}
\end{equation}

{\tt Corollary 2.1.}  The condition 1 of Theorem 2.1 is necessary and
sufficient for system (\ref{2.1}) to possess an infinite set of hydrodynamic
symmetries, generated by the Lie equations
\begin{equation}
s_{\tau}=\tilde\phi(x,t,s,r)s_x, \qquad
r_{\tau}=\tilde\psi(x,t,s,r)r_x,
\label{2.20}
\end{equation}
which are linear homogeneous in derivatives. The coefficients
$\tilde\phi,
\tilde\psi $ of these equations are determined by the following
formulae

if $\beta \neq 0$
\begin{equation}
\begin{array}{c}
\tilde\phi(x,t,s,r)=a(s,r) \exp\left\{ \beta\left[
x+\int\limits_0^t\phi(s,r,t)dt\right]    \right\}+ c_0, \\
\tilde\psi(x,t,s,r)= c(s,r) \exp\left\{ \beta\left[
x+\int\limits_0^t\psi(s,r,t)dt\right]    \right\}+ c_0,
\end{array}
\label{2.21}
\end{equation}

and if $\beta = 0$
\begin{equation}
\begin{array}{c}
\tilde\phi(x,t,s,r)=a(s,r) + c_0 \left[ x +
\int\limits_0^t\phi(s,r,t)dt\right],
\\
\tilde\psi(x,t,s,r)=c(s,r) + c_0\left[ x +
\int\limits_0^t\psi(s,r,t)dt\right].
\end{array}
\label{2.22}
\end{equation}

{\it Remark 2.3.}  The condition 1 of Theorem 2.1 with $\beta=0$ is met
in particular for system 2.1 with the coefficients $\phi(s,r),
\psi(s,r)$ explicitly independent of $t$.  Such a system always has an
infinite set of hydrodynamic symmetries with a functional
arbitrariness. In this case coefficients 2.22 of Lie equations have the
form
\begin{equation}
\begin{array}{c}
\tilde\phi(x,t,s,r)=a(s,r) + c_0 \left[ x + t\phi(s,r)\right], \\
\tilde\psi(x,t,s,r)=c(s,r) + c_0\left[ x + t\psi(s,r) \right].
\end{array}
\label{2.23}
\end{equation}

{\it Remark 2.4.}  The condition 2 of Theorem 2.1 is that additional
constraint which provides the existence of symmetries with the Lie
equations (\ref{2.14}) linear inhomogeneous in derivatives. Every nontrivial
solution of equations (\ref{2.13}) generates such symmetries.

\subsection{Infinite-dimensional Lie algebra of hydrodynamic
symmetries and recursions of symmetries}

Let the system (\ref{2.1}) possess two one-parameter symmetry groups
generated by Lie equations (\ref{2.14}) with a parameter $\tau$ and by
Lie equations of the same form with a parameter $\bar\tau$
\begin{equation}
\begin{array}{c}
s_{ \bar\tau } =\bar f = \bar\phi(x,t,s,r)s_x + \bar{b}(s), \\
r_{ \bar\tau } =\bar g= \bar\psi(x,t,s,r)r_x + \bar{d}(r).
\end{array}
\label{2.24}
\end{equation}
Here the coefficients $\bar\phi,\bar\psi$ are determined by analogy
with the formulae (\ref{2.15}) or (\ref{2.16})

if $\beta \neq 0$
\begin{equation}
\begin{array}{c}
\bar\phi(x,t,s,r)= \bar{a}(s,r) \exp\left\{ \beta\left[
x+\int\limits_0^t\phi(s,r,t)dt\right]    \right\}+
\dfrac{1}{\beta}
\hat{\bar\Phi}(s,r,t),  \\
\bar\psi(x,t,s,r)= \bar{c}(s,r) \exp\left\{ \beta\left[
x+\int\limits_0^t\psi(s,r,t)dt\right]    \right\}+
\dfrac{1}{\beta}
\hat{\bar\Theta}(s,r,t),
\end{array}
\label{2.25}
\end{equation}

and if $\beta = 0$
\begin{equation}
\begin{array}{rl}
\bar\phi(x,t,s,r)=& \bar{a}(s,r)
+\int\limits_0^t\hat{\bar\phi}(s,r,t)dt-\\
      & -\hat{\bar\Phi}(s,r) \left[ x +
\int\limits_0^t\phi(s,r,t)dt\right],
\\
\bar\psi(x,t,s,r)=& \bar{c}(s,r)
+\int\limits_0^t\hat{\bar\psi}(s,r,t)dt-\\
      & -\hat{\bar\Theta}(s,r) \left[ x +
\int\limits_0^t\psi(s,r,t)dt\right].
\end{array}
\label{2.26}
\end{equation}

The functions $\hat{\bar\Phi},\hat{\bar\Theta}$ are defined by
analogy with the formulae (\ref{2.9})
\begin{equation}
\begin{array}{c}
\hat{\bar\Phi}   = \bar{b}(s)\Phi_s + \bar{d}(r)
\Phi_r +\bar\Phi_0(s),   \\
\hat{\bar\Theta}= \bar{b}(s)\Theta_s + \bar{d}(r)
\Theta_r +\bar\Theta_0(r),
\end{array}
\label{2.27}
\end{equation}
and the functions $\hat{\bar\phi},\hat{\bar\psi}$ are defined by
analogy with the formulae (\ref{2.10})
\begin{equation}
\hat{\bar\phi} = \bar{b}(s)\phi_s + \bar{d}(r)\phi_r, \quad
\hat{\bar\psi}= \bar{b}(s)\psi_s + \bar{d}(r) \psi_r.
\label{2.28}
\end{equation}

The functions $\hat{\bar\Phi},\hat{\bar\Theta}$ must satisfy the
condition 2 of the theorem 2.1
\begin{equation}
\hat{\bar\Phi}_r=\Phi_r(\hat{\bar\Phi}-\hat{\bar\Theta}),
\quad
\hat{\bar\Theta}_s=\Theta_s(\hat{\bar\Theta} -\hat{\bar\Phi})
\label{2.29}
\end{equation}

The functions $\bar{a}(s,r),\bar{c}(s,r)$ constitute arbitrary smooth
solution of the same linear system (\ref{2.17})
\begin{equation}
\begin{array}{c}
\bar{a}_r(s,r)=\Phi_r(s,r,0)(\bar{a}-\bar{c}), \\
\bar{c}_s(s,r)=\Theta_s(s,r,0)(\bar{c}-\bar{a}).
\end{array}
\label{2.30}
\end{equation}

Let $\sigma = (f,g)$ and $\bar\sigma = (\bar{f},\bar{g})$ be characteristics
of the symmetries (\ref{2.14}) and (\ref{2.24}). A canonical
Lie-B\"{a}cklund symmetry operator with the characteristic  $\sigma $ is
defined as follows~\cite{ibreng}
\begin{equation}
\begin{array}{c}
\hat{X}_{\sigma} = f \pd{ }{s} + g\pd{ }{r} +(D_t[f])\pd{ }{s_t} +
(D_t[g])\pd{ }{r_t} +
(D_x[f])\pd{ }{s_x} + \\ [2mm]
+(D_x[g])\pd{ }{r_x} + (D_x^2[f])\pd{ }{s_{xx}} +
(D_x^2[g])\pd{ }{r_{xx}} + \dots \\ [2mm]
\dots + (D_x^N[f])\pd{ }{s_x^{(N)}} +
(D_x^N[g])\pd{ }{r_x^{(N)}} + \dots
\end{array}
\label{2.31}
\end{equation}
Here operator $D_t$ of the total derivative with respect to $t$ is calculated
with the use of equations (\ref{2.1}). The formula for $\hat{X}_{\bar\sigma}$
is obtained by a substitution of $(\bar{f},\bar{g})$ for $(f,g)$ to the
formula (\ref{2.31}).

The usual Lie commutator
\begin{equation}
[\hat{X}_{\bar\sigma}, \; \hat{X}_{\sigma}]
=\hat{X}_{[\![\sigma,\bar\sigma]\!]} \equiv
\hat{X}_{\tilde\sigma}
\label{2.32}
\end{equation}
generates the commutator of symmetry characteristics
\begin{equation}
\tilde\sigma=[\![\sigma,\bar\sigma]\!]=\sigma'[\bar\sigma]-
\bar\sigma'[\sigma].
\label{2.33}
\end{equation}
Here $\sigma'$ denotes the operator of the Frech\'et derivative
\cite{fuchfok})
\begin{equation}
(\sigma')_{\alpha\beta}=\sum\limits_{j=0}^{\infty}
\pd{\sigma_{\alpha}}{u_x^{\beta(j)}} D_x^j \quad
(\alpha,\beta=1,2),
\label{2.34}
\end{equation}
where $\sigma_1=f,\sigma_2=g,u^1=s,u^2=r.$ The commutator (\ref{2.33}) of
symmetry characteristics for system (\ref{2.1}) is again a characteristic
$\tilde\sigma = (\tilde{f},\tilde{g})$ of some symmetry for this
system generated by the Lie equations with the parameter $\tilde\tau$:
$s_{\tilde\tau}=\tilde{f},r_{\tilde\tau}=\tilde{g}$ where
\begin{equation}
\begin{array}{rl}
\tilde{f}=[\![\sigma,\bar\sigma]\!]_1=&
\bar{f}\pd{f}{s} -f\pd{\bar{f}}{s} +\bar{g}\pd{f}{r} -g\pd{\bar{f}}{r}+
\\ [2mm]
  &+D_x[\bar{f}]\pd{f}{s_x} -D_x[f]\pd{\bar{f}}{s_x},
\end{array}
\label{2.35}
\end{equation}
\begin{equation}
\begin{array}{rl}
\tilde{g}=[\![\sigma,\bar\sigma]\!]_2=&
\bar{f}\pd{g}{s} -f\pd{\bar{g}}{s} +\bar{g}\pd{g}{r} -
g\pd{\bar{g}}{r}+\\ [2mm]
   &+D_x[\bar{g}]\pd{g}{r_x} -D_x[g]\pd{\bar{g}}{r_x},
\end{array}
\label{2.36}
\end{equation}

{\bf Theorem 2.2.} (see \cite{sh93b}).

A commutator of hydrodynamic symmetries (\ref{2.24}) and (\ref{2.14}) with
the characteristics $\bar\sigma$ and $\sigma$ is again a hydrodynamic
symmetry of system (\ref{2.1}) with the  characteristic $\tilde\sigma
=[\![\sigma,\bar\sigma]\!]= (\tilde{f},\tilde{g}),$ which is
generated by Lie equations
\begin{equation}
\begin{array}{c}
s_{ \tilde\tau } =\tilde f = \tilde{\bar\phi}(x,t,s,r)s_x +
\tilde{\bar{b}}(s), \\
r_{ \tilde\tau } =\tilde g= \tilde{\bar\psi}(x,t,s,r)r_x +
\tilde{\bar{d}}(r)
\end{array}
\label{2.37}
\end{equation}
where the coefficients are determined by the following formulae

if $\beta \neq 0$
\begin{equation}
\begin{array}{rl}
\tilde{\bar\phi}(x,t,s,r)=& \tilde{\bar{a}}(s,r) \exp\left\{
\beta\left[ x +\int\limits_0^t\phi(s,r,t)dt\right]  \right\}+
\\
  & +\dfrac{1}{\beta} \hat{\tilde\Phi}(s,r,t),  \\
\tilde{\bar\psi}(x,t,s,r)=& \tilde{\bar{c}}(s,r) \exp\left\{
\beta\left[ x+\int\limits_0^t\psi(s,r,t)dt\right]    \right\}+
\\
  & + \dfrac{1}{\beta} \hat{\tilde\Theta}(s,r,t),
\end{array}
\label{2.38}
\end{equation}

and if $\beta = 0$
\begin{equation}
\begin{array}{rl}
\tilde{\bar\phi}(x,t,s,r)=& \tilde{\bar{a}}(s,r)
+\int\limits_0^t\hat{\tilde\phi}(s,r,t)dt-      \\
    & -\hat{\tilde\Phi}(s,r) \left[ x +
\int\limits_0^t\phi(s,r,t)dt\right],
\\
\tilde{\bar\psi}(x,t,s,r)=& \tilde{\bar{c}}(s,r)
+\int\limits_0^t\hat{\tilde\psi}(s,r,t)dt-\\
    & -\hat{\tilde\Theta}(s,r) \left[ x +
\int\limits_0^t\psi(s,r,t)dt\right]
\end{array}
\label{2.39}
\end{equation}
and for any value of $\beta$
\begin{equation}
 \tilde{\bar{b}}(s)=\left| \begin{array}{cc}  \bar{b}(s) & b(s) \\
 \bar{b'}(s) & b'(s) \end{array}  \right|, \quad
\tilde{\bar{d}}(r)=\left| \begin{array}{cc}  \bar{d}(r) & d(r) \\
 \bar{d'}(r) & d'(r) \end{array}  \right|.
\label{2.40}
\end{equation}
Here the following notation is used
\begin{equation}
\begin{array}{c}
\hat{\tilde\phi}(s,r,t)= \tilde{\bar{b}}(s)\phi_s(s,r,t)  +
\tilde{\bar{d}}(r) \phi_r(s,r,t),   \\
\hat{\tilde\psi}(s,r,t)= \tilde{\bar{b}}(s)\psi_s(s,r,t) +
\tilde{\bar{d}}(r) \psi_r(s,r,t).
\end{array}
\label{2.41}
\end{equation}
\begin{equation}
\begin{array}{c}
\hat{\tilde\Phi}(s,r,t)=
\bar{b}(s)\hat\Phi_s + \bar{d}(r) \hat{\Phi}_r -
b(s)\hat{\bar\Phi}_s -d(r)
\hat{\bar\Phi}_r,   \\
\hat{\tilde\Theta}(s,r,t)=
\bar{b}(s)\hat\Theta_s + \bar{d}(r)\hat\Theta_r -
b(s)\hat{\bar\Theta}_s
-d(r)\hat{\bar\Theta}_r,
\end{array}
\label{2.42}
\end{equation}
\begin{equation} \begin{array}{rl}
\tilde{\bar{a}}(s,r)=&
\bar{b}(s) [a_s(s,r)-\Phi_s a(s,r)] -b(s)[\bar{a}_s(s,r)-\Phi_s
\bar{a}(s,r)]  + \\
& + \Phi_r [d(r)\bar{c}(s,r)-\bar{d}(r) c(s,r)] +
\Phi_0(s)\bar{a}(s,r)-\bar{\Phi}_0(s) a(s,r), \\ [4mm]
\tilde{\bar{c}}(s,r)=&
\bar{d}(r) [c_r(s,r)-\Theta_r c(s,r)] -d(r)[\bar{c}_r(s,r)-\Theta_r
\bar{c}(s,r)]  +
\\
& + \Theta_s [b(s)\bar{a}(s,r)-\bar{b}(s) a(s,r)] +
\Theta_0(r)\bar{c}(s,r)-\bar{\Theta}_0(r) c(s,r),
\end{array}
\label{2.43}
\end{equation}

$\Phi=\Phi(s,r,0),\Theta=\Theta(s,r,0)$ if $\beta \neq 0$ and
$\Phi=\Phi(s,r),\Theta=\Theta(s,r)$ if $\beta =0.$

{\tt Corollary 2.2.}  The formulae (\ref{2.42}) and (\ref{2.43}) determine
recursions of solutions for the linear systems (\ref{2.13}) and (\ref{2.17})
respectively.  They map any pair of solutions of the corresponding system
again into its solution.

Consider now a special case
\begin{equation}
b(s)=d(r)=0, \quad
\hat\Phi=\hat\Theta=c_0=0, \quad
(\bar{b}(s),\bar{d}(r)) \neq (0,0).
\label{2.44}
\end{equation}

{\tt Corollary 2.3.}  Let the condition 1 of the theorem 2.1 be satisfied
for the system (\ref{2.1}) and inhomogeneous in derivatives hydrodynamic
symmetry of the form (\ref{2.24}) exist. Then the Lie commutator of the
symmetry (\ref{2.24}) and a homogeneous in derivatives hydrodynamic symmetry
of the form (\ref{2.20})--(\ref{2.22}) with $c_0=0$ is again a homogeneous
symmetry of the same form

\begin{equation}
\begin{array}{c}
s_{ \tilde\tau } =\tilde f = \tilde{\bar\phi}(x,t,s,r)s_x , \\
r_{ \tilde\tau } =\tilde g= \tilde{\bar\psi}(x,t,s,r)r_x
\end{array}
\label{2.45}
\end{equation}
where for any real value of $\beta$
\begin{equation}
\begin{array}{c}
\tilde{\bar\phi}(x,t,s,r)= \tilde{\bar{a}}(s,r) \exp\left\{
\beta\left[ x +\int\limits_0^t\phi(s,r,t)dt\right]  \right\},
\\
\tilde{\bar\psi}(x,t,s,r)= \tilde{\bar{c}}(s,r) \exp\left\{
\beta\left[ x+\int\limits_0^t\psi(s,r,t)dt\right]    \right\},
\end{array}
\label{2.46}
\end{equation}
and
\begin{equation}
\begin{array}{rl}
\tilde{\bar{a}}(s,r)=&
\bar{b}(s) a_s(s,r) -[\bar{b}(s)\Phi_s(s,r,0) + \bar{\Phi}_0(s)] a(s,r) -
\\
 & -\bar{d}(r)\Phi_r(s,r,0) c(s,r), \\
\tilde{\bar{c}}(s,r)=&
\bar{d}(r) c_r(s,r) -[\bar{d}(r)\Theta_r(s,r,0) + \bar{\Theta}_0(r)]
c(s,r) -\\
  & -\bar{b}(s)\Theta_s(s,r,0) a(s,r),
\end{array}
\label{2.47}
\end{equation}

Thus, the Lie commutator with the inhomogeneous in derivatives symmetry
(\ref{2.24}) is a linear operator acting on the space of homogeneous
hydrodynamic symmetries with $c_0=0$.  It generates the recursion
(\ref{2.47}) of solutions for the linear system (\ref{2.12}).

{\it Remark 2.5.} Inhomogeneous in derivatives hydrodynamic
symmetry generates via Lie commutator a recursion operator for homogeneous
hydrodynamic symmetries.

{\tt Corollary 2.4.}  If the system (\ref{2.1}) satisfies the condition 1 of
the theorem 2.1, then it possesses an infinite set of mutually commuting
hydrodynamic symmetries with a functional arbitrariness which are
homogeneous in derivatives with $c_0=0$ and have the form (\ref{2.45}),
(\ref{2.46}).

In particular, if for $\beta=0$ the coefficients $\phi(s,r),\psi(s,r)$ of the
system (\ref{2.1}) do not depend explicitly upon $t,$ then we reproduce
Tsarev's result \cite{tsar85} about the commutability of
hydrodynamic-type flows without explicit dependence on $t,x$.

\subsection{Hydrodynamic symmetries of diagonal systems with \protect\\
explicit spatial dependence}

Consider  a diagonal two-component hydrodynamic-type system with
explicit dependence on $x$
\begin{equation}
s_t= \phi^{\ast}(s,r,x)s_x, \quad
r_t=\psi^{\ast}(s,r,x) r_x
\label{2.48}
\end{equation}

Define functions $\Phi(s,r,x),\Theta(s,r,x)$ by the following
equations
\begin{equation} \begin{array}{c}
\Phi_r(s,r,x)=[{\phi^{\ast}}^{-1}(s,r,x)]_r /
[{\phi^{\ast}}^{-1}(s,r,x)-{\psi^{\ast}}^{-1}(s,r,x)], \\
\Theta_s(s,r,x)=[{\psi^{\ast}}^{-1}(s,r,x)]_s /
[{\psi^{\ast}}^{-1}(s,r,x)-{\phi^{\ast}}^{-1}(s,r,x)]
\end{array}
\label{2.49}
\end{equation}

{\bf Theorem 2.3.}  A diagonal generic hydrodynamic-type system
(\ref{2.48}) which may explicitly depend on a space coordinate $x$
possesses an
infinite set of hydrodynamic symmetries with a functional arbitrariness
locally parametrized by two arbitrary functions $c_1(s),c_2(r)$ of one
variable iff the two conditions are satisfied

1. coefficients $\phi^{\ast},\psi^{\ast}$ of the system (\ref{2.48}) satisfy
the equalities
\begin{equation} \begin{array}{c}
\Phi_{rx}(s,r,x)= \beta ({\phi^{\ast}}^{-1}(s,r,x))_r \; ,\\
\Theta_{sx}(s,r,x)=\beta ({\psi^{\ast}}^{-1}(s,r,x))_s
\end{array}
\label{2.50}
\end{equation}
with arbitrary real constant $\beta$;

2. there exist such four functions of one variable
$b(s),d(r),\Phi_0(s),\Theta_0(r)$ which satisfy the equations
\begin{equation}
\begin{array}{c}
\hat\Phi_r(s,r,x)=\Phi_r(s,r,x)(\hat\Phi -\hat\Theta), \\
\hat\Theta_s(s,r,x)=\Theta_s(s,r,x)(\hat\Theta -\hat\Phi)
\end{array}
\label{2.51}
\end{equation}
where the functions $\hat\Phi(s,r,x),\hat\Theta(s,r,x)$ are defined
by the formulae
\begin{equation}
\begin{array}{c}
\hat\Phi(s,r,x)=b(s)\Phi_s(s,r,x)+
d(r)\Phi_r(s,r,x)+\Phi_0(s),\\
\hat\Theta(s,r,x)=
b(s)\Theta_s(s,r,x)+ d(r)\Theta_r(s,r,x)+\Theta_0(r).
\end{array}
\label{2.52}
\end{equation}
These symmetries are generated by the Lie equations
\begin{equation}
\begin{array}{c}
s_{\tau } =\tilde\phi^{\ast}(x,t,s,r) s_x +b(s),
r_{\tau } =\tilde\psi^{\ast}(x,t,s,r) r_x +d(r)
\end{array}
\label{2.53}
\end{equation}
where the functions $\tilde\phi^{\ast},\tilde\psi^{\ast}$ are defined as
follows

if $\beta \neq 0$
\begin{equation}
\begin{array}{rl}
\tilde\phi^{\ast}(x,t,s,r)= & \phi^{\ast}(s,r,x) \bigg\{ a(s,r)
\exp\big[ \beta\big( t+\\
& + \int\limits_0^x{\phi^{\ast}}^{-1}(s,r,x)dx\big) \big] +
       +\dfrac{1}{\beta} \hat{\Phi}(s,r,x)\bigg\},  \\
\tilde\psi^{\ast}(x,t,s,r)= & \psi^{\ast}(s,r,x)\bigg\{c(s,r)
\exp\big[
\beta\big(t+ \\
 &+ \int\limits_0^x{\psi^{\ast}}^{-1}(s,r,x)dx\big) \big]+
     \dfrac{1}{\beta} \hat{\Theta}(s,r,x)\bigg\}
\end{array}
\label{2.54}
\end{equation}

and if $\beta = 0$
\begin{equation}
\begin{array}{c}
\tilde\phi^{\ast}(x,t,s,r)=
\phi^{\ast}(s,r,x)\bigg\{
a(s,r)+\int\limits_0^x\hat{\phi}(s,r,x)dx-\\
-\hat{\Phi}(s,r) \left[ t + \int\limits_0^x{\phi^{\ast}}^{-
1}(s,r,x)dx\right]
\bigg\}, \\
\tilde\psi^{\ast}(x,t,s,r)=\psi^{\ast}(s,r,x)\bigg\{c(s,r)
+\int\limits_0^x
\hat\psi(s,r,x)dx -  \\
-\hat{\Theta}(s,r) \left[ t +
\int\limits_0^x{\psi^{\ast}}^{-1}(s,r,x)dx\right]\bigg\}.
\end{array}
\label{2.55}
\end{equation}
where the functions $\hat{\phi},\hat\psi$ are defined as follows
\begin{equation}
\begin{array}{c}
\hat\phi(s,r,x)=b(s)[{\phi^{\ast}}^{-1}(s,r,x)]_s +
+ d(r)[{\phi^{\ast}}^{-1}(s,r,x)]_r, \\
\hat\psi(s,r,x)=b(s)[{\psi^{\ast}}^{-1}(s,r,x)]_s +
+ d(r)[{\psi^{\ast}}^{-1}(s,r,x)]_r.
\end{array}
\label{2.56}
\end{equation}
Here the functions $a(s,r),c(s,r)$ form an arbitrary smooth solution
of the linear system (\ref{2.17}) with the coefficients $\Phi_r(s,r,0),
\Theta_s(s,r,0)$ which are obtained from the formulae (\ref{2.49}) at $x=0$.

\subsection{Invariant solutions and linearizing transformations for
systems with an explicit dependence on $t$ or $x$}

Let the system (\ref{2.1}) which may explicitly depend on t satisfy the
condition 1 of the theorem 2.1 and hence possess an infinite set of
homogeneous in derivatives hydrodynamic symmetries (\ref{2.20}). For the
corresponding invariant solutions we have $s_{\tau}=r_{\tau}=0$ and taking
into account the condition $s_x r_x \neq 0$ we obtain $\tilde\phi(x,t,s,r)=0,
\tilde\psi(x,t,s,r)=0.$ Using here the formulae (\ref{2.21}), (\ref{2.22})
for $\tilde\phi,\tilde\psi$  with $c_0=1$ we obtain the following
conditions for invariant solutions

if $ \beta\neq 0$
\begin{equation} \begin{array}{c}
a(s,r) +\exp\left\{ -\beta\left[ x
+\int\limits_0^t\phi(s,r,t)dt\right]
\right\}=0,  \\
c(s,r) +\exp\left\{ -\beta\left[
x+\int\limits_0^t\psi(s,r,t)dt\right]
\right\}=0,
\end{array}
\label{2.57}
\end{equation}

and if $\beta=0$
\begin{equation} \begin{array}{c}
a(s,r) + x +\int\limits_0^t\phi(s,r,t)dt=0,  \\ [2mm]
c(s,r) + x+\int\limits_0^t\psi(s,r,t)dt=0.
\end{array}
\label{2.58}
\end{equation}

{\bf Theorem 2.4}  (see \cite{sh93b}). Let the coefficients
$\phi,\psi$ of diagonal system (\ref{2.1}) satisfy the condition 1 of the
theorem 2.1.
Then any solution of the system (\ref{2.57}) for $\beta \neq 0 $ or of the
system (\ref{2.58}) for $\beta = 0 $ is also a solution of the system
(\ref{2.1}) if the following conditions are satisfied

for $\beta \neq 0$
\begin{equation}
\bigg\{\big[\ln
|a(s,r)|+\beta\int\limits_0^t\phi(s,r,t)dt\big]_s\cdot
\big[\ln
|c(s,r)|+\beta\int\limits_0^t\psi(s,r,t)dt\big]_r\bigg\}
\bigg|_{(\ref{2.57})}\!\ne 0,
\label{2.59}
\end{equation}

and for $\beta = 0$
\begin{equation}
\bigg\{\big[a(s,r)  +\int\limits_0^t\phi(s,r,t)dt\big]_s\cdot
\big[c(s,r) +
\int\limits_0^t\psi(s,r,t)dt\big]_r\bigg\}\bigg|_{(2.58)}\neq 0.
\label{2.60}
\end{equation}

And vice versa any smooth solution $s(x,t),r(x,t)$ of the system (\ref{2.1})
can be
obtained from the systems (\ref{2.57}) or (\ref{2.58}) if the condition
(\ref{2.59}) or (\ref{2.60}) is met in the vicinity of any point $(x_0,t_0)$
where the condition $s_x(x_0,t_0)\cdot r_x(x_0,t_0)\neq 0$ is satisfied.

{\it Remark 2.6} (see \cite{sh93b}). Equations (\ref{2.57}) for $\beta
\neq 0 $ and (\ref{2.58}) for $\beta=0$ determine in an implicit form the
linearizing point transformation of the nonlinear system (\ref{2.1}) with
explicit $t$-dependence which satisfies the condition 1 of the theorem 2.1.
Indeed, a search for solutions $s(x,t), r(x,t)$ of the system
(\ref{2.1}) reduces to a search for solutions $a(s,r), c(s,r)$ of the linear
system (\ref{2.17}) with variable coefficients. In particular, if the
coefficients $\phi(s,r), \psi(s,r)$ of the system (\ref{2.1}) are not
explicitly $t$-dependent then the condition 1 of the theorem 2.1 is satisfied
with $\beta=0$ and the equations (\ref{2.58}) coincide with the classical
hodograph transformation \cite{rozhd}
\begin{equation}
a(s,r)+x+t\phi(s,r)=0, \qquad
c(s,r)+x+t\psi(s,r)=0
\label{2.61}
\end{equation}

Consider now the system (\ref{2.48}) with explicit $x$-dependence. Let it
satisfy
the condition (\ref{2.50}) of the theorem 2.3 for possessing an infinite set
of the hydrodynamic
symmetries (\ref{2.53}) with $b(s)=d(r)=0.$ Then for invariant solutions
determined by the equations $s_{\tau}=r_{\tau}=0$ with the condition
$s_x r_x \neq 0$ we obtain
\begin{equation}
\tilde\phi^{\ast}(x,t,s,r)=0, \qquad
\tilde\psi^{\ast}(x,t,s,r)=0.
\label{2.62}
\end{equation}
Here the functions $\tilde\phi^{\ast},\tilde\psi^{\ast}$ are determined
by the formulae (\ref{2.54}) for $\beta\neq 0$ and (\ref{2.55}) for $\beta=0$
with $\hat\Phi=\Phi_0(s),\; \hat\Theta=\Theta_0(r),\;
\hat\phi=\hat\psi=0.$ Let the existence conditions for the implicit
vector-function determined by the equations (\ref{2.62}) be met. They are
similar to the conditions (\ref{2.59}), (\ref{2.60}).

Then the equations (\ref{2.62}) determine a linearizing transformation for
the system (\ref{2.48}) with explicit dependence on $x$ and a theorem
analogous to the theorem 2.4 is obviously valid.

\subsection{Higher symmetries of diagonal two-component systems}

Higher symmetries of the second order are generated by the Lie equations
(\ref{2.3}) with $N=2$
\begin{equation}
\begin{array}{c}
s_{\tau}=f(x,t,s,r,s_x,r_x,s_{xx},r_{xx}), \\
r_{\tau}=g(x,t,s,r,s_x,r_x,s_{xx},r_{xx})
\end{array}
\label{2.63}
\end{equation}
which are compatible with the system (\ref{2.1}).

\underline{\bf Definition 2.2.} We call the system (\ref{2.1}) generic
system with respect to second order symmetries if its coefficients
$\phi,\psi$ do not satisfy the constraints
\begin{equation}
\Phi(s,r,t)=\ln\left|\dfrac{\phi_s}{c(s,t)\phi+d(s,t)}\right|,
\quad
\Theta(s,r,t)=\ln\left|\dfrac{\psi_r}{G(r,t)\psi+H(r,t)}\right|,
\label{2.64}
\end{equation}
\begin{equation}
\Phi_t(s,r,t)=A(s,t)\phi+B(s,t), \quad
\Theta_t(s,r,t)=E(r,t)\psi+F(r,t)
\label{2.65}
\end{equation}
with arbitrary smooth functions $A,B,c,d,E,F,G,H$ and the functions
$\Phi,\Theta$ defined by the equations (\ref{2.8}).

The analysis of the symmetries (\ref{2.63}) for the generic system
(\ref{2.1}) produces the following results.

{\bf Theorem 2.5.}  A necessary existence condition for second order
symmetries of the generic system (\ref{2.1}) coincides with the necessary and
sufficient condition for the system (\ref{2.1}) to possess an infinite set of
homogeneous in derivatives hydrodynamic symmetries, {\it i.e.} with the
condition 1 of the theorem 2.1.

Define the function $\Lambda(s,r)$ by the equation
\begin{equation}
\Lambda_{sr}(s,r)=-\Phi_r(s,r,t)\Theta_s(s,r,t).
\label{2.66}
\end{equation}
The fact that the function $\Lambda$ has no explicit independence of $t$
is a consequence of the equations (\ref{2.8}), (\ref{2.12}) and (\ref{2.66}).

Let $A(s),C(r),b(s),d(r),\Phi_0(s),\Theta_0(r)$ be arbitrary smooth
functions of one variable. Define the functions
$\hat\Phi,\hat\Theta$
\begin{equation}
\begin{array}{rl}
\hat\Phi(s,r,t)\!=\!&\!\! A(s)(\Phi_s^2 -\Phi_{ss}+2\Lambda_{ss}) +
A'(s)\Lambda_s + C(r)(2\Phi_r\Theta_r +\Phi_{rr}-\Phi_r^2)
\\
\!\! &\!\! +C'(r)\Phi_r + b(s)\Phi_s + d(r)\Phi_r+ \Phi_0(s). \\
\hat\Theta(s,r,t)\!=\!&\!\! A'(s)\Theta_s\!+\! A(s)(2\Theta_s\Phi_s +
\Theta_{ss} -\Theta_s^2) + C(r)(\Theta_r^2 -\Theta_{rr} +
2\Lambda_{rr}) \\
\!\!&\!\! + C'(r)\Lambda_r(s,r)+b(s)\Theta_s + d(r)\Theta_r + \Theta_0(r).
\end{array}
\label{2.67}
\end{equation}
If $\beta=0$ in the equations (\ref{2.12}), then in virtue of the condition
(\ref{2.18})
the functions $\Phi(s,r),\Theta(s,r)$ do not dependent explicitly on $t.$
Then define the functions $\hat\phi,\hat\psi$
\begin{equation}
\begin{array}{rl} \hat\phi(s,r,t)=&A(s)(2\Phi_s\phi_s -\phi_{ss}) + C(r)[
2\Theta_r
\phi_r +\phi_{rr} -2 \Phi_r(\phi_r -\psi_r)] + \\
    & + C'(r)\phi_r + b(s)\phi_s + d(r)\phi_r,\\
\hat\psi(s,r,t)=&A(s)[2\Phi_s\psi_s +\psi_{ss}-2\Theta_s(\psi_s-
\phi_s)] +
A'(s)\psi_s +\\
 & +C(r)(2\Theta_r \psi_r -\psi_{rr}) + b(s)\psi_s + d(r)\psi_r.
\end{array}
\label{2.68}
\end{equation}

{\bf Theorem 2.6.} (see \cite{sh94b}). A diagonal two-component
generic hydrodynamic-type system (\ref{2.1}) which may explicitly depend on
time possesses an infinite set of second order higher symmetries with a
functional arbitrariness which is locally parametrized by two arbitrary
functions $c_1(s),c_2(r)$ of one variable iff the following two conditions
are satisfied

1. the coefficients $\phi,\psi$ of the system (\ref{2.1}) satisfy the
equations (\ref{2.12}) with an arbitrary real constant $\beta$ where the
functions $\Phi,\Theta$ are defined by equations (\ref{2.8}) and partial
derivatives with respect to $t$ are taken at constant values of $s$ and $r;$

2. there exist six functions $A(s), C(r), b(s), d(r), \Phi_0(s),
\Theta_0(r)$ of one variable which satisfy the equalities
\begin{equation}
\hat\Phi_r=\Phi_r \cdot (\hat\Phi -\hat\Theta), \qquad
\hat\Theta_s=\Theta_s\cdot (\hat\Theta -\hat\Phi)
\label{2.69}
\end{equation}
with the functions $\hat\Phi(s,r,t),\hat\Theta(s,r,t)$ defined by the
formulae (\ref{2.67}).

These symmetries are generated by the Lie equations
\begin{equation}
\begin{array}{rl}
s_{\tau}=f=& A(s)\dfrac{s_{xx}}{s_x^2} +
\Phi_r\left[A(s)\dfrac{r_x}{s_x} + C(r)\dfrac{s_x}{r_x}\right]+
\\
[2mm]
  & +\beta\dfrac{A(s)}{s_x} + s_x\nu(x,t,s,r)+ 2 A(s)\Phi_s + b(s),
\\ [4mm]
r_{\tau}=g=& C(r)\dfrac{r_{xx}}{r_x^2} +
\Theta_s\left[A(s)\dfrac{r_x}{s_x} + C(r)\dfrac{s_x}{r_x}\right]+
\\ [2mm]
  & +\beta\dfrac{C(r)}{r_x} + r_x\rho(x,t,s,r)+ 2 C(r)\Theta_r +
d(r)
\end{array}
\label{2.70}
\end{equation}
with the coefficients $\nu,\rho$ defined by the formulae

if $\beta \neq 0$
\begin{equation}
\begin{array}{c}
\nu(x,t,s,r)=a(s,r) \exp\left\{ \beta\left[
x +\int\limits_0^t\phi(s,r,t)dt\right]    \right\}+
\dfrac{1}{\beta}
\hat{\Phi}(s,r,t),  \\
\rho(x,t,s,r)= c(s,r) \exp\left\{ \beta\left[
x+\int\limits_0^t\psi(s,r,t)dt\right]    \right\}+
\dfrac{1}{\beta}
\hat{\Theta}(s,r,t),
\end{array}
\label{2.71}
\end{equation}

and if $\beta = 0$
\begin{equation}
\begin{array}{rl}
\nu(x,t,s,r)= & a(s,r) +\int\limits_0^t\hat{\phi}(s,r,t)dt-\\
    & -\hat{\Phi}(s,r) \left[ x +
\int\limits_0^t\phi(s,r,t)dt\right], \\
\rho(x,t,s,r)= & c(s,r) +\int\limits_0^t\hat{\psi}(s,r,t)dt-\\
    & -\hat{\Theta}(s,r) \left[ x +
\int\limits_0^t\psi(s,r,t)dt\right].
\end{array}
\label{2.72}
\end{equation}
where the functions $\hat{\phi},\hat{\psi}$ are defined by the formulae
(\ref{2.68}).  Here the integrals with respect to $t$ are taken at
constant values of $s,r.$ The functions $a(s,r),c(s,r)$ form an arbitrary
smooth solution of the linear system (\ref{2.17}).

{\it Remark 2.7.}  For $A(s)=C(r)=0$ the
higher symmetries (\ref{2.70}) reduce to the hydrodynamic symmetries
(\ref{2.14}) and the coefficients $\nu,\rho$ coincide with
$\tilde\phi,\tilde\psi$.

\subsection{First order recursion operators}

An effective description of an infinite set of symmetries of any order is
obtained by means of recursion operators which by their definition map any
symmetry again into a symmetry. Here we consider recursion operators
which belong to the class of matrix-differential operators
\begin{equation}
R=A_N D_x^N + A_{N-1} D^{N-1}_x + \dots +A_1 D_x +A_0,
\label{2.73}
\end{equation}
where $A_i$ are $n\times n$ matrices for $n$-component system and
here $n=2$. The matrices $A_i$ may depend on $s,r$ and their derivatives
of a finite order with respect to $x$.

\underline{\bf Definition 2.3.}  If $A_N \neq 0$, then the integer
$N$ is called {\it order} of the recursion operator (\ref{2.73}).

Here we consider the case $N=1$.

Define the functions
\begin{equation} \begin{array}{c}
S(s,r)=
\hat\Phi(s,r,0)=A(s)\Phi_s(s,r,0)+C(r)\Phi_r(s,r,0)+\Phi_0(s), \\
T(s,r)=
\hat\Theta(s,r,0)=A(s)\Theta_s(s,r,0)+C(r)\Theta_r(s,r,0)+
\Theta_0(r)
\end{array}
\label{2.74}
\end{equation}
where $A(s),C(r),\Phi_0(s),\Theta_0(r)$ are arbitrary smooth
functions of one variable.

{\bf Theorem 2.7} (see \cite{sh94b}). Let the system (\ref{2.1}) satisfy the
condition 1 of the theorem 2.1. Then a first order recursion operator for the
system (\ref{2.1}) exists iff there exist such functions $A(s)$ and $C(r)$
which satisfy the conditions
\begin{equation}
S_r(s,r)=\Phi_r(s,r,0)(S-T), \qquad
T_s(s,r)=\Theta_s(s,r,0)(T-S)
\label{2.75}
\end{equation}
with the functions $ S(s,r), T(s,r)$ defined by the formulae (\ref{2.74})
with $\Phi_0(s)=\Theta_0(r)=0$.\footnote{It is equivalent to the redefinition
$\Phi_s(s,r,0),\Theta_r(s,r,0)$.}
This recursion operator is given by the following formula
\begin{equation} \begin{array}{c}
R=\left\{ \left( \begin{array}{cc} A(s) & 0 \\ 0 & C(r)
\end{array}
\right)\cdot (D_x -\beta)- \right. \\ [3mm]
-\left. \left( \begin{array}{cc}
         A(s)D_x[\Phi(s,r,t)],       & -\Phi_r(s,r,t)[A(s)r_x-C(r)s_x]
\\
         \Theta_s(s,r,t)[A(s)r_x-C(r)s_x] & C(r)D_x[\Theta(s,r,t)]
\end{array} \right) \right\}\cdot \\ [3mm]
\cdot \left(\begin{array}{cc} 1/s_x & 0 \\ 0 & 1/r_x
\end{array} \right).
\end{array}
\label{2.76}
\end{equation}

A first example of first order recursion operator for a class of
two-component systems
was given in the author's paper \cite{sh83}).

Theorem 2.7 generalizes Teshukov's results \cite{tesh}
for explicitly $t$-dependent systems.

{\bf Theorem 2.8.} For the system (\ref{2.1}) which meets the condition 1 of
the theorem 2.1 there exists a first order recursion operator $R$ iff there
exists a recursion operator $\hat{R}$ which acts on a subspace of
homogeneous in derivatives hydrodynamic symmetries of the form
(\ref{2.20})--(\ref{2.22}) with $C_0= 0.$ An action of the operator $\hat{R}$
is defined by the appropriate restriction of $R$
\begin{equation} \begin{array}{c}
R \left(\begin{array}{c}
a(s,r)\exp\big\{\beta\big[x+\int\limits_0^t\phi(s,r,t)dt\big]
\big\} s_x
\\
c(s,r)\exp\big\{\beta\big[x+\int\limits_0^t\psi(s,r,t)dt\big]
\big\} r_x
\end{array} \right) =\\
= \left(\begin{array}{c}
\hat{a}(s,r)\exp\big\{\beta\big[x+\int\limits_0^t\phi(s,r,t)dt
\big]\big
\} s_x \\
\hat{c}(s,r)\exp\big\{\beta\big[x+\int\limits_0^t\psi(s,r,t)dt
\big]\big
\} r_x
\end{array} \right)
\end{array}
\label{2.77}
\end{equation}
with functions $\hat{a}(s,r),\hat{c}(s,r)$ defined by the formulae
\begin{equation} \begin{array}{c}
\hat{a}(s,r)=A(s)[a_s(s,r)-\Phi_s(s,r,0)a(s,r)]-\Phi_0(s)a(s,r)-
\\ \mbox{}-C(r)\Phi_r(s,r,0)c(s,r), \\
\hat{c}(s,r)=C(r)[c_r(s,r)-\Theta_r(s,r,0)c(s,r)]-\Theta_0(r)c(s,r)-\\
\mbox{}-A(s)\Theta_s(s,r,0)a(s,r).
\end{array}
\label{2.78}
\end{equation}

{\tt Corollary 2.5.} For any smooth solution $a(s,r), c(s,r)$ of the linear
system (\ref{2.17}) the functions $\hat{a}(s,r),\hat{c}(s,r)$ defined by the
formulae (\ref{2.78}) are also a solution of this system iff the condition
(\ref{2.75}) is met.  Formulae (\ref{2.78}) determine a recursion of
solutions of the system (\ref{2.17}).

{\it Remark 2.8.} In virtue of the linearizing  transformations (\ref{2.57})
or (\ref{2.58}) a search for any "nonsingular" solutions of the nonlinear
system (\ref{2.1}) reduces to a search for solutions $a(s,r), c(s,r)$ of the
linear system (\ref{2.17}) with variable coefficients. However, the
integration of the equations (\ref{2.17}) is also a problem.  The existence
of a recursion operator and hence of the recursion (\ref{2.78}) which allows
to multiply solutions of the linear equations (\ref{2.17}) is the important
property of the system (\ref{2.1}).  It allows us to go over from
linearization of the system (\ref{2.1}) to its integration. For this to be
true its coefficients $\phi,\psi $ have to meet the condition 1 of the
theorem 2.1. and the conditions (\ref{2.75}).

{\it Remark 2.9.}  Linear system (\ref{2.17}) has two trivial solutions
$ (a,c)=(-1,-1)$ and $(a(s,r),c(s,r))=(\phi(s,r,0),\psi(s,r,0))$ (see the
equations (\ref{2.8})). The recursion formulae (\ref{2.78}) map them to
nontrivial solutions $\hat{a}(s,r),\hat{c}(s,r)$ of the equations
(\ref{2.17}).

In particular, the first solution is mapped to the solution (\ref{2.74}):
$\hat{a}=S(s,r),\hat{c}=T(s,r)$. Substituting these new solutions for
$a,c$ to the equations (\ref{2.78}) we obtain new nontrivial solutions of the
system (\ref{2.17}).  Thus, we construct two infinite series of its solutions.
Then the linearizing transformations (\ref{2.57}) and (\ref{2.58}) generate
two infinite series of exact solutions of the equations (\ref{2.1}) in
explicit form.

{\tt Corollary 2.6.} The transformation (\ref{2.78}) maps any smooth solution
$(a,c)$ of the system (\ref{2.17}) again into a solution $(\hat{a},\hat{c})$
of this system iff it maps the trivial solution $(a=1,\;c=1)$ into some
solution of the equations (\ref{2.17}).

{\tt Corollary 2.7.}  The recursion (\ref{2.47}) for solutions of the linear
system (\ref{2.17}) generated by the Lie commutator with the inhomogeneous
in derivatives hydrodynamic symmetry coincides with the recursion
(\ref{2.78}) generated by the first order recursion operator (\ref{2.76}).
The existence conditions for the recursion operator and for the inhomogeneous
in derivatives hydrodynamic symmetry of the system (\ref{2.1}) coincide. If
there exist several such symmetries of the form (\ref{2.24}), then
there exist several corresponding recursion operators of the form (\ref{2.76})
with $A(s)=\bar{b}(s),C(r)=\bar{d}(r).$

Recursion operator $R$ of the form (\ref{2.76}) generates an infinite set
of higher symmetries
(\ref{2.3}) of any order $N$. Second order symmetries are generated by a
twofold action of the operator $R$ on homogeneous in derivatives
hydrodynamic symmetries (\ref{2.20})--(\ref{2.22}) with $c_0\neq 0$ where we
can put
$c_0=1,\;a(s,r)=c(s,r)=0.$ For instance, in the case $\beta=0$ the formulae
(\ref{2.20}) and (\ref{2.22})
give the following form of this initial symmetry
\begin{equation}
s_{\tau}=s_x\left[ x + \int\limits_0^t\phi(s,r,t)dt \right],
\quad
r_{\tau}=r_x\left[ x + \int\limits_0^t\psi(s,r,t)dt \right]
\label{2.79}
\end{equation}

\subsection{Second order recursion operators and higher symmetries}

\underline{\bf Definition 2.4.}  We call the system (\ref{2.1}) generic with
respect to second order recursion operators if its coefficients do not
satisfy any of the following constraints
\begin{equation}
\dfrac{\Phi_{rs}}{\Phi_r} + \Theta_s=c_1(s)e^{\Phi}, \quad
\dfrac{\Theta_{sr}}{\Theta_s} + \Phi_r=c_2(r)e^{\Theta}
\label{2.80}
\end{equation}
with arbitrary smooth functions $c_1(s),c_2(r).$

Define the functions
\begin{equation} \begin{array}{rl}
S(s,r)=&A(s)(\Phi_s^2 -\Phi_{ss}+2\Lambda_{ss}) +
A'(s)\Lambda_s +
C(r)(2\Phi_r\Theta_r +\Phi_{rr}-\Phi_r^2)  \\
 & +C'(r)\Phi_r + b(s)\Phi_s + d(r)\Phi_r+ \Phi_0(s), \\
T(s,r)=&A'(s)\Theta_s+ A(s)(2\Theta_s\Phi_s + \Theta_{ss} -
\Theta_s^2) +
C(r)(\Theta_r^2 -\Theta_{rr} + 2\Lambda_{rr}) \\
& + C'(r)\Lambda_r+b(s)\Theta_s + d(r)\Theta_r + \Theta_0(r).
\end{array}
\label{2.81}
\end{equation}
with arbitrary smooth functions
$A(s),C(r),b(s),d(r),\Phi_0(s),\Theta_0(r)$ and the functions\newline
$\Phi(s,r,t),\Theta(s,r,t),\Lambda(s,r)$ defined by the formulae (\ref{2.8})
and (\ref{2.66}).

{\bf Theorem 2.9.}  (see \cite{sh94b}). Let the generic system (\ref{2.1})
satisfy the condition 1 of the theorem 2.1. Then a second order recursion
operator of the form (\ref{2.73}) with $N=2$ exists for the system
(\ref{2.1}) iff there exist six functions
$A(s),C(r),b(s),d(r),\Phi_0(s),\Theta_0(r)$ of one variable which satisfy the
conditions
\begin{equation}
S_r(s,r)=\Phi_r(s,r,0)(S-T), \quad
T_r(s,r)=\Theta_s(s,r,0)(T-S).
\label{2.82}
\end{equation}

This recursion operator is determined by the formula
\begin{equation}
R=(AD_x^2+BD_x+F)
\left(
\begin{array}{cc}
   \dfrac{1}{s_x} & 0 \\
     0 &\dfrac{1}{r_x}
\end{array}
\right)
\label{2.83}
\end{equation}
with $2 \times 2$ matrices $A,B$ defined as follows
\begin{equation}
A=
\left(
\begin{array}{cc}
   \dfrac{A(s)}{s_x} & 0 \\
     0 &\dfrac{C(r)}{r_x}
\end{array}
\right)
\label{2.84}
\end{equation}
\begin{equation}
\begin{array}{rl}
B=&
\left(
\begin{array}{c}
   -\left[
    A(s)\dfrac{s_{xx}}{s^2_x}+2A(s)
    \left(  \Phi_s+\Phi_r \dfrac{r_x}{s_x}+\dfrac{\beta}{s_x}
    \right) +b(s)
\right], \\
-\Theta_s(s,r,t)
\left[A(s)\dfrac{r_x}{s_x} -C(r) \dfrac{s_x}{r_x} \right],
\end{array}
\right. \\
& \left.
\begin{array}{c}
\Phi_r(s,r,t)
\left[A(s)\dfrac{r_x}{s_x} -C(r) \dfrac{s_x}{r_x} \right] \\
-\left[C(r)\dfrac{r_{xx}}{r^2_x}+2C(r)\left(\Theta_s\dfrac{s_x}{r_x
}
+\Theta_r +\dfrac{\beta}{r_x} \right) +d(r)\right]
\end{array}
\right),
\end{array}
\label{2.85}
\end{equation}
and the elements $f_{ij}$ of the matrix $F$ are defined by the equations
\begin{equation} \begin{array}{c}
f_{12}=A(s)\left[-\Phi_r \dfrac{r_x}{s_x}\left(\dfrac{s_{xx}}{s_x}-
\dfrac{r_{xx}}{r_x}
\right) +(\Phi_{rr}-\Phi_r^2)\dfrac{r^2_x}{s_x}
\right] -\\
-C(r)\Lambda_{sr}\dfrac{s^2_x}{r_{x}} + \left\{
A(s)[2(\Phi_{rs}-\Phi_r\Phi_s) -\Lambda_{sr}] -b(s) \Phi_r
\right\}r_x +
\\
+\left\{ C(r)(2\Phi_r \Theta_r +\Phi_{rr}-\Phi_r^2)+
[C'(r)+d(r)]\Phi_r \right\} s_x -\\
-\beta \Phi_r\left[A(s)\dfrac{r_x}{s_x} -C(r) \dfrac{s_x}{r_x}
\right],
\\ [4mm]
f_{21}=C(r)\left[ \Theta_s
\dfrac{s_x}{r_x}\left(\dfrac{s_{xx}}{s_x}-\dfrac{r_{xx}}{r_x} \right)
+(\Theta_{ss}-\Theta_s^2)\dfrac{s^2_x}{r_x}
\right] -\\
-A(s)\Lambda_{sr}\dfrac{r^2_x}{s_{x}} + \left\{
C(r)[2(\Theta_{sr}-\Theta_s\Theta_r) -\Lambda_{sr}] -d(r) \Theta_s
\right\}s_x +
\\
+\left\{ A(s)(2\Theta_s\Phi_s  +\Theta_{ss}-\Theta_s^2)+
[A'(s)+b(s)]\Theta_s \right\} r_x +
\\
+\beta \Theta_s\left[A(s)\dfrac{r_x}{s_x} -C(r) \dfrac{s_x}{r_x}
\right],
\end{array}
\label{2.86}
\end{equation}
\begin{equation} \begin{array}{c}
f_{11}+f_{12}=
\hat\Phi(s,r,t)s_x+\beta\big\{A(s)\left[\dfrac{s_{xx}}{s^2_x} +
2\Phi_s(s,r,t)+\dfrac{\beta}{s_x}\right] +
\\
+ \Phi_r(s,r,t)\left[A(s)\dfrac{r_x}{s_x} + C(r) \dfrac{s_x}{r_x}
\right] +b(s) \big\},
\\ [3mm]
f_{21}+f_{22}=
\hat\Theta(s,r,t)r_x+\beta\big\{C(r)\left[\dfrac{r_{xx}}{r^2_x} +
2\Theta_r(s,r,t)+\dfrac{\beta}{r_x}\right] +
\\
+ \Theta_s(s,r,t)\left[A(s)\dfrac{r_x}{s_x} + C(r) \dfrac{s_x}{r_x}
\right] +d(r) \big\}.
\end{array}
\label{2.87}
\end{equation}
Here $\Phi=\Phi(s,r,t),\Theta=\Theta(s,r,t)$ and the functions
$\hat\Phi(s,r,t),\hat\Theta(s,r,t)$ are defined by the formulae (\ref{2.67}).

{\bf Theorem 2.10.}  For the system (\ref{2.1}) which satisfies the
condition 1 of the theorem 2.1 there exists a second order recursion operator
$R$ of the form (\ref{2.83}) iff there
exists the recursion operator $\hat{R}$ which acts on a subspace of
homogeneous in derivatives hydrodynamic symmetries of the form
(\ref{2.20})--(\ref{2.22}) with $c_0=0.$ The action of the operator $\hat{R}$
is defined as an appropriate restriction of $R$  by the formula (\ref{2.77})
where the functions $\hat{a}(s,r),\hat{c}(s,r)$ are defined as follows
\begin{equation} \begin{array}{c}
\hat{a}=A(s)a_{ss}-[2A(s)\Phi_s+b(s)]a_s-C(r)\Phi_r c_r+
\\
+ [A'(s)\Lambda_s+A(s)(\Phi^2_s-\Phi_{ss}+2\Lambda_{ss})
+ b(s)\Phi_s+\Phi_0(s)]a+
\\
+[C'(r)\Phi_r+C(r)(2\Phi_r\Theta_r+\Phi_{rr}-
\Phi^2_r)+d(r)\Phi_r]c,
\\ [4mm]
\hat{c}=C(r)c_{rr} -[2C(r)\Theta_r+d(r)]c_r -A(s)\Theta_s a_s+
\\
+ [C'(r)\Lambda_r+C(r)(\Theta^2_r-\Theta_{rr}+2\Lambda_{rr})
+ d(r)\Theta_r+\Theta_0(r)]c+
\\
+[A'(s)\Theta_s+A(s)(2\Theta_s\Phi_s+\Theta_{ss}-
\Theta^2_s)+b(s)\Theta_s]a.
\end{array}
\label{2.88}
\end{equation}

{\tt Corollary 2.8.} For any smooth solution  $a(s,r),c(s,r)$ of the linear
system (\ref{2.17}) the functions $\hat{a}(s,r),\hat{c}(s,r)$ defined by the
formulae (\ref{2.88}) form also a solution of this system iff the condition
(\ref{2.82}) for the functions (\ref{2.81}) is satisfied. The formulae
(\ref{2.88}) determine a recursion of solutions for the system (\ref{2.17}).

The Remarks 2.8  and 2.9 and the Corollary 2.6 are still valid for the
recursion (\ref{2.88}).  Thus, again we can construct two infinite series of
solutions of the system (\ref{2.1}) starting from a trivial solution of the
system (\ref{2.17}).

{\bf Theorem 2.11.}  Existence conditions (\ref{2.82}) with the notation
(\ref{2.81}) for the second order recursion of solutions of the system
(\ref{2.17}) are less restrictive than the existence conditions (\ref{2.75})
with the notation (\ref{2.74}) for the first order recursion (\ref{2.78}).

Higher symmetries of the system (\ref{2.1}) are generated by the action of
the recursion operator (\ref{2.83}) on the homogeneous in derivatives
hydrodynamic symmetries (\ref{2.20})--(\ref{2.22}) with $c_0 \neq 0.$

{\bf Theorem 2.12.}  (see \cite{sh94b}). Second order symmetries (\ref{2.70})
of the system (\ref{2.1}) obtained as a general solution of determining
equations coincide with the second order symmetries obtained by the action
of the second order recursion operator (\ref{2.83}) on the hydrodynamic
symmetries (\ref{2.20})--(\ref{2.22}) with $c_0 \neq 0.$ Existence conditions
for the second order higher symmetries and for the second order recursion
operator also coincide.

{\tt Corollary 2.9.}  All the second order symmetries may be
obtained by an action of the second order recursion operator on the
homogeneous in derivatives hydrodynamic symmetries.

{\it Remark 2.10.}  The method of calculation of higher order recursion
operators developed in \cite{sh94a} is much more simple than a
straightforward calculation of higher symmetries from determining equations.
Thus, with the suitable extension of the corollary 2.9 we see that this
easier way of calculation of symmetries by means of the $N$th-order recursion
operators gives all higher symmetries of the same order.  In particular, a
squared first order recursion operator does not produce a general form of
second order symmetries.

\section{Generalized gas dynamics equations}
\setcounter{equation}{0}

\subsection{Symmetries of one-dimensional isoentropic \protect\\
            gas dynamics equations}

We consider one-dimensional gas dynamics equations for the isoentropic
plane-parallel gas flow
\begin{equation}
\begin{array}{c}
u_t+uu_x+\alpha^2(\rho)\rho\rho_x=0, \\
\rho_t + \rho u_x +  u \rho_x =0.
\end{array}
\label{3.89}
\end{equation}
Here $u(x,t),\rho(x,t)$ are gas velocity and density at the point $x$
at the moment $t,$ $c=\rho \alpha(\rho)$ is the speed of sound,
$\alpha(\rho)$ is an arbitrary smooth function. In practice
$\alpha(\rho)$
is determined by the gas state equation $p=P(\rho),$ where $p$ is a gas
pressure:
$\alpha(\rho)=(1/\rho) \sqrt{ P'(\rho) }. $

{\bf Theorem 3.1.} (see \cite{rozhd}).
The system (\ref{3.89}) can be reduced to diagonal form (\ref{2.1})
\begin{equation} \begin{array}{c}
s_t=\phi(s,r)s_x,\quad r_t=\psi(s,r)r_x, \\ [4mm]
\phi(s,r)= -\left[\dfrac{s+r}{2} -\rho \alpha(\rho)\right],
\quad
\psi(s,r)= -\left[\dfrac{s+r}{2} + \rho \alpha(\rho)\right]
\end{array}
\label{3.90}
\end{equation}
by the transformation to Riemann invariants $s,r$
\begin{equation}
s=u-\int\limits_{\rho_0}^{\rho} \alpha(\rho)d\rho, \quad
r=u+\int\limits_{\rho_0}^{\rho} \alpha(\rho)d\rho
\label{3.91}
\end{equation}
where $\rho_0$ is an arbitrary fixed constant. The inverse transformation
is given by the formulae
\begin{equation}
u=\dfrac{r+s}{2}, \qquad
\int\limits_{\rho_0}^{\rho} \alpha(\rho)d\rho=\dfrac{r-s}{2}
\label{3.92}
\end{equation}
where the last equality determines $\rho$ as an implicit function
$\rho=\rho(r-s)$ for any fixed choice of the function
$\alpha(\rho).$

The determining equation for symmetries of system (\ref{3.90}) has the form
\begin{equation}
(ID_t+A)(f,g)^T=(0,0)^T
\label{3.93}
\end{equation}
where $ I= \left(\begin{array}{cc} 1 & 0 \\ 0 & 1 \end{array}
\right),$
$T$ means a transposed matrix, the operator of the total derivative $D_t$ is
calculated
with the use of the equations (\ref{3.90}) and the operator $A$ has the form
\begin{equation} \begin{array}{c}
A=
\left(
\begin{array}{cc}
\dfrac{s+r}{2} -\rho \alpha(\rho), & 0 \\
0 & \dfrac{s+r}{2} + \rho \alpha(\rho)
\end{array}
\right) D_x + \\
+ \left( \begin{array}{cc} s_x & 0 \\  0 & r_x \end{array} \right)
\left(I+ \dfrac{\rho \alpha'(\rho) }{ 2 \alpha(\rho)}
\left(\begin{array}{cc} 1 &
-1 \\ -1 & 1 \end{array}\right)\right).
\end{array}
\label{3.94}
\end{equation}

Canonical Lie-B\"{a}cklund symmetries (\ref{2.3}), (\ref{2.4}) of $N$th order
for $N=1,2,3$ were obtained as general solutions of the determining equation
(\ref{3.93}) for a generic $\alpha(\rho)$ (see \cite{sh82,sh83}) explicitly
independent of $x,t$

$N=1:$
\begin{equation}
f_1=a(s,r)s_x, \qquad
g_1=c(s,r)r_x
\label{3.95}
\end{equation}
where $a(s,r), c(s,r)$ is an arbitrary smooth solution of the linear system
\begin{equation}
a_r(s,r)=\dfrac{\alpha'(\rho)}{4 \alpha^2} (a-c), \qquad
c_s(s,r)=\dfrac{\alpha'(\rho)}{4 \alpha^2} (a-c),
\label{3.96}
\end{equation}

$N=2:$
\begin{equation}
\begin{array}{c}
f_2=-\dfrac{s_{xx}}{s^2_x} -\dfrac{\alpha'(\rho)}{4 \alpha^2}
\dfrac{(s_x-r_x)^2}{s_x r_x}, \\ [4mm]
g_2=-\dfrac{r_{xx}}{r^2_x} + \dfrac{\alpha'(\rho)}{4 \alpha^2}
\dfrac{(s_x-r_x)^2}{s_x r_x}\,,
\end{array}
\label{3.97}
\end{equation}

$N=3:$
\begin{equation}
\begin{array}{c}
f_3=-\left( \dfrac{s_{xxx}}{s^3_x} -\dfrac{3 s^2_{xx}}{s^4_x}\right)
-\dfrac{\alpha'(\rho)}{4 \alpha^2}
\left(\dfrac{1}{s^3_x} -\dfrac{1}{r^3_x} \right) s_x r_{xx} -\\
-\dfrac{ 3 \alpha'(\rho)}{4 \alpha^2}
\dfrac{s_{xx}}{s^3_x}(s_x-r_x)-\left[\dfrac{1}{2} \left(\dfrac{\alpha'}{
\alpha^2}
\right)^2
\left(\dfrac{1}{s_x} +\dfrac{1}{r_x} \right)-\left(\dfrac{\alpha'}{
\alpha^2} \right)'\dfrac{1}{\alpha s_x} \right] \dfrac{(s_x-r_x)^3}{8 s_x
r_x}, \\
[4mm]
g_3=-\left(\dfrac{r_{xxx}}{r^3_x} -\dfrac{3
r^2_{xx}}{r^4_x}\right)-\dfrac{\alpha'(\rho)}{4 \alpha^2}
\left(\dfrac{1}{s^3_x} -\dfrac{1}{r^3_x} \right)r_x s_{xx} -\\
-\dfrac{ 3 \alpha'(\rho)}{4 \alpha^2}
\dfrac{r_{xx}}{r^3_x}(s_x-r_x)+\left[\dfrac{1}{2} \left(\dfrac{\alpha'}{
\alpha^2}
\right)^2
\left(\dfrac{1}{s_x} + \dfrac{1}{r_x} \right)-\left(\dfrac{\alpha'}{
\alpha^2} \right)'\dfrac{1}{\alpha r_x} \right] \dfrac{(s_x-r_x)^3}{8
s_x r_x}.
\end{array}
\label{3.98}
\end{equation}

Here every $N$th-order symmetry is presented up to addition of lower-order
symmetries. Hence all the higher symmetries have a functional
arbitrariness
determined by the linear system (\ref{3.96}) since we can add to them the
terms (\ref{3.95}).

Analysis of the determining equation (\ref{3.93}) for second order $(N=2)$
symmetries produces the following special choices of $\alpha(\rho)$
which lead to extensions of the set of symmetries.

1. Function $\alpha(\rho)$ satisfies the differential equation
\begin{equation}
\left[(c\rho+b)\alpha^2(\rho)/\alpha'(\rho)\right]'_{\rho}=-
\dfrac{A}{2}\alpha
(\rho)
\label{3.99}
\end{equation}
with arbitrary constants $A,b,c.$ The second order symmetries are
determined by the formula
\begin{equation} \begin{array}{c}
f_2=(A s + \bar{A}) \dfrac{s_{xx}}{s^2_x} + \dfrac{\alpha'(\rho)}{4
\alpha^2} \left[ (A s + \bar{A}) \left(\dfrac{r_x}{s_x} -2\right)+
(A r + \bar{C})\dfrac{s_x}{r_x}\right] + b
\\[2mm] \mbox{}+ a(s,r)s_x, \\[4mm]
g_2=(A r + \bar{C}) \dfrac{r_{xx}}{r^2_x} -\dfrac{\alpha'(\rho)}{4
\alpha^2} \left[ (A r + \bar{C})\left(\dfrac{s_x}{r_x} -2\right)+
(A s + \bar{A})\dfrac{r_x}{s_x}\right] + b
\\[2mm] \mbox{}+ c(s,r)r_x
\end{array}
\label{3.100}
\end{equation}
with arbitrary constants $A,\bar{A},\bar{C},b$ and functions
$a(s,r),c(s,r)$ satisfying the linear system (\ref{3.96}).

In particular, the equation (\ref{3.99}) is satisfied for $c=0$ by the
physically interesting state equation of a polytropic gas
\begin{equation}
P(\rho)=a^2\rho^{\gamma}, \qquad
\alpha(\rho)=a \sqrt{\gamma}\rho^{(\gamma-3)/2}
\label{3.101}
\end{equation}
where $a,\gamma$ are constants .

2. Function $\alpha(\rho)$ satisfies the condition
$\alpha'(\rho)=0.$
This is a polytropic gas with $\gamma=3.$ Then the second order
symmetries
\begin{equation}
f_2=s_x\psi_1\left( \dfrac{s_{xx}}{s^3_x},s\right), \qquad
g_2=r_x\psi_2\left( \dfrac{r_{xx}}{r^3_x},r\right)
\label{3.102}
\end{equation}
depend on two arbitrary smooth functions $\psi_1,\psi_2.$

The gas dynamics equations (\ref{3.90}) have the form
\begin{equation}
s_t=-s s_x, \qquad
r_t= -r r_x.
\label{3.103}
\end{equation}
Their general solution
\begin{equation}
x-st=F(s), \qquad
x-rt=G(r)
\label{3.104}
\end{equation}
depends on two arbitrary smooth functions $F$ and $G.$
Hence $s_{\tau}=r_{\tau}=0$ and the solution manifold consists solely of
invariant solutions.

3. The function $\alpha(\rho)$ satisfies the condition
$\alpha'(\rho)=-(2/\rho)\alpha(\rho).$ This is the Chaplygin gas \cite{ovs}
with the state equation
\begin{equation}
P(\rho)=P_0 -\dfrac{a^2}{\rho}, \qquad
\alpha(\rho)=\dfrac{a}{\rho^2} \qquad (P_0 > 0)
\label{3.105}
\end{equation}
with the constants $P_0,a.$ Then the set of second order symmetries
depends again on two arbitrary functions $\psi_1(s,s_q),\psi_2(r,r_q)$
where $q$  is the Lagrangian coordinate defined by the equation \cite{rozhd}
\begin{equation}
dq=\rho dx-\rho u dt.
\label{3.106}
\end{equation}

The diagonal form (\ref{3.90}) of gas dynamics equations after the
transformation to the Lagrangian coordinates $q,t$  becomes
\begin{equation}
s_t=a s_q, \qquad
r_t=-a r_q.
\label{3.107}
\end{equation}
Its general solution
\begin{equation}
s=F(q+at), \qquad
r=G(q-at)
\label{3.108}
\end{equation}
depends on two arbitrary smooth functions $F$ and $G.$
Since $x_{\tau}=t_{\tau}=0$ and hence $q_{\tau}=0$, we have
$s_{\tau}=r_{\tau}=0$ as well and the solution manifold again consists solely
of invariant solutions.

In the last two cases the reason for the gas dynamics equations to be
integrable in explicit form is that the extent of arbitrariness
of the set of
invariant solutions and of the general solution manifold turns out to be the
same, {\it i.e.}  two arbitrary functions of one variable. Hence all the
(nonsingular) solutions are invariant.

\subsection{Recursion operators for gas dynamics equations}

{\bf Theorem 3.2.} A first order recursion operator for symmetries of the
gas dynamics equations (\ref{3.90}) is given by the formula
\begin{equation}
R=\left(ID_x -\dfrac{\alpha_x(\rho)}{2\alpha}
\left( \begin{array}{cc} 1 & -1 \\  -1 & 1 \end{array} \right)
\right)
\left( \begin{array}{cc}  \dfrac{1}{s_x} & 0 \\ 0 & \dfrac{1}{r_x}
\end{array}  \right).
\label{3.109}
\end{equation}

It commutes with the operator of the determining equation (\ref{3.93}) on the
solution manifold of the equation (\ref{3.93})
\begin{equation}
[ ID_t+A,R]=0
\label{3.110}
\end{equation}
where the operator $D_t$ is calculated with the use of the gas dynamics
equations.

{\tt Corollary 3.1.}  Operator (\ref{3.109}) raises the order of higher
symmetries by one unit according to the recursion formula
\begin{equation}
R(f_{{\scriptscriptstyle N}},g_{{\scriptscriptstyle N}})^T
=(f_{{\scriptscriptstyle
N+1}},g_{{\scriptscriptstyle N+1}})^T, \quad (N=2,3,\dots)
\label{3.111}
\end{equation}
and thus, generates an infinite countable set containing Lie-B\"{a}cklund
symmetries of any order.

We note the equalities
\begin{equation}
\left(
\begin{array}{c} f_2 \\ g_2 \end{array}
\right) = R
\left(
\begin{array}{c} 1 \\ 1 \end{array}
\right)= R^2 x \left(
\begin{array}{c} s \\ r \end{array}
\right)_x.
\label{3.112}
\end{equation}

{\tt Corollary 3.2.} A solution of the recursion relation (\ref{3.111}) has
the form
\begin{equation}
\left(
\begin{array}{c} f_{{\scriptscriptstyle N}} \\ g_{{\scriptscriptstyle N}}
\end{array}
\right) = R^{N-1}
\left(
\begin{array}{c} 1 \\ 1 \end{array}
\right)= R^N x \left(
\begin{array}{c} s \\ r \end{array}
\right)_x \qquad (N=2,3,\dots).
\label{3.113}
\end{equation}

Formula (\ref{3.109}) gives us a general form of the first order recursion
operator for a generic function $\alpha(\rho).$ If the function
$\alpha(\rho)$ satisfies the equation
\begin{equation}
(\dfrac{\alpha'}{\alpha^2})'[A(r-s)+\bar{c}-\bar{a}]=-2 A
\dfrac{\alpha'}{\alpha}
\label{3.114}
\end{equation}
with arbitrary constant $A,\bar{a},\bar{c}$, then we obtain a more general
form of a recursion operator. In particular, the equation (\ref{3.114}) is
satisfied by the  state equation (\ref{3.101}) of the polytropic gas if the
constants satisfy the equations
\begin{equation}
\begin{array}{cc}
A=-(\gamma-1)(\bar{a}-\bar{c})/(4a\sqrt{\gamma}\rho_0^{(\gamma-
1)/2})  &
(\rho_0 \neq 0), \\ [4mm]
\bar{a}=\bar{c} & (\rho_0 = 0).
\end{array}
\label{3.115}
\end{equation}

{\bf Theorem 3.3.}  For equations of the polytropic gas there exists a
first order recursion operator which depends upon one essential arbitrary
constant $\bar{a}/\bar{c} \quad (\rho_0 \neq 0)$ or $\bar{a}/A \quad
(\rho_0 = 0)$ and has the form
\begin{equation} \begin{array}{c}
R=\left\{
\left(
\begin{array}{cc} As+\bar{a} & 0 \\ 0 & Ar+\bar{c} \end{array}
\right)
D_x - \right. \\ [3mm]
-\dfrac{\alpha'(\rho)}{4 \alpha^2}
\left. \left(
\begin{array}{cc}
(As+\bar{a})(r_x-s_x) & -[(As+\bar{a})r_x -(Ar+\bar{c})s_x] \\
-[(As+\bar{a})r_x -(Ar+\bar{c})s_x] & (Ar+\bar{c})(r_x-s_x)
\end{array}\! \right)\! \right\}\! \cdot \\ [3mm]
\cdot
\left(
\begin{array}{cc} \dfrac{1}{s_x} & 0 \\ 0 & \dfrac{1}{r_x}
\end{array}
\right)
\end{array}
\label{3.116}
\end{equation}
where $\alpha'(\rho)/(4
\alpha^2)=(\gamma-3)/[8a\sqrt{\gamma}\rho^{(\gamma-1)/2}]$ and
the constants $A,\bar{a},\bar{c}$ satisfy the equations (\ref{3.115}).

Now consider the action of the recursion operator (\ref{3.109}) on the
subspace of hydrodynamic symmetries (\ref{3.95}) subject to conditions
(\ref{3.96})
\begin{equation}
R \left( \begin{array}{c} a(s,r)s_x \\ c(s,r)r_x \end{array}
\right)
= \left( \begin{array}{c} a_1(s,r)s_x \\ c_1(s,r)r_x \end{array}
\right).
\label{3.117}
\end{equation}
Here the functions $a_1,c_1$ are generated from $a,c$ by the
transformation
\begin{equation} \begin{array}{c}
a_1(s,r)=a_s(s,r)+[\alpha'(\rho)/(4 \alpha^2)](a-c), \\
c_1(s,r)=c_r(s,r)+[\alpha'(\rho)/(4 \alpha^2)](a-c).
\end{array}
\label{3.118}
\end{equation}

{\tt Corollary 3.3.}  The transformation (\ref{3.118}) determines a
recursion for solutions of the linear system (\ref{3.96}), {\it i.e.}
if $a(s,r),c(s,r)$ is its solution, then $a_1(s,r),c_1(s,r)$ is also the
solution.

From the equation (\ref{3.117}) we obtain
\begin{equation}
R^N \left( \begin{array}{c} a(s,r)s_x \\ c(s,r)r_x \end{array}
\right)
= \left( \begin{array}{c} a_N(s,r)s_x \\ c_N(s,r)r_x \end{array}
\right)
\quad (N=1,2,\dots)
\label{3.119}
\end{equation}
where $a_N(s,r),c_N(s,r)$ is a result of $N$-fold application of the
transformation (\ref{3.118}) to the solution $a(s,r),c(s,r)$ of the equations
(\ref{3.96}).  Then $a_N(s,r),c_N(s,r)$ is also a solution of the
equations (\ref{3.96}).

\subsection{Generalized gas dynamics equations, their symmetries \protect
\newline and recursion operators}

In the Lie equations for hydrodynamic symmetries (\ref{3.95}) of
gas dynamics equations (\ref{3.90})
\begin{equation}
s_{\tau}=a(s,r)s_x, \qquad r_{\tau}=c(s,r)r_x
\label{3.120}
\end{equation}
subject to the conditions (\ref{3.96}) we consider the group parameter $\tau$
as a new time variable. Then we change the notation $a(s,r),c(s,r)$ to
$\phi(s,r),\psi(s,r).$ We obtain systems of the form (\ref{2.1}) subject to
the additional constraints
\begin{equation} \begin{array}{c}
s_t=\phi(s,r)s_x, \qquad r_t=\psi(s,r)r_x, \\ [4mm]
\phi_r(s,r)=\psi_s(s,r)=[\alpha'(\rho)/4 \alpha^2](\phi-\psi)
\end{array}
\label{3.121}
\end{equation}
{\it i.e.} $(\phi,\psi)$ is an arbitrary smooth solution of the linear system
(\ref{3.96}).

Equations (\ref{3.121}) appeared for the first time in the papers of the
author \cite{sh83,sh86} and later Olver and Nutku
had called them generalized gas dynamics (GGD) equations \cite{olvnut}.
They also have pointed out many interesting applications of these
equations in physics.

The determining equation for symmetries of any of GGD systems (\ref{3.121})
has the form similar to the equation (\ref{3.93})
\begin{equation}
(ID_t + A_{(\phi,\psi)})(f,g)^T=(0,0)^T
\label{3.122}
\end{equation}
where the operator $A_{(\phi,\psi)}$ is defined by the formula
\begin{equation} \begin{array}{c}
A_{(\phi,\psi)}=-\left(
\begin{array}{cc}
\phi(s,r) & 0 \\
0 & \psi(s,r)
\end{array}
\right)D_x+ \\
+ \left(\begin{array}{cc} s_x & 0 \\  0 & r_x \end{array}  \right)
\left\{ -\left(
\begin{array}{cc}
\phi_1(s,r) & 0 \\ [4mm]
0 & \psi_1(s,r)
\end{array}
\right) + \dfrac{\alpha'(\rho)}{4 \alpha^2}(\phi-\psi)
\left(\begin{array}{cc} 1 & -1 \\  -1 & 1 \end{array} \! \right)
\!\right\}.
\end{array}
\label{3.123}
\end{equation}
The functions $\phi_1,\psi_1$ are generated from $\phi,\psi$ by
the transformation (\ref{3.118})
\begin{equation}  \begin{array}{c}
\phi_1(s,r)=\phi_s(s,r)+[\alpha'(\rho)/(4 \alpha^2)](\phi-\psi),
\\
\psi_1(s,r)=\psi_r(s,r)+[\alpha'(\rho)/(4 \alpha^2)](\phi-\psi)
\end{array}
\label{3.124}
\end{equation}
and hence the functions $\phi_1,\psi_1$ satisfy the equations (\ref{3.121}).

{\bf Theorem 3.4.}  For the GGD equations (\ref{3.121}) all the hydrodynamic
symmetries homogeneous in derivatives and explicitly independent of
$x,t$ are generated by the Lie equations (\ref{3.120}) with the coefficients
$a(s,r),c(s,r)$ satisfying the linear system (\ref{3.96}) and hence coincide
with the hydrodynamic symmetries of gas dynamics equations.

Thus for any GGD system hydrodynamic symmetries are generated by
the Lie equations which belong to the same GGD hierarchy but have a
different time variable $\tau.$

{\bf Theorem 3.5.}  Operator $R$ defined by the formula (\ref{3.109}) is
a recursion operator for symmetries  of the whole infinite GGD hierarchy
(\ref{3.121}).
It commutes with the operator of determining equation (\ref{3.122}) on
the solution manifold of the equation (\ref{3.122})
\begin{equation}
[ID_t+ A_{(\phi,\psi)},R]=0
\label{3.125}
\end{equation}
where the operator $D_t$ is calculated with the use of GGD equations.

{\bf Theorem 3.6.}  For generalized gas dynamics equations (\ref{3.121})
with the coefficients $\phi(s,r),\psi(s,r)$ all the hydrodynamic symmetries
are generated by the Lie equations
\begin{equation} \begin{array}{c}
s_{\tau}=a(s,r)s_x-\lambda[x+t\phi(s,r)]s_x
+c_0[1+t\phi_1(s,r)s_x],
\\ [3mm]
r_{\tau}=c(s,r)r_x-\lambda[x+t\psi(s,r)]r_x
+c_0[1+t\psi_1(s,r)r_x]
\end{array}
\label{3.126}
\end{equation}
where the functions $a(s,r),c(s,r)$ satisfy the equations (\ref{3.96}) and
$\lambda,c_0$ are arbitrary constants.

{\bf Theorem 3.7.} Generalized gas dynamics equations with the coefficients
$\phi(s,r)$,$\psi(s,r)$ have an infinite series of higher symmetries
of any order $N=2,3,\dots$ with a functional arbitrariness and an explicit
$t$-dependence. Symmetries of the order $N=2$ are generated by the Lie
equations
\begin{equation} \begin{array}{c}
s_{\tau}=a_1(s,r)s_x-\lambda[1+t\phi_1(s,r)s_x]
+c_0[t\phi_2(s,r)s_x
+ f_2], \\
r_{\tau}=c_1(s,r)r_x-\lambda[1+t\psi_1(s,r)r_x]
+c_0[t\psi_2(s,r)r_x
+
g_2].
\end{array}
\label{3.127}
\end{equation}
Here the functions $a_1,c_1$ and $\phi_1,\psi_1$ are obtained by the
transformations (\ref{3.118}) and (\ref{3.124}) from the functions $a,c$
and $\phi,\psi$ respectively.
The functions $\phi_2,\psi_2$ are obtained by twofold application of the
transformation (\ref{3.124}) to the functions $\phi,\psi$ and the functions
$a_1(s,r),c_1(s,r)$ form an arbitrary smooth solution of the equations
(\ref{3.96}). The functions $f_2,g_2$ are defined by the equations
(\ref{3.97}).  $(N+1)$th-order symmetries $(N+1\geq 3)$ are generated by the
Lie equations
\begin{equation} \begin{array}{c}
\left( \begin{array}{c} s \\ r \end{array} \right)_{\tau}=
\left(
\begin{array}{c}
a_N(s,r)s_x \\  c_N(s,r)r_x
\end{array} \right) -\lambda\left\{ \left( \begin{array}{c} f_N
\\
g_N \end{array} \right) + t \left( \begin{array}{c}
\phi_N(s,r)s_x \\  \psi_N(s,r)r_x
\end{array} \right) \right\} + \\ [3mm]
+ c_0 \left\{\left( \begin{array}{c} f_{N+1} \\ g_{N+1}
\end{array}
\right) + t \left( \begin{array}{c}
\phi_{N+1}(s,r)s_x \\  \psi_{N+1}(s,r)r_x
\end{array} \right) \right\} \quad (N=2,3,\dots).
\end{array}
\label{3.128}
\end{equation}
Here the subscript $N$ denotes $N$-fold application of the transformations
(\ref{3.118}) and (\ref{3.124}), the functions $f_N,g_N$ are defined by the
formula (\ref{3.113}) with their explicit form for $N=2,3$ given by the
formulae (\ref{3.97}), (\ref{3.98}) and $\lambda, c_0$ are arbitrary
constants.

\subsection{Noncommutative Lie-B\"{a}cklund algebra associated with \protect
\newline gas dynamics equations}

Denote by $\hat{X}_{(a,c)}$ canonical Lie-B\"{a}cklund operators of
hydrodynamic symmetries generated by the Lie equations (\ref{3.120}) subject
to the condition (\ref{3.96}). Let $\hat{X}_N$ denote canonical
Lie-B\"{a}cklund operators of the $N$th-order higher symmetries (\ref{3.113})
$(N=2,3,\dots)$ for gas dynamics equations.

{\bf Theorem 3.8.} Hydrodynamic and higher symmetries of gas dynamics
equations generate infinite-dimensional noncommutative Lie-B\"{a}cklund
algebra in which the hydrodynamic symmetries form an infinite-dimensional
commutative ideal
\begin{equation} \begin{array}{c}
\left[\hat{X}_{(a,c)},\hat{X}_{(\tilde{a},\tilde{c})} \right]=0,
\qquad
\left[\hat{X}_{N},\hat{X}_{(a,c)} \right]=\hat{X}_{(a_N,c_N)},
\\
\left[\hat{X}_{M},\hat{X}_{N} \right]=0 \qquad
(M,N=2,3,\dots).
\end{array}
\label{3.129}
\end{equation}

Here the functions $a(s,r),c(s,r)$ and $\tilde{a}(s,r),\tilde{c}(s,r)$
satisfy the equations (\ref{3.96}).

{\bf Theorem 3.9.}  Let $N=2,3,\dots$ .  The generalized gas dynamics
equations have a common infinite series of higher symmetries (\ref{3.128}) of
$(N+1)$th or larger order iff the right-hand sides of these equations
differ only by a term belonging to a kernel of the operator $R^N,$
{\it i.e.}  $\phi_N,\psi_N$ coincide for all these equations.  If the
right-hand sides $\phi s_x, \psi r_x$ of GGD equations (\ref{3.121}) belong
themselves to a kernel of the operator $R^N,$ then for all
such equations the operators $\hat{X}_{(\phi,\psi)}$ commute with all the
operators $\hat{X}_n$ of higher symmetries (\ref{3.113}) of the
order $n=N,N+1,\dots$ .

{\tt Corollary 3.4.}  Commutative symmetry subalgebras for gas dynamics
equations are generated by the operators $\hat{X}_{(a,c)}$ of
those hydrodynamic symmetries whose characteristics $(as_x,cr_x)$
belong to a kernel of some integer degree $R^N$ of the recursion
operator $(N \geq 2),$ and by the operators $\hat{X}_n$ of higher
symmetries (\ref{3.113}) of the order $n$ larger or equal to $N.$

Thus, a problem of constructive description of the kernel of operator
$R^N$ naturally arises. It is solved by means of the inverse recursion
operator $R^{-1}$
\begin{equation}
R^{-1}=\left(\!\! \begin{array}{cc} s_x & 0 \\ 0 & r_x \end{array}
\right)
\left\{ I + \dfrac{\alpha}{2}
\left( \begin{array}{cc} 1 & -1 \\ -1 & 1  \end{array}\!\!\right)
D_x^{-1} \dfrac{\alpha_x}{\alpha^2} \right\} D_x^{-1}
\label{3.130}
\end{equation}
where $D_x^{-1}=\int\limits dx$ is the operator of integration with
respect to $x$
at a constant value of $t$ with the integration "constant" $c(t)$
depending upon $t.$ Hence a kernel of the operator $R$ has the form
\begin{equation}
R^{-1} \left( \begin{array}{c}  0 \\ 0  \end{array} \right) =
c_1(t) \left(\begin{array}{c} s_x  \\  r_x \end{array} \right) +
c_2(t) \alpha(\rho) \left( \begin{array}{c} -s_x  \\  r_x
\end{array}
\right).
\label{3.131}
\end{equation}

A kernel of the operator $R^2$ is given by the formula
\begin{equation} \begin{array}{c}
R^{-2} \left(\! \begin{array}{c}  0 \\ 0  \end{array} \! \right) =
c_1(t) \left(\!\begin{array}{c} (u-\rho \alpha(\rho))s_x  \\ (u+\rho
\alpha(\rho)) r_x \end{array}\! \right)\! + \!
c_2(t) \!\vec{\!\left[\int \alpha^2(\rho) d\rho-
u\alpha(\rho)\right]\!s_x}{\!\left[\int\!
\alpha^2(\rho) d\rho + u \alpha(\rho)\right]\!r_x\!}\!  \\ [3mm]
+ c_3(t)\vec{s_x}{r_x}+ c_4(t)\alpha(\rho)\vec{-s_x}{r_x}.
\end{array}
\label{3.132}
\end{equation}

Here $c_i(t)$ are arbitrary smooth functions.

\subsection{Lax representation and invariant solutions of generalized
gas dynamics equations}

Formula (\ref{3.110}) gives the Lax representation \cite{lax} of the
gas dynamics equations
\begin{equation}
\partial R / \partial t = [ R,A].
\label{3.133}
\end{equation}
Here the recursion operator (\ref{3.109}) and the "stationary part"
(\ref{3.94}) of the operator of the determining equation form the Lax pair of
the Ibragimov-Shabat type \cite{ibrshab80a,ibrshab80b,mikhshab}.
Equation (\ref{3.125}) also gives Lax representation for any generalized
gas dynamics equations (\ref{3.121})
\begin{equation}
\partial R / \partial t = [ R,A_{(\phi,\psi)}]
\label{3.134}
\end{equation}
with the recursion operator (\ref{3.109}) as L-operator and the operator
(\ref{3.123}) as A-operator.

In 1982 L.D. Faddeev and P.P. Kulish in a private communication pointed out
to the author that these Lax pairs were degenerate, {\it i.e.} the mapping of
a potential to scattering data was singular, so that the method of inverse
scattering transform could not be applied.  However, it turned out that it is
not a deficiency of GGD equations but the indication to a possibility of
linearizing these equations in a classical sense by the hodograph
transformation and not by the inverse scattering transform.

Nevertheless, the Lax representation for these equations turns out to be
useful for construction of their invariant solutions  by means of the inverse
recursion operator (\ref{3.130}) \cite{sh83}.

At first consider the solutions of the gas dynamics equations (\ref{3.89})
which are invariant with respect to the higher symmetries given by the
formula (\ref{3.113})
\begin{equation}
(f_N,g_N)^T=R^{N-1}(1,1)^T = 0  \qquad (N=2,3,\dots)
\label{3.135}
\end{equation}

Define the matrix $U$
\begin{equation}
U= \left(
\begin{array}{cc}  u & \int\limits_0^{\rho}\alpha^2 (\rho) d \rho
\\
\rho & u \end{array}
\right)
\label{3.136}
\end{equation}
and the matrix-integral operator $K$
\begin{equation}
K=D_x^{-1} U_x =D_x^{-1} \left(
\begin{array}{cc}  u_x & \alpha^2 (\rho) \rho_x \\
\rho_x & u_x \end{array}
\right).
\label{3.137}
\end{equation}

Define the operator $K$ and its degrees $K^n$ with an integer $n$ so that all
the constants $c_i$ of integrations $D_x^{-1}= \int\limits dx$ do not depend
on $t$.

{\bf Theorem 3.10.}  For any $N=2,3,\dots$ the equalities
\begin{equation} \begin{array}{c}
K^{N-1} \vec{c_1}{c_2} \equiv \vec{c_{2N-1}}{c_{2N}} + c_{2N-2}
\vec{\int\limits\alpha^2 (\rho) d \rho}{u} + c_{2N-3}
\vec{u}{\rho} +  \\ [3mm]
+ c_{2N-4} \vec{ u \int\limits\alpha^2 (\rho)
d\rho}{\dfrac{u^2}{2} + \int\limits
d\rho \int\limits\alpha^2 (\rho) d \rho }  + c_{2N-5}
\vec{\dfrac{u^2}{2}
+\int\limits\alpha^2 (\rho)\rho d \rho\!}{u\rho\!} + \dots = \\ [3mm]
= \vec{x-ut}{-\rho t}
\end{array}
\label{3.138}
\end{equation}
determine an infinite series of exact solutions $u=u(x,t), \rho=\rho(x,t)$
of gas dynamics equations (\ref{3.89}) which are invariant with respect to
higher symmetries (\ref{3.113}). They are given in the form of an implicit
function $t=t(u,\rho), x=x(u,\rho).$

For $N=1$ the formula (\ref{3.138}) have the form
\begin{equation}
\vec{c_1}{c_2}=\vec{x-ut}{-\rho t}
\label{3.139}
\end{equation}
and gives a trivial solution for which $u_t+u u_x=0, \rho_t+\rho u_x=0,$
{\it i.e.} the velocity and density $u$ and $\rho$ are constant in the
Lagrangian frame which moves together with a gas particle.

For $N=2$ the formula (\ref{3.138}) can be taken in the form
\begin{equation} \begin{array}{c}
u-u_0= \bar\rho(t-t_0), \\
\bar\rho(t-t_0)^2-\int\limits^{\bar\rho}_{\bar\rho_0}
P'(\lambda
\bar\rho ) \dfrac{1}{\bar\rho^2} d \bar\rho= x-x_0 -u_0(t-t_0),
\end{array}
\label{3.140}
\end{equation}
where $\bar\rho=\rho/\lambda $ and
$\lambda,u_0,x_0,t_0,\bar\rho_0$ are constants. It corresponds to
the motion of a piston in a gas flow after the explosion.

Now we consider invariant solutions of generalized gas dynamics equations
expressed through variables $u,\rho$
\begin{equation} \begin{array}{c}
u_t=w(u,\rho)u_x + v(u,\rho)\alpha^2(\rho)\rho_x, \\
\rho_t=v(u,\rho)u_x + w(u,\rho)\rho_x
\end{array}
\label{3.141}
\end{equation}
where the coefficients $v,w$ satisfy the linear system
\begin{equation}
w_u=v_{\rho}, \qquad w_{\rho}=\alpha^2(\rho)v_u.
\label{3.142}
\end{equation}
The corollary 3.4 implies that the higher symmetries (\ref{3.113}) of gas
dynamics equations of the orders $N_0,N_0+1,...$ are also higher symmetries
of those GGD equations (\ref{3.121}) whose right-hand sides belong to a
kernel of the recursion operator $R^{N_0},$ {\it i.e.}
$\phi_{N_0}=\psi_{N_0}=0.$ We give now their explicit form allowing them to
depend explicitly on $t$
\begin{equation} \begin{array}{c}
D_t \vec{u}{\rho}\! =\! \left(\!
\begin{array}{cc} u_x & \alpha^2(\rho)\rho_x \\
\rho_x & u_x
\end{array} \!\right)\!
K^{N_{0}-1} \vec{a_1(t)}{a_2(t)}\! \equiv \!\left(\!
\begin{array}{cc} u_x & \alpha^2(\rho)\rho_x \\
\rho_x & u_x
\end{array}
\! \right)\! \cdot \\[4mm]
\cdot\!
\left\{\! \vec{\!a_{2N_0-1}(t)\!}{\!a_{2N_0}(t)\!}\! + \! a_{2N_0-2}(t)\!
\vec{\!\int\!\alpha^2 (\rho) d \rho\!}{\!u\!}\! +\! a_{2N_0-3}(t)\!
\vec{\!u\!}{\!\rho\!}\!+\! a_{2N_0-4}(t)\cdot\! \right. \\[4mm]
\left. \!\!\cdot\vec{u\int\!\alpha^2\,(\rho)d\rho}{\dfrac{u^2}{2}
+ \int\! d\rho\int\!\alpha^2 (\rho)\,d\rho }  +  a_{2N_0-5}(t)
\vec{\dfrac{u^2}{2} +\int\!\alpha^2 (\rho)\rho\,d\rho}{u\rho} +
\! \dots\! \right\}.
\end{array}
\label{3.143}
\end{equation}
Here in the definition (\ref{3.137}) of operator $K$ the "constants" $a_i(t)$
of the integration with respect to $x$ may depend on $t$ and are assumed to
be arbitrary smooth functions.

We consider the solutions of GGD equations (\ref{3.143}) which are invariant
with respect to the higher symmetries (\ref{3.113}).

{\bf Theorem 3.11.}   For any $N=N_0,N_0+1,\dots (N\geq2)$ the
equations
\begin{equation} \begin{array}{c}
K^{N-1} \vec{c_1}{c_2} \equiv \vec{c_{2N-1}}{c_{2N}} + c_{2N-2}
\vec{\int \alpha^2 (\rho) d \rho}{u} + c_{2N-3}
\vec{u}{\rho} +  \\[2mm]
+ c_{2N-4} \vec{ u \int \alpha^2 (\rho)
d\rho}{\dfrac{u^2}{2} + \int\limits
d\rho \int \alpha^2 (\rho) d \rho }  + c_{2N-5}\!\vec{\!\dfrac{u^2}{2}
+\int \alpha^2 (\rho)\rho d \rho\!}{\!u\rho\!} + \dots = \\[2mm]
= \vec{x}{0} + \int\limits_0^t\vec{a_{2N_0-1}(t)}{a_{2N_0}(t)}dt +
\vec{\int \alpha^2 (\rho) d \rho}{u} \int\limits_0^t
a_{2N_0-2}(t)dt +
\\[2mm] + \vec{u}{\rho}\int\limits_0^t a_{2N_0-3}(t)dt  +  \vec{ u
\int \alpha^2
(\rho) d\rho}{\dfrac{u^2}{2} + \int  d\rho
\int \alpha^2 (\rho) d \rho
}
\int\limits_0^t a_{2N_0-4}(t)dt +  \\[2mm]
+ \vec{\dfrac{u^2}{2} +\int \alpha^2 (\rho)\rho d
\rho}{u\rho}
\int\limits_0^t
a_{2N_0-5}(t) dt + \dots
\end{array}
\label{3.144}
\end{equation}
determine an infinite series of exact solutions
$u=u(x,t),\rho=\rho(x,t)$ of any generalized gas dynamics equations of the
form (\ref{3.143}) (with arbitrarily fixed integer $N_0$) which may be
explicitly $t$-dependent. These solutions are invariant with respect to the
$N$th-order higher symmetries (\ref{3.113}) of these equations. Here the
functions $a_i(t)$ and the constants $c_i$ with $i\leq0$ must be put equal to
zero. The definition of the operator $K$ is the same as in the formula
(\ref{3.138}).

For the gas dynamics equations (\ref{3.89}) we have $a_{2N_0-3}(t)=-1,
a_i(t)=0$ for $i \neq 2N_0-3.$

\section{Separable two-component Hamiltonian systems}
\setcounter{equation}{0}

\subsection{Hamiltonian structure of generalized gas dynamics\protect\newline
equations}

We consider two-component hydrodynamic-type Hamiltonian systems of the form
\begin{equation}
D_t \vec{u}{\rho}= \sigma_1 D_x\vec{H_u(u,\rho)}{H_{\rho}(u,\rho)}
\label{4.145}
\end{equation}
where $\sigma_1 = \left( \begin{array}{cc} 0 & 1 \\ 1 & 0 \end{array}\right)$
is the Pauli matrix. Here $H(u,\rho)$ is Hamiltonian density of
the hydrodynamic-type which corresponds to the Hamiltonian
$ {\cal H}= \int\limits_{-\infty}^{\infty}H(u,\rho)dx.$ For short
we call $H(u,\rho)$ also a Hamiltonian.

Equations (\ref{4.145}) take the form of the Hamilton equations
\begin{equation}
D_t \vec{u}{\rho}= \left\{ \vec{u}{\rho},H \right\}
\label{4.146}
\end{equation}
with the Poisson bracket of the hydrodynamic-type \cite{dubnov83}
\begin{equation}
\{ H,h \}=(h_u,h_{\rho})D_x \sigma_1(H_u,H_{\rho})^T .
\label{4.147}
\end{equation}

Define the Hamiltonian matrix
\begin{equation}
\hat{H}=\left( \begin{array}{cc} H_{\rho u} & H_{\rho \rho} \\
H_{u u} & H_{u \rho} \end{array} \right).
\label{4.148}
\end{equation}

Then the equations (\ref{4.145}) take the form
\begin{equation}
\vec{u}{\rho}_t = \hat{H}\vec{u}{\rho}_x \; \Longleftrightarrow \;
(ID_t -\hat{H} D_x)\vec{u}{\rho} = \vec{0}{0}.
\label{4.149}
\end{equation}

The gas dynamics equations (\ref{3.89}) have the Hamiltonian form
(\ref{4.145}), (\ref{4.149}) with the Hamiltonian
\begin{equation}
H(u,\rho)=-\left[ \rho u ^2 / 2 + \int\limits_0^{\rho}
d\rho\int\limits_0^{\rho}
\alpha^2(\rho)\rho d\rho \right].
\label{4.150}
\end{equation}
The same is valid for the generalized gas dynamics equations (\ref{3.143})
whose Hamiltonians $H(u,\rho,t)$ are defined by the equality
\begin{equation} \vec{H_{\rho }}{H_{ u}}
=K^{N_0}\vec{a_1(t)}{a_2(t)}
\label{4.151}
\end{equation}
with the operator $K$ defined by the formula (\ref{3.137}) and they may
depend explicitly upon $t.$ For example the equation (\ref{4.151}) for
$N_0=2$ becomes
\begin{equation} \begin{array}{c}
H(u,\rho,t)= a_1(t)\left[ \rho u ^2 / 2 + \int\limits_0^{\rho}
d\rho\int\limits_0^{\rho} \alpha^2(\rho) \rho d\rho \right] +  \\
+ a_2(t)\left[  u^3/3! + u \int\limits_0^{\rho}
d\rho\int\limits_0^{\rho} \alpha^2
(\rho)d\rho \right] + \\
+ a_3(t)u \rho + a_4(t)\left[  u ^2 / 2 + \int\limits_0^{\rho}
d\rho\int\limits_0^{\rho} \alpha^2(\rho) d\rho \right] + \\
+ a_5(t)\rho + a_6(t) u.
\end{array}
\label{4.152}
\end{equation}

\subsection{Separable Hamiltonian systems}

\underline{\bf Definition 4.1.}  We say that Hamiltonians
$\int\limits_{-\infty}^{\infty}H \, dx$ and $\int\limits_{-\infty}^{\infty}
h\,dx$ commute
if the Poisson bracket (\ref{4.147}) of their densities is the exact
derivative with respect to $x$ of some function
\begin{equation}
\{ H(u,\rho,t),h(u,\rho,t)\}= D_x[Q(u,\rho,t)].
\label{4.153}
\end{equation}

Then we also say that Hamiltonians $H$ and $h$ commute and the Hamiltonian
matrices $\hat{H}$ and $\hat{h}$ defined by the equation (\ref{4.148})
commute also: $[\hat{H},\hat{h}]=0.$

{\bf Theorem 4.1.}  For any hydrodynamic-type Hamiltonian $H(u,\rho)$
there exists an infinite set of Hamiltonians $h(u,\rho)$ which commute
with it and with each other. They are arbitrary smooth solutions of the
wave equation
\begin{equation}
H_{\rho\rho} h_{uu} -h_{\rho\rho} H_{uu} =0
\label{4.154}
\end{equation}
or for $H_{\rho\rho} \neq 0$
\begin{equation}
h_{uu} -V(u,\rho) h_{\rho\rho}=0
\label{4.155}
\end{equation}
with $V(u,\rho) =H_{uu}/H_{\rho\rho}.$

\underline{\bf Definition 4.2.}  Those systems (\ref{4.149}) for which the
wave equation (\ref{4.155}) admits a separation of variables $u,\rho$,
{\it i.e.}
$V(u,\rho) = \beta^2(u)/\alpha^2(\rho)$, are called separable Hamiltonian
systems \cite{sh86,olvnut}.  Their
Hamiltonians $H$ satisfy the equation $(H_{uu}H_{\rho\rho} \neq 0)$
\begin{equation}
(1/ \beta^2(u))H_{uu}=(1/\alpha^2(\rho))H_{\rho\rho}
\label{4.156}
\end{equation}
and just the same for all the Hamiltonians $h$ commuting with $H$
\begin{equation}
(1/ \beta^2(u))h_{uu}=(1/\alpha^2(\rho))h_{\rho\rho}.
\label{4.157}
\end{equation}
Gas dynamics equations (\ref{3.89}) and GGD equations (\ref{3.121}) with
the Hamiltonians (\ref{4.150}) and (\ref{4.151}) are examples of separable
systems with $\beta^2(u)=1.$

\underline{\bf Definition 4.3.}  If $H(u,\rho)$ is a Hamiltonian
and the ratio $H_{uu}/H_{\rho\rho}= 1/\alpha^2(\rho)$ is a function
of $\rho$ only, then $H$ is called Hamiltonian of generalized
gas dynamics. In this case $\beta(u)=1.$

Many interesting physical applications of the separable Hamiltonian
systems are given in the paper \cite{olvnut}.

\subsection{Second order recursion operator and Lax representation for
separable Hamiltonian  systems}

We show that separable Hamiltonian  systems obtained by the Manin's
construction possess higher symmetries, a recursion operator, the
Lax representation of the Ibragimov-Shabat type and good integrability
properties \cite{manin,sh86}.

We introduce the notation
\begin{equation}
\partial_u=\partial/\partial u, \quad
\partial_{\rho}=\partial/\partial \rho, \quad
\partial^{-1}_u=\int\limits_0^u du, \quad
\partial^{-1}_{\rho}=\int\limits_0^{\rho} d\rho
\label{4.158 }
\end{equation}
where the integration with respect to one variable is performed at a constant
value of another variable. The Manin's construction gives rise to two
fundamental series of mutually commuting Hamiltonians \cite{sh86}
\begin{equation} \begin{array}{c}
H^{(2m)}(1,0)= \sum\limits^m_{n=0}(\partial^{-2}_u
\beta^2(u))^{m-n}
(\partial^{-2}_\rho \alpha^2(\rho))^{n} \rho, \\
H^{(2m-1)}(1,0)= \sum\limits^m_{n=0}(\partial^{-2}_u
\beta^2(u))^{m-n-1}
(\partial^{-2}_\rho \alpha^2(\rho))^{n} u\rho,
\end{array}
\label{4.159}
\end{equation}
\begin{equation} \begin{array}{c}
H^{(2m)}(0,1)= \sum\limits^m_{n=0}(\partial^{-2}_u
\beta^2(u))^{m-n}
(\partial^{-2}_\rho \alpha^2(\rho))^{n} u, \\
H^{(2m-1)}(0,1)= \sum\limits^m_{n=0}(\partial^{-2}_u
\beta^2(u))^{m-n}
(\partial^{-2}_\rho \alpha^2(\rho))^{n}
\end{array}
\label{4.160}
\end{equation}
with $m=0,1,2,\dots$ and $H^{(-1)}(1,0)=0,\; H^{(-1)}(0,1)=1.$
Here every operator $\partial^{-1}_u,\partial^{-1}_\rho$ acts on all the
factors standing to the right of it, e.g.
\begin{equation}
(\partial^{-2}_u\beta^2(u) )^2= \int\limits_0^u du
\int\limits_0^u\left[ \beta^2(u)
\int\limits_0^u du \int\limits_0^u\beta^2(u)du \right] du.
\label{4.161}
\end{equation}
For arbitrary constants $c_1,c_2$ we define
\begin{equation}
H^{(N)}(c_1,c_2)=c_1 H^{(N)}(1,0) +c_2 H^{(N)}(0,1).
\label{4.162}
\end{equation}
The basic Hamiltonians for the GGD equations (\ref{3.143}) are obtained at
$\beta(u)=1$
\begin{equation}
H^{(N)}(1,0)=\sum\limits_{n=0}^{[\frac{N}{2}]} \dfrac{u^{N-2n}}{(N-2n)!}
(\partial^{-2}_\rho \alpha^2(\rho))^{n} \rho,
\label{4.163}
\end{equation}
\begin{equation}
H^{(N)}(0,1)=\sum\limits_{n=0}^{[\frac{N+1}{2}]} \dfrac{u^{N+1-
2n}}{(N+1-2n)! }
(\partial^{-2}_\rho \alpha^2(\rho))^{n}.
\label{4.164}
\end{equation}

We list explicitly several Hamiltonians from the series
(\ref{4.159}), (\ref{4.160})
\begin{equation} \begin{array}{c}
H^{(0)}(1,0)=\rho, \qquad H^{(1)}(1,0)=u\rho, \\
H^{(2)}(1,0)=(\partial^{-2}_u \beta^2(u)+\partial^{-2}_\rho
\alpha^2(\rho) ) \rho, \\
 H^{(3)}(1,0)=(\partial^{-2}_u \beta^2(u)+\partial^{-2}_\rho
\alpha^2(\rho) )u\rho\,;
\end{array}
\label{4.165}
\end{equation}
\begin{equation} \begin{array}{c}
H^{(0)}(0,1)=u, \qquad  H^{(1)}(0,1)=\partial^{-2}_u
\beta^2(u)+\partial^{-2}_\rho \alpha^2(\rho), \\
H^{(2)}(0,1)=(\partial^{-2}_u \beta^2(u)+\partial^{-2}_\rho
\alpha^2(\rho) )u, \\
H^{(3)}(0,1)=(\partial^{-2}_u \beta^2(u))^2 +\partial^{-2}_u
\beta^2(u)
\partial^{-2}_\rho \alpha^2(\rho) +(\partial^{-2}_\rho\alpha^2(\rho))^2.
\end{array}
\label{4.166}
\end{equation}

In particular, for the Hamiltonians (\ref{4.163}), (\ref{4.164})
of the generalized gas dynamics we obtain \cite{olvnut}
\begin{equation}
H^{(2)}(1,0)=\rho u^2/2 + \tilde{G}_1(\rho), \quad
H^{(3)}(1,0)=\rho
u^3/6 + u \tilde{G}_1(\rho)
\label{4.167}
\end{equation}
where
\begin{equation}
\tilde{G}_1(\rho)=\int\limits_0^\rho(\rho -\sigma)\sigma
\alpha^2(\sigma)
d \sigma
\label{4.168}
\end{equation}
and
\begin{equation}
H^{(1)}(0,1)=u^2/2 + G_1(\rho), \quad H^{(2)}(0,1)=u^3/6 + u
G_1(\rho)
\label{4.169}
\end{equation}
where
\begin{equation}
G_1(\rho)=\int\limits_0^\rho(\rho -\sigma) \alpha^2(\sigma)
d \sigma.
\label{4.170}
\end{equation}

Consider at first the special case of the generalized gas dynamics
Hamiltonians,
{\it i.e.} $\beta(u)=1$. Define the integral-matrix operator $K_0$ using
the matrix $U$  defined by the formula (\ref{3.136})
\begin{equation} \begin{array}{c}
K_0=\partial^{-1}_x U_x= \int\limits_0^x dx
\left( \begin{array}{cc} u_x & \alpha^2(\rho) \rho_x \\
\rho_x & u_x \end{array} \right) = \\
= \int\limits_{(0,0)}^{(u,\rho)} U_u du + U_\rho
d\rho=\partial^{-1}_u I +
\partial^{-1}_\rho\left( \begin{array}{cc} 0 & \alpha^2(\rho)  \\
1 & 0 \end{array} \right).
 \end{array}
\label{4.171}
\end{equation}
Here the integral with respect to $x$ is taken at a constant value of $t$ and
the integral operator in the second line is the curvilinear integral
not depending on the integration path in the $(u,\rho)$ plane.

{\bf Theorem 4.2.}  For $\beta(u)=1$ the generalized gas dynamics
Hamiltonians of the series (\ref{4.163}), (\ref{4.164})
and their combinations $H^{(N)}(c_1,c_2)$ defined by the formula
(\ref{4.162}) satisfy the recursion relation
\begin{equation}
\left(H^{(N+1)}_\rho,H^{(N+1)}_u\right)^T=K_0
\left(H^{(N)}_\rho,H^{(N)}_u\right)^T
\label{4.172}
\end{equation}
for $N=0,1,2,\dots$ with the recursion operator $K_0$ defined by the
formula (\ref{4.171}).

The solution of the equation (\ref{4.172}) is given by the formula
\begin{equation}
\left(H^{(N)}_\rho,H^{(N)}_u\right)^T=K_0^N
\left(H^{(0)}_\rho,H^{(0)}_u\right)^T=K_0^N(c_1,c_2)^T
\label{4.173}
\end{equation}
where $H^{(N)}=H^{(N)}(c_1,c_2).$

For arbitrary functions $\beta(u),\alpha(\rho)$ we define the matrices
\begin{equation}
U_1=\left( \begin{array}{cc}
u & \partial^{-1}_\rho \alpha^2(\rho) \\
\rho & \partial^{-1}_u \beta^2(u) \end{array} \right), \quad
U_2=\left( \begin{array}{cc}
 \partial^{-1}_u \beta^2(u) & \partial^{-1}_\rho \alpha^2(\rho)
\\
\rho & u  \end{array}  \right)
\label{4.174}
\end{equation}
and the matrix-integral operators
\begin{equation} \begin{array}{c}
K_1=\partial^{-1}_x U_{1x}=\int\limits_0^x dx \left(
\begin{array}{cc}
u_x &\alpha^2(\rho)  \rho_x \\
\rho_x & \beta^2(u)  u_x
\end{array}
\right) = \\
=\int\limits_{(0,0)}^{(u,\rho)} U_{1u} du + U_{1\rho } d\rho =
\partial^{-1}_u
\left( \begin{array}{cc}
 1 & 0 \\
0 &  \beta^2(u)  \end{array}  \right) + \partial^{-1}_\rho
\left( \begin{array}{cc}
 0 & \alpha^2(\rho)\\
1 &  0 \end{array}  \right),
\end{array}
\label{4.175}
\end{equation}
\begin{equation} \begin{array}{c}
K_2=\partial^{-1}_x U_{2x}=\int\limits_0^x dx \left(
\begin{array}{cc}
 \beta^2(u)  u_x &\alpha^2(\rho)  \rho_x \\
\rho_x & u_x\end{array}  \right) = \\
=\int\limits_{(0,0)}^{(u,\rho)} U_{2u} du + U_{2\rho } d\rho =
\partial^{-1}_u
\left( \begin{array}{cc}
 \beta^2(u) & 0 \\
0 &  1  \end{array}  \right) + \partial^{-1}_\rho
\left( \begin{array}{cc}
0 & \alpha^2(\rho)\\
1 &  0  \end{array}  \right)
\end{array}
\label{4.176}
\end{equation}
with the operator $\partial^{-1}_x$ defined as in the equation (\ref{4.171}).
Here again in the second line we have curvilinear integrals which do
not depend on the integration path.

Define also the operator
\begin{equation}
K=K_1 K_2=\partial^{-1}_x U_{1x} \partial^{-1}_x U_{2x}.
\label{4.177}
\end{equation}

{\bf Theorem 4.3.} The separable Hamiltonians (\ref{4.159}), (\ref{4.160})
and their combinations $H^{(N)}(c_1,c_2)$ satisfy the recursion relation
\begin{equation}
(H^{(N+2)}_\rho,H^{(N+2)}_u)^T=K (H^{(N)}_\rho,H^{(N)}_u)^T
\label{4.178}
\end{equation}
for $N=0,1,2,\dots$ with the recursion operator $K$ defined by the equation
(\ref{4.177}).

This equation has the solutions
\begin{equation}
(H^{(2m)}_\rho,H^{(2m)}_u)^T=K^m
(H^{(0)}_\rho,H^{(0)}_u)^T=K^m(c_1,c_2)^T
\label{4.179}
\end{equation}
for $m=0,1,2,\dots$ and
\begin{equation}
(H^{(2m-1)}_\rho,H^{(2m-1)}_u)^T = K^{m-1}
(H^{(1)}_\rho,H^{(1)}_u)^T
= K^{m-1} U_1
(c_1,c_2)^T
\label{4.180}
\end{equation}
for $m=1,2,\dots$  where $H^{(N)}=H^{(N)}(c_1,c_2).$

Thus, all the Hamiltonians of the Manin's series (\ref{4.159}), (\ref{4.160})
are generated by the  recursion operator (\ref{4.177}) (see \cite{sh86}).

Consider now the inverse recursion operator for these Hamiltonians
\begin{equation}
K^{-1}=K^{-1}_2 K^{-1}_1 =(U_{2x})^{-1} D_x (U_{1x})^{-1} D_x
\label{4.181}
\end{equation}
with the matrices
\begin{equation}
(U_{1x})^{-1}= \dfrac{1}{\beta^2 u^2_x -\alpha^2 \rho^2_x }
\left( \begin{array}{cc} \beta^2 u_x  &  -\alpha^2 \rho_x \\ -\rho_x
&  u_x
\end{array} \right),
\label{4.182}
\end{equation}
\begin{equation}
(U_{2x})^{-1}= \dfrac{1}{\beta^2 u^2_x -\alpha^2 \rho^2_x }
\left( \begin{array}{cc} u_x  &  -\alpha^2 \rho_x \\ -\rho_x  &
\beta^2 u_x
\end{array} \right).
\label{4.183} \
\end{equation}

Define the second order matrix-differential operator
\begin{equation}
L=D_x K^{-1} \partial^{-1}_x = D_x (U_{2x})^{-1} D_x (U_{1x})^{-1}.
\label{4.184}
\end{equation}

Consider the Lie equations for Lie-B\"{a}cklund symmetries of the
Hamiltonian system (\ref{4.149}) with the Hamiltonian $H$ and the
Hamiltonian matrix $\hat{H}$
\begin{equation}
\vec{u}{\rho}_{\tau}=
\vec{ f(x,t,u,\rho,u_x,\rho_x,\dots,u_x^{(N)},\rho_x^{(N)})}
{ g(x,t,u,\rho,u_x,\rho_x,\dots,u_x^{(N)},\rho_x^{(N)})} \equiv
\vec{f_n}{g_n}
\label{4.185}
\end{equation}
assuming that $x_{\tau}=t_{\tau}=0.$ These symmetries satisfy the determining
equation
\begin{equation}
(I D_t -D_x\hat{H}) \vec{f}{g}=\vec{0}{0}\,.
\label{4.186}
\end{equation}

{\bf Theorem 4.4.}  Any separable Hamiltonian system (\ref{4.149}) possesses
the second order
matrix-differential recursion operator of the form (\ref{4.184})
which satisfies the recursion relation
\begin{equation}
L(f_n,g_n)^T =(f_{n+2},g_{n+2})^T \qquad (n=2,3,\dots).
\label{4.187} \
\end{equation}

{\tt Corollary 4.1.}  On the solution manifolds of the systems (\ref{4.149})
the following equality is satisfied
\begin{equation}
[ID_t -D_x \hat{H}, L] = 0.
\label{4.188}
\end{equation}

Thus, we obtain the Lax representation for these systems
\begin{equation}
\partial L / \partial t = [L,A]
\label{4.189}
\end{equation}
with the Lax pair of the Ibragimov-Shabat type \cite{ibrshab80a,ibrshab80b}
where $A= -D_x \hat{H}$ and the operator $ L$ is defined by the equation
(\ref{4.184}).

{\tt Corollary 4.2.}  For generalized gas dynamics equations we must put
$\beta(u)=1$,
$U_1=U_2=U$ and the second order recursion operator (\ref{4.184}) is equal to
the squared first order recursion operator $L_0$
\begin{equation}
L=L_0^2, \qquad
L_0=D_x (U_x)^{-1}.
\label{4.190}
\end{equation}
The operator $L_0$ coincides with the recursion operator (\ref{3.109})
transformed from the
Riemann invariants $s,r$ to the separable variables $u, \rho$  and
satisfies the relation
\begin{equation}
L_0(f_n,g_n)^T =(f_{n+1},g_{n+1})^T \qquad (n=2,3,\dots).
\label{4.191}
\end{equation}

{\bf Theorem 4.5.}  For the separable Hamiltonian systems (\ref{4.149}) with
the generic functions $\beta(u),\alpha(\rho)$ the first order recursion
operator does not exist and the second order recursion operator (\ref{4.184})
is not reduced to a square of a first order recursion operator or to a
product of different first order operators.

\subsection{Hydrodynamic symmetries and higher symmetries}

{\bf Theorem 4.6}  Separable systems (\ref{4.149}) with the Hamiltonian
$H(u,\rho)$ which do not depend explicitly on $t$ possess an infinite set
of homogeneous in derivatives hydrodynamic symmetries. All such
symmetries with or without an explicit dependence upon $x,t$ are
generated by the Lie equations
\begin{equation}
\vec{u}{\rho}_\tau = (xI + t \hat{H} + \hat{h})\vec{u}{\rho}_x
\label{4.192}
\end{equation}
or
\begin{equation}
\vec{u}{\rho}_\tau = \hat{h}\vec{u}{\rho}_x \equiv \sigma_1
D_x
\vec{h_u(u,\rho)}{h_\rho(u,\rho)}
\label{4.193}
\end{equation}
respectively.
Here $h(u,\rho)$ is an arbitrary smooth solution of the equation
(\ref{4.157}), $\hat{H}$ and $\hat{h}$ are Hamiltonian matrices of the form
(\ref{4.148}).

{\tt Corollary 4.3.} All the hydrodynamic symmetries (\ref{4.193}) without
an explicit dependence on $x,t$ are separable Hamiltonian systems with the
Hamiltonians $h(u,\rho)$ which mutually commute and also commute with the
Hamiltonian $H(u,\rho)$ of the system (\ref{4.149}).

Thus, any separable Hamiltonian system (\ref{4.149}) is included into the
infinite hierarchy of commuting separable Hamiltonian flows.

{\bf Theorem 4.7.}  Hydrodynamic symmetries (\ref{4.193}) of the system
(\ref{4.149}) without an explicit dependence on $x,t$ form an invariant
subspace for the second order recursion operator $L$. This operator generates
the recursion relation for the hydrodynamic symmetries
\begin{equation}
L \hat{h} \vec{u}{\rho}_x =\sigma_1 D_x
\vec{h_{1u}(u,\rho)}{h_{1\rho}(u,\rho)} \equiv
\hat{h}_1\vec{u}{\rho}_x
\label{4.194}
\end{equation}
preserving the Hamiltonian structure of the Lie equations (\ref{4.193})
and generates the recursion of their Hamiltonians
\begin{equation}
h_{1}(u,\rho)=\dfrac{1}{\beta^2(u)}h_{uu}(u,\rho)=\dfrac{1}{\alpha^2
(\rho)} h_{\rho \rho}(u,\rho).
\label{4.195}
\end{equation}

{\tt Corollary 4.4.} The hydrodynamic symmetries (\ref{4.193}) of the
separable Hamiltonian system (\ref{4.149}) which do not depend explicitly on
$x,t$ form an infinite-dimensional commutative Lie-B\"{a}cklund algebra. Its
elements have a functional arbitrariness, {\it i.e.} depend on an arbitrary
smooth solution $h(u,\rho)$ of the wave equation (\ref{4.157}).

Denote by $h_m(u,\rho)$ the result of $m$-fold application of the
transformation (\ref{4.195}) to $h(u,\rho)$
\begin{equation}
h_m(u,\rho)=(\beta^{-2}(u) \partial^2 / \partial u^2)^m h(u,\rho)
= (\alpha^{-2}(\rho) \partial^2 / \partial \rho^2)^m h(u,\rho).
\label{4.196}
\end{equation}
Then the formula (\ref{4.194}) gives the result
\begin{equation}
L^m \hat{h} \vec{u}{\rho}_x = \hat{h}_m\vec{u}{\rho}_x.
\label{4.197}
\end{equation}

{\bf Theorem 4.8.} For any separable Hamiltonian system (\ref{4.149}) all
its higher symmetries of the even order $n=2m$ which are generated by the
action of the recursion operators $L^m$ upon the hydrodynamic symmetries
(\ref{4.192}) are given by the formula
\begin{equation}
\vec{f_{2m}}{g_{2m}}=L^m x \vec{u}{\rho}_x + ( t\hat{H}_m+
\hat{h}_m) \vec{u}{\rho}_x
\label{4.198}
\end{equation}
for $m=1,2,\dots$ . Here $H_m(u,\rho) $ is obtained by the
transformation (\ref{4.196}) out of $H(u,\rho).$  For $m=1$ this formula
gives all second order symmetries of the equations (\ref{4.149}).

{\bf Theorem 4.9.} The separable system (\ref{4.149}) with the Hamiltonian
$H(u,\rho)$ possesses an infinite series of higher symmetries of even
orders $2m$ without the explicit dependence on $t,x$ if for some integer
$N$ the Hamiltonian $H$ meets the condition
\begin{equation}
L^N \hat{H} \vec{u}{\rho}_x = \vec{0}{0} \; \Longleftrightarrow
\; \hat{H}_N = 0
\label{4.199}
\end{equation}
{\it i. e.} the vector $\hat{H}(u_x,\rho_x)^T$ belongs to the kernel of the
operator $L^N,$ and then $m$ is larger or equal to $N.$ These
symmetries are determined by the formula (\ref{4.198}) with $\hat{H}_m = 0$
and all of them are common symmetries of all the systems (\ref{4.149})
with the right-hand side belonging to the kernel of the operator $L^N.$

For $\hat{h}_m = 0$ we obtain the special form of higher symmetries
subject to the condition (\ref{4.199}) with $m \geq N$
\begin{equation}
\vec{u}{\rho}_\tau= L^m x \vec{u}{\rho}_x =
\vec{f_{2m}(u,\rho,u_x,\rho_x,\dots,u_x^{(2m)},\rho_x^{(2m)})}%
{g_{2m}(u,\rho,u_x,\rho_x,\dots,u_x^{(2m)},\rho_x^{(2m)})}.
\label{4.200}
\end{equation}

For $\beta(u)=1$ we have $L=L_0^2$ where $L_0$ is the first order
recursion operator (\ref{4.190}). Then the equations (\ref{4.149}) form the
GGD system of the section 4.2 and have the higher symmetries of any order
$n\geq 2$ larger or equal to $N$
\begin{equation}
\vec{u}{\rho}_\tau= L^n_0 x \vec{u}{\rho}_x =
\vec{f_n}{g_n}
\label{4.201}
\end{equation}
if the condition
\begin{equation}
L_0^N \hat{H}\vec{u}{\rho}_x = \vec{0}{0}
\label{4.202}
\end{equation}
is met.

For the generic function $\beta(u)$ we shall denote by $\hat{X}_h$ and
$\hat{X}_{2m}$ the canonical Lie-B\"{a}cklund operators of the hydrodynamic
symmetries (\ref{4.193}) and of the higher symmetries (\ref{4.200})
respectively.

{\bf Theorem 4.10.} Hydrodynamic and higher symmetries of the separable
Hamiltonian systems (\ref{4.149}) which do not depend explicitly on $x$ and
$t$ form the infinite-dimensional noncommutative Lie-B\"{a}cklund algebra
\begin{equation} \begin{array}{c}
\left[ \hat{X}_h, \; \hat{X}_{\tilde{h}} \right] = 0 , \qquad
\left[ \hat{X}_{2m}, \; \hat{X}_h \right] = \hat{X}_{h_m}, \\
\left[ \hat{X}_{2m}, \; \hat{X}_{2n} \right] = 0
\end{array}
\label{4.203}
\end{equation}
in which the hydrodynamic symmetries $\hat{X}_h$ form an infinite-dimensional
commutative ideal.

{\tt Corollary 4.5.}  Higher and hydrodynamic symmetries of the
theorem 4.10 commute iff $\hat{h}_m=0$ where $h(u,\rho)$ are the
Hamiltonians of the hydrodynamic symmetries (\ref{4.193}).

The theorem 4.9 poses a problem of obtaining the kernel of the operator
$L^N$ to solve the equation (\ref{4.199}). In order to obtain its general
solution we have to allow an explicit dependence of the Hamiltonians on
$t$ and use the Manin's series (\ref{4.159}), (\ref{4.160}) of
Hamiltonians $H^{(N)}(1-\nu,\nu)$ with $\nu=0,1.$

We denote
\begin{equation} \begin{array}{c}
H^{[N]}(u,\rho,t)= \sum\limits_{k=0}^N [a_{4(N-k)+1}(t)
H^{(2k)}(1,0) + a_{4(N-k)+2}(t)H^{(2k)}(0,1) + \\
+  a_{4(N-k)+3}(t) H^{(2k-1)}(1,0) + a_{4(N-k)+4}(t)H^{(2k-1)}(0,1)]
\\
(N=1,2,\dots)
\end{array}
\label{4.204}
\end{equation}
with arbitrary smooth functions $a_i(t).$

Consider the Hamiltonian system (\ref{4.149}) with the
 Hamiltonians $H^{[N]} \; (N=1,2,\dots)$ explicitly dependent on $t$
\begin{equation}
\vec{u}{\rho}_t=\sigma_1 D_x
\vec{H^{[N]}_u(u,\rho,t)}{H^{[N]}_\rho(u,\rho,t)} \equiv
\hat{H}^{[N]}(u,\rho,t)\vec{u}{\rho}_x\,.
\label{4.205}
\end{equation}

{\bf Theorem 4.11.}  The kernel of the operator $L^N$ coincides with the
right-hand side of the Hamiltonian system (\ref{4.205})
\begin{equation}
L^{-N} \vec{0}{0} = \hat{H}^{[N]}(u,\rho,t)\vec{u}{\rho}_x
\label{4.206}
\end{equation}
where
\begin{equation}
L^{-1}= U_{1x} D_x^{-1} U_{2x} D_x^{-1}.
\label{4.207}
\end{equation}

{\tt Corollary 4.6.}  Let $N$ be some integer. Any separable Hamiltonian
system which has higher symmetries (\ref{4.198}) with $\hat{H}_m=0$
(with no explicit dependence on $t,x$) of the order $2m\ge 2N$
has the form (\ref{4.205}). In particular, such a system has the
higher symmetries (\ref{4.200}) which meet the condition $m \geq N.$

{\bf Theorem 4.12.}  Any separable Hamiltonian system (\ref{4.149}) with the
explicitly $t$-dependent Hamiltonian $H(u,\rho,t)$ possesses an
infinite set of the explicitly dependent on $t,x$ hydrodynamic
symmetries
\begin{equation}
\vec{u}{\rho}_\tau =\left[ x I + \int\limits_0^t
\hat{H}(u,\rho,t) dt + \hat{h}(u,\rho) \right] \vec{u}{\rho}_x
\label{4.208}
\end{equation}
with the functional arbitrariness determined by an arbitrary
smooth solution $h(u,\rho)$ of the wave equation (\ref{4.157}). Infinite
series of higher symmetries of any even order $2m$ for such a system is
generated by the action of recursion operators $L^m$ on the symmetries
(\ref{4.208})
\begin{equation}
\vec{u}{\rho}_\tau = L^m x \vec{u}{\rho}_x + \left[
\int\limits_0^t \hat{H}_m(u,\rho,t) dt + \hat{h}(u,\rho)
\right]\vec{u}{\rho}_x
\label{4.209}
\end{equation}
and have the same extent of arbitrariness as the symmetries (\ref{4.208}). In
the formulae (\ref{4.208}) and (\ref{4.209}) the integration with respect to
$t$ is performed at constant values of $u,\rho$.

\subsection{Invariant solutions and linearization}

We are looking for series of exact solutions of any separable Hamiltonian
system (\ref{4.205}) that are invariant with respect to higher symmetries
(\ref{4.209}) which have no explicit dependence on $t,x$ due to the
condition $\hat{H}_m^{[N]}(u,\rho,t)=0$ for $m \geq N.$

Invariance conditions for these solutions have the form
\begin{equation}
L^m x \vec{u}{\rho}_x + \hat{h}(u,\rho) \vec{u}{\rho}_x =
\vec{0}{0}
\label{4.210}
\end{equation}
for $m=N,N+1,\dots$ with $h(u,\rho)$ satisfying the equation (\ref{4.157}).
For obtaining these solutions we suggest the method which uses the Lax
representation (\ref{4.188}) and the inverse recursion operator (\ref{4.207})
\cite{sh83,sh86}.

Define the Hamiltonian $H^{[m]}(u,\rho)$ by the formula (\ref{4.204}) with
$N=m$ and arbitrary constants $c_i$ substituted for functions $a_i(t)$
\begin{equation} \begin{array}{c}
H^{[m]}(u,\rho)=\sum\limits_{k=0}^m [ c_{4(m-k)+1} H^{(2k)}(1,0)
+ c_{4(m-k)+2} H^{(2k)}(0,1) +  \\
+ c_{4(m-k)+3} H^{(2k-1)}(1,0) + c_{4(m-k)+4} H^{(2k-1)}(0,1)].
\end{array}
\label{4.211}
\end{equation}

Let $H(u,\rho)$ denote an arbitrary solution of the wave equation
(\ref{4.156}).

{\bf Theorem 4.13.}  Invariant solutions of the system (\ref{4.205}) with the
Hamiltonian \newline $H^{[N]}(u,\rho,t )$ which meet the invariance condition
(\ref{4.210}) are given by the formula
\begin{equation}
\vec{H_{\rho u}(u,\rho)}{H_{ uu}(u,\rho)} = \vec{x}{0} +
\int\limits_0^t \vec{H^{[N]}_{\rho u}(u,\rho,t)}{H^{[N]}_{uu}
(u,\rho,t)} dt
\label{4.212}
\end{equation}
if the existence conditions are met for the implicit function
$(u(x,t),\rho(x,t))$ determined by the equation (\ref{4.212}).

This equation defines the linearizing transformation which reduces the
solution of the nonlinear separable Hamiltonian system (\ref{4.205}), which
is explicitly $t$-dependent, to the solution of the linear wave equation
(\ref{4.156}) for $H(u,\rho).$

{\tt Corollary 4.7.}
The formula
\begin{equation}
\vec{H^{[m]}_{\rho u}(u,\rho)}{H^{[m]}_{ uu}(u,\rho)} =
\vec{x}{0} + \int\limits_0^t \vec{H^{[N]}_{\rho u}(u,\rho,t)}
{H^{[N]}_{uu} (u,\rho,t)} dt
\label{4.213}
\end{equation}
for $m=N,N+1,\dots$ gives the infinite series of exact invariant
solutions of the equations (\ref{4.205}) subject to the condition
(\ref{4.210}) with $h(u,\rho)=0$, {\it i. e.} invariant with respect to the
higher symmetries (\ref{4.200}).  To obtain an explicit form of these
solutions, the expression (\ref{4.211}) for $H^{[m]}(u,\rho)$ must be
used.

In the formulae (\ref{4.212}) and (\ref{4.213}) integrations with respect to
$t$ are performed at constant values of $u$ and $\rho.$

{\tt Corollary 4.8.}  For the system (\ref{4.205}) with the Hamiltonian
$H^{[N]}(u,\rho)$ without an explicit dependence on $t$ the formula
(\ref{4.212}) gives the hodograph transformation
\begin{equation}
\vec{H_{\rho u}(u,\rho)}{H_{ uu}(u,\rho)} = \vec{x}{0} + t
\vec{H^{[N]}_{\rho u}(u,\rho)} {H^{[N]}_{uu} (u,\rho)}
\label{4.214}
\end{equation}
which interchanges the roles of independent and dependent variables
$(x,t)$ and $(u,\rho).$

{\it Remark 4.1.}
Explicitly $t$-dependent Hamiltonian system (\ref{4.205}) is not linearizable
by the common hodograph transformation (\ref{4.214}). The formula
(\ref{4.212}) linearizes all systems of the form (\ref{4.205}) and presents
a generalization of the hodograph transformation for explicitly
time-dependent systems.

{\bf Theorem 4.14.}  Consider any separable system of the form (\ref{4.149})
with explicitly $t$-dependent Hamiltonian $H(u,\rho,t)$ satisfying the
equation (\ref{4.156}). Its invariant solutions with respect to the
hydrodynamic symmetries (\ref{4.208}) are given by the equalities
\begin{equation} \begin{array}{c}
x + \int\limits_0^t H_{\rho u}(u,\rho,t)dt = h_{\rho u}(u,\rho),
\\
\int\limits_0^t H_{uu}(u,\rho,t)dt = h_{uu}(u,\rho) \;
\Longleftrightarrow \;
\int\limits_0^t H_{\rho\rho}(u,\rho,t)dt =
h_{\rho\rho}(u,\rho)
\end{array}
\label{4.215}
\end{equation}
where $h(u,\rho)$ is an arbitrary smooth solution of the linear equation
(\ref{4.157}). These formulae reduce the solution of any separable
time-dependent system (\ref{4.149}) to the solution of the linear equation
(\ref{4.157}) for $h(u,\rho).$ This is the linearizing transformation which
generalizes the hodograph transformation for explicitly time-dependent
separable Hamiltonian systems (\ref{4.149}).

\section{Semi-Hamiltonian equations}
\setcounter{equation}{0}

\subsection{Geometry of semi-Hamiltonian systems and \protect\newline
hydrodynamic symmetries}

Consider a system of quasi-linear first order equations homogeneous in
derivatives
\begin{equation}
u^i_t=v_i(u)u^i_x \qquad (i=1,2,\dots,n)
\label{5.216}
\end{equation}
with the diagonal $n \times n$  matrix $V(u)={\rm diag}(v_i(u)).$
In the rest of the article no summation on repeated indices will be assumed.
Here $ u=(u^1,u^2,\dots,u^n)$ is $n$-vector, $u^i$ are the Riemann invariants
\cite{rozhd}.

Lie-B\"{a}cklund symmetries of the $N$th order
of the system (\ref{5.216}) are generated by the Lie equations
\begin{equation}
u^i_\tau=\eta_i(x,t,u,u_x,\dots,u_x^{(N)}) \qquad (i=1,2,\dots,n)
\label{5.217}
\end{equation}
where $\tau$ is the group parameter and we assume that
$x_\tau=t_\tau=0.$  Denote $\eta=(\eta_1,\dots,\eta_n)^T.$

Consider the operators $D_x,D_t$ of the total derivatives with respect to
$x,t$ and for the calculation of $D_t$ use the system (\ref{5.216})
\begin{equation}
D_x= \pd{ }{x} +\sum\limits_{j=1}^n \left( u^j_x \pd{ }{u^j} +
\sum\limits_{k=1}^\infty u^{j(k+1)}_x \pd{ }{u_x^{j(k)}} \right),
\label{5.218}
\end{equation}
\begin{equation}
D_t= \pd{ }{t} +\sum\limits_{j=1}^n \left( v_j(u) u^j_x \pd{ }{u^j} +
\sum\limits_{k=1}^\infty D_x^k[v_j(u) u^j_x] \pd{ }{u_x^{j(k)}} \right)
\label{5.219}
\end{equation}
where $\partial /\partial x $ and $\partial /\partial t $ are calculated
at constant values of $u,u_x^{(k)}$.

The compatibility conditions $u^i_{\tau t}=u^i_{t \tau}$ of the systems
(\ref{5.216}) and (\ref{5.217}) take the form of the determining equation for
the symmetry characteristic $\eta$
\begin{equation}
\left( ID_t -V D_x -U_x\left(\pd{V}{U}\right)\right)[\eta]=0.
\label{5.220}
\end{equation}
Here $I$ is the unit matrix, $U_x={\rm diag}(u^i_x)$, $\partial V/\partial
U = (v_{i,u^j})$ is the Jacobian matrix of the mapping $ v= v(u).$

Hydrodynamic symmetries are obtained if we choose $N=1$ in the Lie equations
(\ref{5.217}). Here we consider only hydrodynamic symmetries
linear homogeneous in derivatives
\begin{equation}
u^i_\tau=W_i(u,x,t) u^i_x \qquad (i=1,2,\dots,n).
\label{5.221}
\end{equation}

Impose the condition
\begin{equation}
v_i(u) \neq v_j(u) \qquad (i \neq j).
\label{5.222}
\end{equation}
Define the symmetrical connection coefficients associated with the system
(\ref{5.216}) \cite{tsar85}
\begin{equation}
\Gamma^i_{ij}(u)=\Gamma^j_{ji}(u)=v_{i,u^j}(u)/(v_j-v_i)  \qquad  (i
\neq j),
\label{5.223}
\end{equation}
\begin{equation}
\Gamma^j_{ii}=-( g_{ii}/g_{jj})\Gamma^i_{ij} \quad (i \neq j), \qquad
\Gamma^i_{jk } =0 \quad (i \neq j \neq k \neq i).
\label{5.224}
\end{equation}

This connection is compatible with the non-degenerate diagonal metric
\begin{equation}
g_{ii}(u)=H^2_i(u)=e^{2 \Phi_i(u)},\qquad g_{ij}=0 \quad (i \neq j),\qquad
\det(g^{ij})\neq 0.
\label{5.225}
\end{equation}
Here $H_i(u)$ are Lam\'e coefficients.

The connection is determined by the metric
\begin{equation}
\Gamma_{ij}^i= (\ln \sqrt{g_{ii}})_{u^j}= (\ln H_i)_{u^j}=\Phi_{i,u^j}(u).
\label{5.226}
\end{equation}

Integrability conditions of the system (\ref{5.226})
$\Gamma_{ij,u^k}^i=\Gamma_{ik,u^j}^i$ with the use of the expressions
(\ref{5.223}) take the form of the Tsarev's conditions \cite{tsar85}
\begin{equation}
[v_{i,u^j} / (v_j-v_i)]_{u^k}=[v_{i,u^k} / (v_k-v_i)]_{u^j} \qquad
(i \neq j \neq k \neq i ).
\label{5.227}
\end{equation}

\underline{\bf Definition 5.1} (see \cite{tsar91}). A diagonal system
(\ref{5.216}) is called semi-Hamiltonian if its coefficients $v_i(u)$ meet
the conditions (\ref{5.222}) and (\ref{5.227}).

{\bf Theorem 5.1} (see \cite{tsar91}). The diagonal system (\ref{5.216})
which meets the condition (\ref{5.222}) possesses an infinite set of
the hydrodynamic symmetries without explicit dependence on $x,t$
\begin{equation}
u^i_\tau =w_i(u) u^i_x \qquad (i = 1,2,\dots,n)
\label{5.228}
\end{equation}
with a functional arbitrariness iff the system (\ref{5.216}) is
semi-Hamiltonian, {\it i. e.} the condition (\ref{5.227}) is satisfied.
These symmetries are generated by the Lie equations (\ref{5.228}) with the
coefficients $w_i(u)$ which form an arbitrary smooth solution of the linear
system
\begin{equation}
w_{i,u^j} =\Gamma^i_{ij}(w_j-w_i) \qquad (i \neq j).
\label{5.229}
\end{equation}

The coefficients $\Gamma^i_{ij}(u) $ are defined by the formula
(\ref{5.223}). All these symmetries mutually commute. The set of these
symmetries is locally parametrized by $n$ arbitrary functions $c_i(u^i)$ of
one variable.

{\bf Theorem 5.2} (see \cite{sh93a}).
Any semi-Hamiltonian system of the form (\ref{5.216}) possesses an infinite
set of explicitly dependent on $x,t$ hydrodynamic symmetries with the same
functional arbitrariness as in the theorem 5.1. These symmetries are
generated by the Lie equations
\begin{equation}
u^i_\tau = [w_i(u) + c(x + t v_i(u))] u^i_x  \qquad (i=1,2,\dots,n)
\label{5.230}
\end{equation}
with the coefficients $w_i(u)$ which form an arbitrary smooth solution of the
linear system (\ref{5.229}). Here $c$ is an arbitrary constant.

{\bf Theorem 5.3} (see \cite{tsar91}). Any semi-Hamiltonian system of the
form (\ref{5.216}) with the non-degenerate metric (\ref{5.225}) is a
Hamiltonian system iff the following components of the Riemann curvature
tensor vanish: $R^i_{jji}=0 \ (i \neq j)$. Then the curvature tensor vanishes
identically and the variables $u^i$ form an orthogonal curvilinear
coordinate system in a flat (pseudo-Euclidean) space.

{\tt Corollary 5.1.} For the semi-Hamiltonian system (\ref{5.216}) the
following components of the curvature tensor vanish
\begin{equation}
R^i_{jkl}=0 \ (i \neq j \neq k \neq l ),  \quad R^i_{ikj}\equiv
\Gamma^i_{ij,u^k}
-\Gamma^i_{ik,u^j}=0,
\label{5.231}
\end{equation}
\begin{equation} \begin{array}{c}
R^i_{jki}\equiv \Gamma^i_{ij,u^k} -[ \Gamma^i_{ik} \Gamma^k_{kj}  +
\Gamma^i_{ij}
(\Gamma^j_{jk}  -\Gamma^i_{ik}) ] =0. \\
 (i \neq j \neq k \neq i )
 \end{array}
\label{5.232}
\end{equation}

{\it Remark 5.1.} The equalities (\ref{5.232}) give compatibility conditions
for the linear system (\ref{5.229}) and they are equivalent to the Tsarev's
conditions (\ref{5.227}).

\subsection{Invariant solutions and linearization of semi-Hamiltonian
systems}

Consider explicitly dependent on $x,t$  hydrodynamic symmetries (\ref{5.230})
with $c=-1.$ Invariance conditions for solutions of the semi-Hamiltonian
system (\ref{5.216}) with respect to these symmetries with $u_x  \neq 0 $
have the form
\begin{equation}
w_i(u) = tv_i(u) + x \qquad (i=1,2, \dots , n).
\label{5.233}
\end{equation}

{\bf Theorem 5.4}  (see \cite{tsar91}). Let the functions $w_i(u)$ in the
equations (\ref{5.233}) form an arbitrary smooth
solution of the linear system (\ref{5.229}). Then any smooth solution
$u^i(x,t)$ of the system (\ref{5.233}) is a solution of the semi-Hamiltonian
system (\ref{5.216}). Vice versa, any solution $u^i(x,t)$ of the system
(\ref{5.216}) may be locally represented as a solution of the system
(\ref{5.233}) in the vicinity of such a point $(x_0,t_0)$ where the
condition $u^i_x(x_0,t_0) \neq 0$ is met for each value of $i=1,2,\dots, n.$

{\it Remark 5.2.}  The equalities (\ref{5.233}) determine a linearizing
transformation for the semi-Hamiltonian system (\ref{5.216}), {\it i.e.}
the solution of the nonlinear system (\ref{5.216}) is reduced to the solution
of the linear system (\ref{5.229}) for the functions $w_i(u).$ It generalizes
the classical hodograph transformation for the case of multi-component systems
and hence is called the generalized hodograph transformation \cite{tsar91}.
We have shown here the { \it group-theoretical origin of the linearizing
transformation}\/: every nonsingular solution of the system (\ref{5.216}) is
its invariant solution with respect to the hydrodynamic symmetries
(\ref{5.230}) and the extent of arbitrariness of the set of these symmetries
and of the set of the corresponding invariant solutions is the same as for
the general solution manifold of the system (\ref{5.216}), {\it i.e.} $n$
arbitrary functions $c_i(u^i)$ of one variable.

Hence to obtain explicit formulae for the invariant solutions of the system
(\ref{5.216}) one must solve the linear system (\ref{5.229}) with variable
coefficients.

\subsection{First order recursion operators}

By definition, a recursion operator $R$ maps any symmetry of the
semi-Hamiltonian system (\ref{5.216}) again into its symmetry, {\it i.e.} any
solution $\eta=(\eta_1,\dots,\eta_n)^T$ of the determining equation
(\ref{5.220}) is mapped again into the solution $R[\eta].$ For this to be
true it is sufficient for the operator $R$ to commute with the operator of
the determining equation
\begin{equation}
\left[ ID_t - VD_x -U_x \left( \pd{V}{U}\right), R \right] = 0
\label{5.234}
\end{equation}
on solution manifolds of the equations (\ref{5.216}) and (\ref{5.220}). If
$N$ is any integer, then $N$th-order recursion operator has the form
(\ref{2.73}).

For the systems (\ref{5.216}) it is convenient to search for recursion
operators in a slightly different form.

In particular, for the first order recursion operators we assume
\begin{equation}
R=(AD_x + B) U_x^{-1}
\label{5.235}
\end{equation}
where $A=A(u), B=B(u,u_x)$ are $n \times n$ matrices.

Define the functions $(i=1,2,\dots,n)$
\begin{equation}
S_i(u)=\sum\limits_{k=1}^n \Gamma^i_{ik}(u)c_k(u^k) + d_i(u^i)
\label{5.236}
\end{equation}
which depend on $2n$ functions $c_i(u^i),d_i(u^i)$ of one variable.

{\bf Theorem 5.5}  (see \cite{tesh}). For the semi-Hamiltonian system
(\ref{5.216}) there exists a first order recursion operator $R$ of the form
(\ref{5.235}) iff there exist $2n$ functions $c_i(u^i),d_i(u^i)$ of one
variable which meet the conditions
\begin{equation}
S_{i,u^j}(u)= \Gamma^i_{ij}(S_j - S_i) \qquad (i \neq j)
\label{5.237}
\end{equation}
with the functions $S_i(u)$ defined by the formula (\ref{5.236}). Matrix
elements of this recursion operator are given by the formula
\begin{eqnarray}
R_{ij}\!\!&\!\! =\!\!&\!\! \biggl[ \delta_{ij} c_j(u^j) \Bigl(D_x +
\sum\limits_{k=1}^n \Gamma^i_{ik}(u)  u^k_x \Bigr) \nonumber \\
 \mbox{} \!\!&\!\!+\!\!&\!\! \Gamma^i_{ij}(u) \bigl( c_j(u^j) u^i_x -
 c_i(u^i) u^j_x\bigr)\biggr] \left(\frac{1}{u^j_x}\right) + d_j(u^j)
 \delta_{ij}
\label{5.238}
\end{eqnarray}
where $\delta_{ij}$ is the Kroneker symbol.

{\bf Theorem 5.6.}  For the semi-Hamiltonian system (\ref{5.216})
homogeneous in derivatives hydrodynamic symmetries (\ref{5.228}) with no
explicit dependence on $x,t$ subject to the condition (\ref{5.229}) form the
invariant subspace for the recursion operator (\ref{5.238}). A restriction of
the operator $R$ to this subspace is determined as follows
\begin{equation}
R \left( \begin{array}{c} w_1(u)u^1_x \\ \dots \\ w_n(u)u^n_x
\end{array} \right) = \left( \begin{array}{c} \hat{w}_1(u)u^1_x \\
\dots \\ \hat{w}_n(u)u^n_x \end{array} \right)
\label{5.239}
\end{equation}
where the functions $\hat{w}_i(u)$ are determined by the formula
\begin{equation}
\begin{array}{cc}
\hat{w}_i(u) = & c_i(u^i) w_{i,u^i}(u) + d_i(u^i) w_i(u) + \\
 & + \sum\limits_{k=1}^n \Gamma^i_{ik}(u) c_k(u^k)w_k(u).
\end{array}
\label{5.240}
\end{equation}

{\tt Corollary 5.2.}  For any solution $\{ w_i(u) \}$ of the linear system
(\ref{5.229}) the functions $\hat{w}_i(u)$ form also a solution of this
system, {\it i.e.} the formula (\ref{5.240}) is a recursion for
solutions of the system (\ref{5.229}), iff the conditions (\ref{5.237}) are
met.

The first order recursion operator (\ref{5.238}) for the multi-component
system (\ref{5.216}) was constructed at first by a straightforward solution
of the equation (\ref{5.234}) \cite{tesh}. A new, more simple method for
constructing the recursion operators, based on the study of symmetries of
the set of hydrodynamic symmetries of the system (\ref{5.216}), was developed
by the author \cite{sh94a}. Thus, the group-theoretical origin of recursion
operators was discovered. A simple geometrical meaning of the existence
conditions (\ref{5.237}) for the first order recursion operators was also
clarified in this paper.

To present the last statement explicitly,
we consider another orthogonal curvilinear coordinates $\{ r^i\},$ which
will be specified later, with the Lam\'e coefficients
\begin{equation}
H_i(r)= \sqrt{g_{ii}(r)} = e^{\Phi_i(r)}.
\label{5.241}
\end{equation}

Consider the rotation coefficients $\beta_{ji}(r)$ of this coordinate system
defined by the equations \cite{egor,tsar91}
\begin{equation}
H_{i,r^j}=\beta_{ji}H_j \qquad (i \neq j).
\label{5.242}
\end{equation}
Let $\rho=\{ (r^k - r^l ) \}$ denote the set of coordinate differences.

{\bf Theorem 5.7} (see \cite{sh94a}).
First order recursion operator for the semi-Hamiltonian system (\ref{5.216})
exists iff the rotation coefficients $\beta_{ji}(r)$ of some curvilinear
orthogonal coordinate system $r^i$ depend only on the coordinate differences
$\rho$: $\beta_{ji}=\beta_{ji}(\rho).$

\subsection{Second order recursion operators}

For second order recursion operators we assume the form
\begin{equation}
R=(AD_x^2 + B D_x + F) U_x^{-1}
\label{5.243}
\end{equation}
where $A=A(u,u_x),\; B=B(u,u_x,u_{xx}), \; F=F(u,u_x,u_{xx},u_{xxx})$ are
$n \times n$ matrices.

Define the ''connection potential'' $V(u)$ by a completely integrable
(in the Frobenius sense) system
\begin{equation}
V_{u^i u^j}(u)= \Gamma^i_{ij}\Gamma^j_{ji} \qquad  (i \neq j).
\label{5.244}
\end{equation}
Its solution $V(u)$ depends upon $n$ arbitrary functions of one variable.
The integrability conditions for the system (\ref{5.244}) are met as a
consequence of the semi-Hamiltonian property (\ref{5.227}).

Define the functions
\begin{equation} \begin{array}{cc}
b_{ik}(u)= & f_k(u^k) \left[  \Gamma^i_{ik} (2\Gamma^k_{kk} -
\Gamma^i_{ik}  ) -  \Gamma^i_{ik,u^k}\right]  +
\\[2mm]
& + [ c_k(u^k) - f'_k(u^k)] \Gamma^i_{ik} \qquad (i \neq k),
\end{array}
\label{5.245}
\end{equation}
\begin{equation} \begin{array}{cc}
b_{ii}(u)= & f_i(u^i) \left[ \Gamma^i_{ii,u^i} +
(\Gamma^i_{ii})^2 - 2 V_{u^iu^i}(u) \right]  -
\\
& -  f'_i(u^i)V_{u^i}(u) + c_i(u^i) \Gamma^i_{ii} + d_i(u^i),
\end{array}
\label{5.246}
\end{equation}
\begin{equation}
B_i(u)=\sum\limits_{k=1}^n b_{ik}(u)
\label{5.247}
\end{equation}
which depend upon $3n$ functions $f_i(u^i),c_i(u^i),d_i(u^i)$ of one
variable.

Let $R_{ik},A_{ik},B_{ik},F_{ik}$ denote matrix elements of the operator $R$
and the matrices $A,B,F$ respectively.

\underline{\bf Definition 5.2.}  Let $H_i(u)$ and $\beta_{ij}(u)$ denote
Lam\'e coefficients and rotation coefficients of the curvilinear orthogonal
system $\{u^i \}$ and $G_i(\hat{u}^i)$ denote a function of $u$
independent of $u^i.$ Semi-Hamiltonian system (\ref{5.216}) is called a
generic system (with respect to the second order recursion operators) if none
of the following special cases is met for $i,j,k=1,2,\dots,n$
\begin{equation}
v_{i,u^i}(u)=0,
\label{5.248}
\end{equation}
\begin{equation}
v_{i,u^i}(u)= \left( F_i(u^i) G_i(\hat{u}^i) /g_{ii} \right) e^{V(u)},
\label{5.249}
\end{equation}
\begin{equation} \begin{array}{c}
[\beta_{ji}/(v_j - v_i)] [(\beta_{ki}/H_i)_{u^i}/v_{i,u^i}]_{u^i} = \\
= [\beta_{ki}/(v_k - v_i)][(\beta_{ji}/H_i)_{u^i}/v_{i,u^i}]_{u^i} \\
(  i \neq j \neq k \neq i ).
\end{array}
\label{5.250}
\end{equation}

{\bf Theorem 5.8} (see \cite{sh94a}).

For the semi-Hamiltonian generic system (\ref{5.216}) a second order
recursion operator $R$ of the form (\ref{5.243}) exists iff there exist $3n$
functions $f_i(u^i),c_i(u^i),d_i(u^i)$ of one variable which
meet the conditions
\begin{equation}
B_{i,u^j}(u)=\Gamma^i_{ij}(B_j - B_i) \qquad (i \neq j)
\label{5.251}
\end{equation}
with the functions $B_i(u)$ defined by the formulae
(\ref{5.245})--(\ref{5.247}).  Its matrix elements are given by the formulae
\begin{equation}
R_{ik}=(A_{ik} D^2_x + B_{ik} D_x + F_{ik})(1/u^k_x),
\label{5.252}
\end{equation}
\begin{equation}
A_{ik}= \delta_{ik} f_i(u^i)/u^i_x,
\label{5.253}
\end{equation}
\begin{equation}
B_{ik}=\Gamma^i_{ik} [ f_k(u^k) (u^i_x/u^k_x) - f_i(u^i) (u^k_x/u^i_x)]
\quad (i \neq k),
\label{5.254}
\end{equation}
\begin{equation} \begin{array}{cc}
B_{ii}= & - f_i(u^i)u^i_{xx} / [(u^i_x)^2] + 2[ f_i(u^i)/u^i_x]
\sum\limits_{j \neq i} \Gamma^i_{ij} u^j_x +
\\
& + 2f_i(u^i) \Gamma^i_{ii}  + c_i(u^i),
\end{array}
\label{5.255}
\end{equation}
\begin{equation} \begin{array}{c}
F_{ik}  = f_i(u^i)\Gamma^i_{ik}(u^k_x/u^i_x)
[(u^i_{xx}/u^i_x) - (u^k_{xx}/u^k_x)] + \\
+ f_k(u^k) [(u^i_x)^2/u^k_x] \Gamma^i_{ik}\Gamma^k_{ki} -
f_i(u^i)[(u^k_x)^2/u^i_x]\Gamma^i_{ik,u^k} + (\Gamma^i_{ik})^2] +
\\
+ \{ f_i(u^i)( \Gamma^i_{ik} \Gamma^k_{ki} - 2\Gamma^i_{ii,u^k}) -
[2f_i(u^i) \Gamma^i_{ii} + c_i(u^i)] \Gamma^i_{ik} \} u^k_x +
\\
+ b_{ik}(u) u^i_x - f_i(u^i) (u^k_x/u^i_x) \sum\limits_{j \neq i,k} u^j_x
(\Gamma^i_{ik} \Gamma^k_{kj} + \Gamma^i_{ij}\Gamma^j_{jk}) +
\\
+ u^i_x \sum\limits_{j \neq i,k} [f_k(u^k)(u^j_x/u^k_x)
\Gamma^i_{ik}\Gamma^k_{kj} - f_j(u^j) (u^k_x/u^j_x) \Gamma^i_{ij}
\Gamma^j_{jk}]\quad  (i \neq k),
\end{array}
\label{5.256}
\end{equation}
\begin{equation}
F_{ii}= B_i(u) u^i_x - \sum\limits_{k \neq i} F_{ik}.
\label{5.257}
\end{equation}

{\bf Theorem 5.9}  (see \cite{sh94a}). For the semi-Hamiltonian
system (\ref{5.216}) the homogeneous in derivatives hydrodynamic symmetries
(\ref{5.228}) with no explicit dependence on $x,t$, subject to the condition
(\ref{5.229}), form the invariant subspace for the second order recursion
operator (\ref{5.252}). Its action on this subspace is determined by the
same formula (\ref{5.239}) from the section 5.3 but with the different
definition of functions
\begin{equation} \begin{array}{c}
\hat{w}_i(u)= f_i(u^i)w_{i,u^i u^i} + [ 2f_i(u^i)\Gamma^i_{ii} + c_i(u^i)]
w_{i,u^i} +
\\
+ \sum\limits_{k \neq i} f_k(u^k)\Gamma^i_{ik} w_{k,u^k} +
\sum\limits_{k=1}^n b_{ik}(u)w_k
\end{array}
\label{5.258}
\end{equation}
where the functions $b_{ik}(u)$ are defined by the formulae (\ref{5.245}),
(\ref{5.246}).

{\tt Corollary 5.3.} For any solution $\{ w_i(u) \}$ of the linear system
(\ref{5.229}) the formula (\ref{5.258}) gives also a solution $\{
\hat{w}_i(u) \}$ of this system  iff the conditions (\ref{5.251}) are met.
Thus, the formula (\ref{5.258}) is a second order recursion for
solutions of the system (\ref{5.229}).

{\bf Theorem 5.10}  (see \cite{sh94a}). If the first order recursion
operator exists then there also exists the second order recursion operator
equal to the squared first order recursion operator. The inverse is not true,
{\it i.e.} the existence conditions (\ref{5.237}) for the first order
recursion operator do not follow from the existence conditions (\ref{5.251})
for the second order recursion operators.

This means that the existence conditions for the second order recursion
operator are less restrictive than for the first order operator.

\subsection{Generation of infinite series of exact solutions}

To obtain explicit formulae for invariant solutions of the system
(\ref{5.216}) one must search for solutions of the linear system
(\ref{5.229}) and substitute these solutions for the set of functions
$\{w_i(u)\}$ in the linearizing transformation (\ref{5.233}). Existence
of a recursion operator for the semi-Hamiltonian system (\ref{5.216}) is the
additional constraint which makes it possible to obtain particular solutions
of the system (\ref{5.229}).

The linear system (\ref{5.229}) has two trivial solutions
\begin{equation}
w_i=1, \quad w_i=v_i(u) \qquad (i=1,2,\dots,n).
\label{5.259}
\end{equation}

They serve as initial elements for the generation of infinite series of
nontrivial solutions by recursion operators.

In particular, assume that the first order recursion operator (\ref{5.238})
exists for the system (\ref{5.216}), {\it i.e.} the conditions (\ref{5.237})
are met. It generates the recursion (\ref{5.240}) for solutions of the system
(\ref{5.229})
\begin{equation}
\hat{w_i}(u)= (\hat{R}_1[w])_i= \sum\limits_{k=1}^n
(\hat{R}_1)_{ik}[w_k]
\label{5.260}
\end{equation}
with the first order recursion operator
\begin{equation}
(\hat{R}_1)_{ik}= \delta_{ik} \left[ c_i(u^i) \pd{ }{u^i} + d_i(u^i) \right]
+ \Gamma^i_{ik}(u) c_k(u^k).
\label{5.261}
\end{equation}

Two trivial solutions (\ref{5.259}) are mapped by the operator $\hat{R}_1$ to
the nontrivial solutions of the system (\ref{5.229})
\begin{equation}
\hat{w}_i(u)=(\hat{R}_1[1])_i = \sum\limits_{k=1}^n \Gamma^i_{ik}(u)
c_k(u^k) + d_i(u^i) \equiv S_i(u),
\label{5.262}
\end{equation}
\begin{equation} \begin{array}{cc}
\hat{w}_i(u)=(\hat{R}_1[v])_i = & c_i(u^i) v_{i,u^i}(u) + d_i(u^i)v_{i}(u) +
\\
& + \sum\limits_{k=1}^n \Gamma^i_{ik}(u) c_k(u^k)v_{k}(u).
\end{array}
\label{5.263}
\end{equation}

Substituting these expressions for $w_i(u)$ to the equations (\ref{5.233})
we obtain the explicit formulae for nontrivial solutions of the system
(\ref{5.216})
\begin{equation}
\sum\limits_{k=1}^n \Gamma^i_{ik}(u)c_k(u^k) + d_i(u^i)= t v_i(u) + x,
\label{5.264}
\end{equation}
\begin{equation} \begin{array}{c}
c_i(u^i) v_{i,u^i}(u) + d_i(u^i)v_{i}(u) + \sum\limits_{k=1}^n
\Gamma^i_{ik}(u) c_k(u^k)v_{k}(u) = \\
= t v_i(u) + x
\qquad (i=1,2,\dots,n).
\end{array}
\label{5.265}
\end{equation}
These equalities determine the exact solutions $u^i=u^i(x,t)$ of the system
(\ref{5.216}) as implicit functions.

The action by powers of the operator $\hat{R}_1$ on the trivial solutions
(\ref{5.259}) generates the explicit formulae for two infinite series of
exact invariant solutions with $N=1,2,\dots$
\begin{equation}
(\hat{R}_1^N[1])_i= t v_i(u) + x,
\label{5.266}
\end{equation}
\begin{equation}
(\hat{R}_1^N[v])_i= t v_i(u) + x.
\label{5.267}
\end{equation}

Assume now that the less restrictive existence conditions (\ref{5.251})
for the second order recursion operator are met. Then there exists the
recursion (\ref{5.258}) for solutions of the system (\ref{5.229})
\begin{equation}
\hat{w}_i(u)=(\hat{R}_2[w])_i = \sum\limits_{k=1}^n
(\hat{R}_2)_{ik}[w_k]
\label{5.268}
\end{equation}
with the second order recursion operator
\begin{equation} \begin{array}{cc}
(\hat{R}_2)_{ik}= & \delta_{ik}  \left\{  f_i(u^i)
 \dfrac{\partial^2}{(\partial u^i)^2} + \left[ f_i(u^i) \Gamma^i_{ii}(u)
 + c_i(u^i) \right] \pd{}{u^i} \right\}  +
\\
& + f_k(u^k)\Gamma^i_{ik}(u)\pd{}{u^k} + b_{ik}(u).
\end{array}
\label{5.269}
\end{equation}
Here the functions $b_{ik}(u)$ are defined by the formulae (\ref{5.245}),
(\ref{5.246}).

The trivial solutions (\ref{5.259}) are mapped by the operator $\hat{R}_2$ to
the nontrivial solutions of the system (\ref{5.229}) and via the equations
(\ref{5.233}) to the corresponding solutions of the system (\ref{5.216})
\begin{equation} \begin{array}{c}
(\hat{R}_2[1])_{i} \equiv \sum\limits_{k=1}^n b_{ik}(u) \equiv
B_i(u)=tv_i(u)+x,
\\ [2mm]
(\hat{R}_2[v])_{i} \equiv f_i(u^i) v_{i,u^iu^i}(u) + \left[
2f_i(u^i)\Gamma^i_{ii}(u)  + c_i(u^i) \right] v_{i,u^i}(u) +
\\
+ \sum\limits_{k \neq i} f_k(u^k)\Gamma^i_{ik}(u) v_{k,u^k}(u) +
\sum\limits_{k=1}^n b_{ik}(u)v_k(u) =tv_i(u)+x.
\end{array}
\label{5.270}
\end{equation}
 The powers $\hat{R}_2^N$ generate two infinite series of exact invariant
solutions out of initial solutions (\ref{5.259}) with $N=1,2,\dots$
\begin{equation}
(\hat{R}_2^N[1])_i=tv_i(u)+x,
\label{5.271}
\end{equation}
\begin{equation}
(\hat{R}_2^N[v])_i=tv_i(u)+x.
\label{5.272}
\end{equation}
We can also use linear combinations of solutions of the two series for the
functions $w_i(u)$ in the equations (\ref{5.233})
\begin{equation}
c_1(\hat{R}_{1,2}^N[1])_i + c_2(\hat{R}_{1,2}^M[v])_i=tv_i(u)+x
\label{5.273}
\end{equation}
with any integers $N,M,$ the operator $\hat{R}_{1}$ or $\hat{R}_{2}$ and
arbitrary constants $c_1,c_2.$

Further generalization is obtained if we substitute for the characteristic
$\eta^d=x+ t v_i(u)$ of the dilatation symmetry group in the formulae
(\ref{5.230}), (\ref{5.233}) the results of the action on $\eta^d$ of the
operators $\hat{R}_{1}^L$ or $\hat{R}_{2}^L$ where $L$ is any integer. Then
we obtain the formula
\begin{equation}
c_1(\hat{R}_{1,2}^N[1])_i + c_2(\hat{R}_{1,2}^M[v])_i =
t(\hat{R}_{1,2}^L[v])_i + x (\hat{R}_{1,2}^L[1])_i.
\label{5.274}
\end{equation}

In the cases when $L>N$ and (or) $L>M$ this formula is equivalent to using
negative powers of the operators $\hat{R}_{1}$ or $\hat{R}_{2}$
in the equations (\ref{5.273}).

{\it Remark 5.3.} The operators $\hat{R}_{1}$ and $\hat{R}_{2}$ coincide
with the first and second order symmetry operators for the linear system
(\ref{5.229}) essential for the separation of variables in these equations
(see \cite{mill,bagr}).  Solution of the system (\ref{5.229}) by means of the
separation of variables would mean solving completely the original nonlinear
system (\ref{5.216}). Therefore the linearizing transformation (\ref{5.233})
may be considered as the extension of the method of separation of variables
to nonlinear systems "rich in symmetries".

\subsection{Higher symmetries of semi-Hamiltonian systems}

{\bf Theorem 5.11} (see \cite{sh93a}). All second order symmetries of the
semi-Hamiltonian system (\ref{5.216}) are generated by the second order
recursion operator (\ref{5.252}) out of the hydrodynamic symmetries
(\ref{5.230}) (with $c=1$). The corresponding Lie equations have the form
\begin{equation} \begin{array}{c}
u^i_\tau= \eta_{2,i} \equiv \sum\limits_{j=1}^n R_{ij} \big[ u^j_x
\big(x + t v_j(u) + w_j(u)\big)\big] \equiv
\\
\equiv - f_i(u^i)u^i_{xx}/[(u^i_{x})^2] +
\big(f_i(u^i)/u^i_{x} \big) \sum\limits_{j \neq i}
\Gamma^i_{ij}u^j_x +
\\
+ u^i_x \sum\limits_{j \neq i} \Gamma^i_{ij} f_j(u^j)/u^j_{x} +
2f_i(u^i)\Gamma^i_{ii} + c_i(u^i) + u^i_x \big[ x B_i(u) + \\
+ t (\hat{R}_{2}[v])_i + \hat{w}_i(u) \big] \qquad (i=1,2,\dots,n)
\end{array}
\label{5.275}
\end{equation}
where the set of functions $\{ \hat{w}_i(u)\}$ is an arbitrary solution of
the linear system (\ref{5.229}). The existence conditions (\ref{5.251}) for
the second order symmetries and for the second order recursion operators
coincide and must be met by a choice of functions
$f_i(u^i),c_i(u^i),d_i(u^i).$ In the case when $B_i(u) \neq 0$ and
$(\hat{R}_{2}[v])_i \neq 0$ these symmetries explicitly depend on $x,t.$

The action of the powers $R^N$ of the recursion operator with $N=1,2,\dots$
on the same hydrodynamic symmetries generates the infinite series of higher
symmetries of the system (\ref{5.216})
\begin{equation}
u^i_\tau= \sum\limits_{j=1}^n (R^N)_{ij}  \big[ u^j_x\big(x + t v_j(u) +
w_j(u)\big)\big].
\label{5.276}
\end{equation}
 If $R$ is a second order recursion operator, then all these symmetries are
of the even order $2N.$

\section{Multi-component diagonal systems explicitly \protect\newline
dependent on $t$ or $x$}
\setcounter{equation}{0}

\subsection{Hydrodynamic symmetries of $t$-dependent systems}

We consider the first order quasi-linear diagonal system of explicitly
$t$-dependent equations with $n \geq 3$
\begin{equation}
u^i_t= v_i (u,t) u^i_x ,\qquad i=1,2,\dots,n
\label{6.277}
\end{equation}
subject to the condition $v_i \neq v_j$ for $i \neq j.$ We search for
hydrodynamic symmetries of this system with the Lie equations
\begin{equation}
 u^i_\tau= \sum\limits_{j=1}^n A^i_j(u,t,x) u^j_x ,\qquad i=1,2,\dots,n
\label{6.278}
\end{equation}
and $x_\tau=t_\tau=0$.

 Define the functions
\begin{equation}
c_{ij}(u,t)= v_{i,u^j}(u,t) / (v_j - v_i) \qquad (i \neq j),
\label{6.279}
\end{equation}
\begin{equation}
\Gamma^i_{ij}(u)=c_{ij}(u,0).
\label{6.280}
\end{equation}

{\bf Theorem 6.1} (see \cite{sh89a,sh89b,grshwi}). Diagonal $n$-component
system (\ref{6.277}) of the hy\-dro\-dy\-namic-type  with an explicit
time dependence possesses an infinite set of hydrodynamic symmetries of the
form (\ref{6.278}) with a functional arbitrariness iff its coefficients meet
the condition (\ref{5.227}) and the condition
\begin{equation}
\big[v_{i,u^j}(u,t) / (v_i(u,t) - v_j(u,t)) \big]_t = \beta v_{i,u^j}(u,t)
\qquad (i \neq j)
\label{6.281}
\end{equation}
with an arbitrary real constant $\beta.$  These symmetries are
generated by the Lie equations
\begin{equation}
 u^i_\tau= A^i(u,t,x) u^i_x ,\qquad i=1,2,\dots,n
\label{6.282}
\end{equation}
with the coefficients $A_i$ defined by the formulae

for $\beta \neq 0$
\begin{equation}
 A_i(u,t,x)= a_i(u)\exp\left\{ \beta \left[x + \int\limits_0^t v_i(u,t)
dt \right] \right\} + C
\label{6.283}
\end{equation}

and for $\beta=0$
\begin{equation}
 A_i(u,t,x)= a_i(u) + C \left[x + \int\limits_0^t v_i(u,t) dt \right]
\label{6.284}
\end{equation}
where $C$ is an arbitrary constant and the integrations with respect to $t$
are performed at a constant value of $u.$ Here the set of functions $\{
a_i(u)\}$ is an arbitrary smooth solution of the linear system
\begin{equation}
a_{i,u^j}(u)= \Gamma^i_{ij}(u)(a_j - a_i) \qquad (i \neq j)
\label{6.285}
\end{equation}
with the coefficients $\Gamma^i_{ij}$ defined by the formulae (\ref{6.279}),
(\ref{6.280}). The solution manifold of the system (\ref{6.285}) depends upon
$n$ arbitrary functions $c_i(u^i)$ of one variable which locally parametrize
the set of hydrodynamic symmetries.

\subsection{Hydrodynamic symmetries of $x$-dependent systems}

Let coefficients of the diagonal system of $(n \geq 3)$ equations
 explicitly depend on the coordinate $x$
\begin{equation}
u^i_t=\tilde{v}_i(u,x) u^i_x ,\qquad i=1,2,\dots,n
\label{6.286}
\end{equation}
and meet the condition $\tilde{v}_i \neq \tilde{v}_j \ (i \neq j).$

Define the functions
\begin{equation}
\tilde{c}_{ij}(u,x)=\tilde{v}_{i,u^j}(u,x) /(\tilde{v}_j-\tilde{v}_i)
\qquad (i \neq j),
\label{6.287}
\end{equation}
\begin{equation}
\Gamma^i_{ij}(u)=\tilde{c}_{ij}(u,0)\tilde{v}_j(u,0)/\tilde{v}_i(u,0).
\label{6.288}
\end{equation}

{\bf Theorem 6.2}  (see \cite{sh89a,sh89b,grshwi}). Diagonal $n$-component
system (\ref{6.286}) of the hydrodynamic type  with an explicit
$x$-dependence possesses an infinite set of hydrodynamic symmetries of the
form (\ref{6.278}) with a functional arbitrariness iff its coefficients meet
the condition (\ref{5.227}) with a substitution of $\tilde{v}_i$ for $v_i$
and the condition
\begin{equation}
\left[ \big(\tilde{v}_{i}^{-1}(u,x)\big)_{u^j} / \big(\tilde{v}_{i}^{-
1}(u,x) - \tilde{v}_{j}^{-1}(u,x)\big)\right]_x= \beta\big(\tilde{v}_{i}^{-
1}(u,x)\big)_{u^j} \qquad (i \neq j)
\label{6.289}
\end{equation}
with an arbitrary real constant $\beta.$ These symmetries are generated by
the Lie equations
\begin{equation}
 u^i_\tau= \tilde{A}_i(u,t,x) u^i_x ,\qquad i=1,2,\dots,n
\label{6.290} \
\end{equation}
with the coefficients $\tilde{A}_i$ defined by the formulae

for $\beta \neq 0$
\begin{equation}
 \tilde{A}_i(u,t,x)= \tilde{v}_i(u,x)\left\{a_i(u)\exp \left[\beta
\left( t + \int\limits_0^x \tilde{v}^{-1}_i(u,x) dx\right) \right] +
C\right\}
\label{6.291}
\end{equation}

and for $\beta=0$
\begin{equation}
 \tilde{A}_i(u,t,x)= \tilde{v}_i(u,x)\left\{a_i(u) + C \left[t +
\int\limits_0^x \tilde{v}^{-1}_i(u,x) dx \right] \right\}
\label{6.292}
\end{equation}
where $C$ is an arbitrary constant and the integrations with respect to $x$
are performed at a constant value of $u.$ The set of functions $\{ a_i(u)\}$
is an arbitrary smooth solution of the linear system (\ref{6.285}) with the
coefficients $\Gamma^i_{ij}$ defined by the formulae (\ref{6.287}),
(\ref{6.288}). The extent of arbitrariness is the same as in the theorem
6.1.

\subsection{Invariant solutions and linearization}

Equations determining the invariant solutions of the systems (\ref{6.277})
and (\ref{6.286}) subject to the constraint $u^i_x \neq 0$ for
$i=1,2,\dots,n$ are obtained from the Lie equations (\ref{6.282}) and
(\ref{6.290}) with the invariance condition $u^i_\tau=0.$ They have
the form $A_i(u,t,x)=0$ and $\tilde{A}_i(u,t,x)=0$ with the functions $A_i$
and $\tilde{A}_i$ defined by the formulae (\ref{6.283}), (\ref{6.284}) and
(\ref{6.291}), (\ref{6.292}) respectively where we put $c=1$ without the loss
of generality. Hence we obtain

for the system (\ref{6.277})
\begin{equation}
a_i(u) + \exp\left\{ -\beta \left[x + \int\limits_0^t v_i(u,t) dt
\right] \right\} =0 \qquad (\beta \neq 0),
\label{6.293} \
\end{equation}
\begin{equation}
 a_i(u) + x + \int\limits_0^t v_i(u,t) dt =0 \qquad (\beta=0),
\label{6.294}
\end{equation}

and for the system (\ref{6.286})
\begin{equation}
a_i(u) + \exp \left[-\beta \left( t + \int\limits_0^x \tilde{v}^{-
1}_i(u,x) dx\right) \right] =0  \qquad (\beta \neq 0),
\label{6.295}
\end{equation}
\begin{equation}
a_i(u) + t + \int\limits_0^x \tilde{v}^{-1}_i(u,x) dx=0 \qquad (\beta=0)
\label{6.296}
\end{equation}
where $i= 1,2,\dots,n.$

Here the set of functions $a_i(u)$ is an arbitrary smooth solution of the
linear system (\ref{6.285}). Thus, the equations (\ref{6.293})--(\ref{6.296})
determine a linearizing transformation for the systems (\ref{6.277}),
(\ref{6.286}), {\it i.e.} the solution of these systems reduces to the
solution of the linear system (\ref{6.285}). These equations determine the
solutions $u^i=u^i(x,t)$ of the original nonlinear system if the conditions
of the implicit-function theorem are met.

  More complete results on the diagonal systems with explicit $t$ or $x$
dependence and the example of a new integrable system of this class can
be found in the recent paper \cite{grshwi}.

\section{Conclusions}

 The existence of an infinite-dimensional group of the hydrodynamic
symmetries for the equations of the hydrodynamic type is an important
property which provides the existence of linearizing transformations. The
reason for this is that the
degree of generality of the set of symmetries coincides with the degree of
generality of the general solution set for these equations. Therefore,
almost all solutions are the invariant solutions and they are obtained
by standard formulae provided the symmetries are already determined. Such a
formula gives a linearizing transformation reducing the original non-linear
problem to the linear problem of determining the symmetries. The additional
property is the existence of the recursion operator which makes it possible
to solve partially the linear problem by constructing infinite series of
its solutions and hence solutions of the original non-linear equations.
The existence of such an operator also has a group-theoretical basis,
since the recursion operator is completely determined by the symmetries
of the determining equations for the hydrodynamic symmetries, {\it i.e.}
by the `symmetries of symmetries".

  This shows a group-theoretical origin of linearizing transformations and
of the integrability property by which we mean a possibility to construct
infinitely many exact solutions. The Hamiltonian structure, if it exists,
does not improve the integrability properties of the equations of the
hydrodynamic type. We have to conclude that the symmetry is the major
necessary property that insures the integrability which was the original idea
of S.~Lie.


\begin{thebibliography}{99}
\bibitem{dubnov83}
B.A. Dubrovin  and S.P. Novikov,
Hamiltonian formalism of one-dimensional systems of hydrodynamic type and the
Bogolyubov-Whitham averaging method, {\it Soviet Math. Dokl.} {\bf 27},
665--669 (1983).
\bibitem{dubnov89}
B.A. Dubrovin  and S.P. Novikov,
Hydrodynamics of weakly deformed soliton lattices. Differential geometry and
Hamiltonian theory, {\it Russ. Math Surveys} {\bf 44}, 35--104 (1989).
\bibitem{tsar85}
S.P. Tsarev,
On Poisson brackets and one--dimensional systems of hydrodynamic type,
{\it Soviet Math. Doklady} {\bf 31}, 488--491 (1985).
\bibitem{tsar91}
S.P. Tsarev,
Geometry of Hamiltonian systems of hydrodynamic type.
Generalized hodograph method, {\it Math. in the USSR Izvestiya} {\bf 37},
397--419 (1991).
\bibitem{rozhd}
B.L. Rozhdestvenskii and N.N. Yanenko,
{\it Systems of Quasilinear Equations and their Applications to Gas
Dynamics} (Amer. Math. Soc., Providence, RI, 1983).
\bibitem{olvnut}
P.J. Olver and Y. Nutku,
Hamiltonian structures for systems of hyperbolic conservation laws,
{\it J.Math.Phys.} {\bf 29}, 1610--1619 (1988).
\bibitem{pavl}
M.V. Pavlov,
Hamiltonian formalism of the electrophoresis equation. Integrable equations
of hydrodynamics (preprint No.~17, Landau Inst. Theoret. Phys., Moscow, 1987)
(in Russian).
\bibitem{fertsar}
E.V. Ferapontov and S.P. Tsarev,
Equations of hydrodynamic type arising in gas chromatography. Riemann
invariants and exact solutions, {\it Mat. Modelirovanie} {\bf 3}, No.~2, 82
(1991) (in Russian).
\bibitem{gumnut}
H. G\"{u}mral and Y. Nutku,
Multi-Hamiltonian structure of equations of hydrodynamic type,
{\it J. Math. Phys.} {\bf 31}, 2606--2611 (1990).
\bibitem{anov}
M. Arik, F. Neyzi, Y. Nutku, P.J. Olver, and J.M. Verosky,
Multi-Hamiltonian structure of the Born-Infeld equation,{\it J. Math. Phys.}
{\bf 30}, 1338--1344 (1989).
\bibitem{men'}
O.F. Men$'$shikh,
On the Cauchy problem and boundary value problems
for a class of systems of quasilinear hyperbolic equations,
{\it Soviet Math. Dokl.} {\bf 42}, 346--350 (1991).
\bibitem{sh93a}
M.B. Sheftel,
Higher integrals and symmetries of semi-Hamiltonian systems,
{\it Differentsial'nye Uravneniya,} {\bf 29}, No.~10, 1782--1795 (1993)
(in Russian); English translation in {\it Differential
Equations,} {\bf 29} No.~10, 1548 (1993).
\bibitem{sh93b}
M.B. Sheftel,
Lie-B\"{a}cklund group analysis of two-component systems of hydrodynamic type,
in {\it Modern Group Analysis and Problems of Mathematical Modeling,}
194--198 (Samara University, Samara, 1993).
\bibitem{sh94a}
M.B. Sheftel,
Group analysis of determining equations --- a method for searching out
recursion operators, {\it Differentsial'nye Uravneniya,}
{\bf 30}, No.~3, 444--456 (1994) (in Russian).
\bibitem{sh94b}
M.B. Sheftel,
{\it Symmetries, Recursions and Integrals for Hydrodynamic-Type Systems,}
Dr. Phys.-Math. Sci. Thesis, Tomsk State University, Tomsk (1994)
(in Russian).
\bibitem{sh82}
M.B. Sheftel,
On the Lie-B\"{a}cklund groups admissible by the equations of one-dimensional
gas dynamics, {\it Vestnik of Leningrad State Univ.} No.~7, 37--41 (1982)
(in Russian).
\bibitem{sh83}
M.B. Sheftel,
On the infinite-dimensional noncommutative Lie-B\"{a}cklund
algebra associated with the equations of one-dimensional gas dynamics,
{\it Theor. Math. Phys.} {\bf 56}, 878--891 (1983).
\bibitem{sh86}
M.B. Sheftel,
Integration of Hamiltonian systems of hydrodynamic type with
two dependent variables with the aid of the Lie-B\"{a}cklund group,
{\it Func. Anal. Appl.} {\bf 20}, 227--235 (1986).
\bibitem{sh89a}
M.B. Sheftel,
Group analysis and linearization of quasilinear systems, in
{\it Modern Group Analysis,} 244--250 (Elm, Baku, 1989) (in Russian).
\bibitem{sh89b}
M.B. Sheftel,
Linearization of homogeneous quasilinear systems by the method
of group-analysis , differential geometry and Hamiltonian formalism, in {\it
Lie-B\"{a}cklund groups and Quasilinear Systems,} (LIIAN, preprint No.~106,
Ch.~1, 4--24, Leningrad, 1989) (in Russian).
\bibitem{sh95}
M.B. Sheftel, Generalized hydrodynamic-type systems, in
{\it CRC Handbook of Lie Group Analysis of Differential Equations}
Vol.~{\bf 3}, Ch.7, 169--189 (Boca Raton, CRC Press, 1995).
\bibitem{grshwi}
A.M. Grundland, M.B. Sheftel, and P. Winternitz,
Invariant solutions of hydrody\-na\-mic-type equations, {\it J. Phys. A: Math.
Gen.} {\bf 33}, 8193--8215 (2000).
\bibitem{andibr}
R.L. Anderson and N.H. Ibragimov
{\it Lie--B\"{a}cklund Transformations in Applications} (SIAM, Philadelphia,
1979).
\bibitem{ibreng}
N.H. Ibragimov,
{\it Transformation Groups Applied to Mathematical Physics,} (Reidel,
Dodrecht--Boston, 1985) (Russian version see \cite{ibrus}).
\bibitem{fuchfok}
B. Fuchssteiner and A.S. Fokas,
Symplectic structures, their B\"{a}cklund transformations and hereditary
symmetries, {\it Physica D,} {\bf 4}, 47--66 (1981).
\bibitem{tesh}
V.M. Teshukov,
Hyperbolic systems, which admit nontrivial Lie--B\"{a}cklund groups, in
{\it Lie--B\"{a}cklund  Groups and Quasilinear Systems},
(LIIAN preprint No.~106, 25--30, Leningrad, 1989) (in Russian).
\bibitem{ovs}
L.V. Ovsiannikov, {\it Lectures on Foundations of Gas Dynamics,}
(Nauka, Moscow, 1981) (in Russian).
\bibitem{lax}
P.D. Lax,
Integrals of nonlinear equations of evolution and solitary waves,
{\it Comm. Pure Appl. Math.} {\bf 21}, 467--490 (1968).
\bibitem{ibrshab80a}
N.H. Ibragimov and A.B. Shabat,
Infinite Lie--B\"{a}cklund algebras, {\it Funct. Anal.Appl.} {\bf 14}, No.~4,
313--315 (1980).
\bibitem{ibrshab80b}
N.H. Ibragimov and A.B. Shabat,
Evolution equations with non-trivial Lie-B\"{a}cklund group,
{\it Func. Anal. Appl.} {\bf 14}, No.~1, 19--28 (1980).
\bibitem{mikhshab}
A.V. Mikhailov, A.B. Shabat, and V.V. Sokolov,
The symmetry approach to classification of integrable equations, in
{\it What is Integrability?}\/, Springer Lecture Notes in Nonlinear Dynamics,
V.E. Zakharov, Ed., 115--184 (Springer-Verlag, New York, 1991).
\bibitem{manin}
Yu.I. Manin,
Algebraic aspects of nonlinear differential equations, {\it J. Soviet Math.}
{\bf 11}, 1--122 (1979).
\bibitem{egor}
D.F. Egorov,
{\it Works on Differential Geometry,} (Nauka, Moscow, 1970) (in Russian).
\bibitem{mill}
W. Miller, Jr.
{\it Symmetry and Separation of Variables,} (Addison-Wesley, Reading, MA,
1977).
\bibitem{bagr}
V.G. Bagrov, B.F. Samsonov, A.V. Shapovalov, and I.V. Shirokov,
Commutative subalgebras of symmetry operators of the wave equation including
one second order operator and separation of variables (preprint No.~27,
Tomsk Scientific Center of the Siberian Branch of Academy of Sciences of USSR,
Tomsk, 1990).
\bibitem{ibrus}
N.H. Ibragimov,
{\it Transformation Groups Applied to Mathematical Physics,} (Nauka, Moscow
1983) (Engl.transl. see \cite{ibreng}).
\end{thebibliography}
\end{document}